\documentclass[11pt,a4paper,openright,fleqn]{book}

\newcommand{\Ra}{\mathrm{Ra}}
\newcommand{\Pran}{\mathrm{Pr}}
\newcommand{\be}{\begin{equation}}
\newcommand{\ee}{\end{equation}}
\newcommand{\ba}{\begin{array}}
\newcommand{\ea}{\end{array}}
\newcommand{\bae}{\begin{eqnarray}}
\newcommand{\eae}{\end{eqnarray}}
\newcommand{\bse}{\begin{subeqnarray}}
\newcommand{\ese}{\end{subeqnarray}}

\newcommand{\parti}{\partial}

\usepackage[dvips]{graphicx}
\usepackage[dvipdfm,
            pdfstartview=FitH,
            CJKbookmarks=true,
            bookmarksnumbered=true,
            bookmarksopen=true,
            colorlinks=true, 
            pdfborder=001,   
            citecolor=blue%
            ]{hyperref}

\usepackage{color}      
\usepackage{subfigure}  
\usepackage{floatflt}   
\usepackage{rotating}   
\usepackage{flafter}    
\usepackage[below]{placeins}    
\usepackage{pstricks}  

\usepackage{rotating}   
\usepackage{longtable}  
\usepackage{tabls}
\usepackage{multirow}   
\usepackage{colortbl}   
\usepackage{dcolumn}    
\usepackage{hhline}     
\usepackage{slashbox}   

\usepackage[top=1.7cm,bottom=2cm,left=2.5cm,right=2.1cm,includehead,includefoot%
            ]{geometry}                 

\usepackage{indentfirst}                
\usepackage[perpage,symbol]{footmisc}   
\usepackage{fancyhdr}                   
\usepackage{lastpage}                   
\usepackage{tabularx}

\usepackage{bm}         
\usepackage{amsmath}    
\usepackage{amssymb}    
\usepackage[amsmath,thmmarks]{ntheorem} 
\usepackage{subeqnarray} 

\usepackage[sort&compress,numbers]{natbib}  

\begin{document}







\setlength{\parindent}{2em}                 
\setlength{\parskip}{3pt plus1pt minus1pt}  
\renewcommand{\baselinestretch}{1.2}        

\setlength{\topsep}{3pt plus1pt minus2pt}           
\setlength{\partopsep}{3pt plus1pt minus2pt}        
\setlength{\itemsep}{3pt plus1pt minus2pt}          
\setlength{\floatsep}{5pt plus 2pt minus 2pt}      
\setlength{\textfloatsep}{10pt plus 2pt minus 2pt}  
\setlength{\intextsep}{5pt plus 2pt minus 2pt}  
\setlength{\abovecaptionskip}{2pt plus1pt minus1pt} 
\setlength{\belowcaptionskip}{3pt plus1pt minus1pt} 
\setlength{\extrarowheight}{3pt}

\renewcommand{\textfraction}{0.15}
\renewcommand{\topfraction}{0.85}
\renewcommand{\bottomfraction}{0.65}
\renewcommand{\floatpagefraction}{0.60}

\setlength{\mathsurround}{2pt}

\setlength{\abovedisplayskip}{7pt plus3pt minus1pt}     
\setlength{\belowdisplayskip}{7pt plus3pt minus1pt}     
\setlength{\arraycolsep}{2pt}   

\allowdisplaybreaks[4]  

\renewcommand{\mathbf}[1]{\boldsymbol{#1}}  

\newcommand{\me}{\mathrm{e}}  
\newcommand{\mi}{\mathrm{i}}
\newcommand{\mj}{\mathrm{j}}
\newcommand{\dif}{\mathrm{d}}

\renewcommand\chaptername{Chapter\space \arabic{chapter}}


\setcounter{secnumdepth}{3}


\renewcommand{\thesection}{\S\hspace{0.2em}\arabic{chapter}.\arabic{section}}
\renewcommand{\thesubsection}{\S\hspace{0.2em}\arabic{chapter}.\arabic{section}.\arabic{subsection}}
\renewcommand{\thesubsubsection}{\arabic{subsubsection}.}

\let\oldtitle=\title
\def\title#1{\oldtitle{\cnbf{#1}}}


\renewcommand\labelenumi{\textcircled{\scriptsize \theenumi}}  
\renewcommand\labelenumii{(\theenumii)}
\renewcommand\labelenumiii{\theenumiii.}
\renewcommand\labelenumiv{\theenumiv.}

\renewcommand{\thetable}{\arabic{chapter}-\arabic{table}}
\renewcommand{\theequation}{\arabic{chapter}-\arabic{equation}}
\renewcommand{\thefigure}{\arabic{chapter}-\arabic{figure}}

\setcounter{tocdepth}{3} \setcounter{secnumdepth}{3}

\newcommand{\makeheadrule}{%
    \makebox[-3pt][l]{\rule[.7\baselineskip]{\headwidth}{0.4pt}}
    \rule[0.85\baselineskip]{\headwidth}{1.5pt}\vskip-.8\baselineskip}
\makeatletter
\renewcommand{\headrule}{%
    {\if@fancyplain\let\headrulewidth\plainheadrulewidth\fi
     \makeheadrule}}
\makeatother

\pagestyle{fancyplain}

\renewcommand{\chaptermark}[1]%
{\markboth{\chaptername \ #1}{}}            
\renewcommand{\sectionmark}[1]%
{\markright{\thesection \ #1}{}}            
\fancyhf{}  


\fancyhead[CO]{\color{green}\small\rightmark}
\fancyhead[CE]{\color{red}\small\leftmark}

\fancyfoot[CE,CO]{\thepage}

\fancypagestyle{plain} {
\fancyhead{}                                    
\renewcommand{\headrulewidth}{0pt}
\fancyfoot{}                                    
\fancyfoot[CE,CO]{\thepage} }

\theoremstyle{plain}
\theoremheaderfont{\normalfont\rmfamily\CJKfamily{hei}}
\theorembodyfont{\normalfont\rm\CJKfamily{song}} \theoremindent0em
\theoremseparator{\hspace{1em}} \theoremnumbering{arabic}
\newtheorem{definition}{\hspace{2em}¶¨Òå}[chapter]
\newtheorem{proposition}{\hspace{2em}ÃüÌâ}[chapter]
\newtheorem{property}{\hspace{2em}ÐÔÖÊ}[chapter]
\newtheorem{lemma}{\hspace{2em}ÒýÀí}[chapter]
\newtheorem{theorem}{\hspace{2em}¶¨Àí}[chapter]
\newtheorem{axiom}{\hspace{2em}¹«Àí}[chapter]
\newtheorem{corollary}{\hspace{2em}ÍÆÂÛ}[chapter]
\newtheorem{exercise}{\hspace{2em}Ï°Ìâ}[chapter]
\theoremsymbol{$\blacksquare$}
\newtheorem{example}{\hspace{2em}Àý}[chapter]

\theoremstyle{nonumberplain}
\theoremheaderfont{\CJKfamily{hei}\rmfamily}
\theorembodyfont{\normalfont \rm \CJKfamily{song}}
\theoremindent0em \theoremseparator{\hspace{1em}}
\theoremsymbol{$\blacksquare$}
\newtheorem{proof}{\hspace{2em}Ö¤Ã÷}
\newtheorem{Floquet}{\hspace{2em}Floquet¶¨Àí}

\newcommand{\upcite}[1]{\textsuperscript{\textsuperscript{\cite{#1}}}}  

\renewcommand{\thefootnote}{\arabic{footnote}}



\newsavebox{\AphorismAuthor}
\newenvironment{Aphorism}[1]
{\vspace{0.2cm}\begin{sloppypar} \slshape
\sbox{\AphorismAuthor}{#1}
\begin{quote}\small\itshape }
{\\ \usebox{\AphorismAuthor}
\end{quote}
\end{sloppypar}\vspace{0.5cm}}


\makeatletter
\def\hlinewd#1{%
  \noalign{\ifnum0=`}\fi\hrule \@height #1 \futurelet
   \reserved@a\@xhline}
\makeatother
\newcommand\vlinewd[1][1pt]{\vrule width #1}

%

\newcounter{saveeqn}%

\newcommand{\alpheqn}{%
\setcounter{saveeqn}{\value{equation}}%
\stepcounter{saveeqn}%
\setcounter{equation}{0}%
\renewcommand{\theequation}{\arabic{chapter}-\arabic{saveeqn}\alph{equation}}}

\newcommand{\reseteqn}{%
\setcounter{equation}{\value{saveeqn}}%
\renewcommand{\theequation}{\arabic{chapter}-\arabic{equation}}}  

\newcounter{savefig}%

\newcommand{\alphfig}{%
\setcounter{savefig}{\value{figure}}%
\stepcounter{savefig}%
\setcounter{figure}{0}%
\renewcommand{\thefigure}{\arabic{chapter}-\arabic{savefig}\alph{figure}}}

\newcommand{\resetfig}{%
\setcounter{figure}{\value{savefig}}%
\renewcommand{\thefigure}{\arabic{chapter}-\arabic{figure}}}  

\newcounter{savetab}%

\newcommand{\alphtab}{%
\setcounter{savetab}{\value{table}}%
\stepcounter{savetab}%
\setcounter{table}{0}%
\renewcommand{\thetable}{\arabic{chapter}-\arabic{savetab}\alph{table}}}

\newcommand{\resettab}{%
\setcounter{table}{\value{savetab}}%
\renewcommand{\thetable}{\arabic{chapter}-\arabic{table}}}  

\newcounter{newlist} 
\newenvironment{mylist}[1][¿É¸Ä±äµÄÁбíÌâÄ¿]{
\begin{list}{\textbf{\hei #1} ({\hei\arabic{newlist}})\,} 
    {
     \usecounter{newlist}
     \setlength{\labelwidth}{22pt} 
     \setlength{\labelsep}{0cm} 
     \setlength{\leftmargin}{0cm} 
     \setlength{\rightmargin}{0cm}
     \setlength{\parsep}{0.5ex plus0.2ex minus0.1ex} 
     \setlength{\itemsep}{0ex plus0.2ex} 
     \setlength{\itemindent}{36pt} 
     \setlength{\listparindent}{18pt} 
    }}
{\end{list}}

\begin{titlepage}
\thispagestyle{empty}
\renewcommand{\baselinestretch}{2}

\newpage
\thispagestyle{empty}
\hspace{5cm}  
\vspace{1.5cm}
\begin{center}
{\bf\huge Stabilities of Parallel Flow \\
and Horizontal Convection }

\vskip 2cm {\bf\Large Liang SUN}

\vskip 2cm {\em\Large Report of Postdoc Research}

\vskip 2cm {\bf\Large
\begin{tabular}{rl}
Academic Advisor£º & Prof. Yun-Fei Fu \\
\end{tabular}
}

\vskip 4cm
{\bf\large LSRSCE, School of Earth and Space Sciences, \\
University of Science and Technology of China,\\
 November, 2007}
\end{center}
\end{titlepage}


\sloppy

\chapter*{Abstract\markboth{Abstract}{Abstract}}
\addcontentsline{toc}{chapter}{Abstract}

In the first part, the stability of two-dimensional parallel flow
is discussed. A more restrictively general stability criterion for
inviscid parallel flow is obtained analytically. First, a
sufficient criterion for stability is found as either
$-\mu_1<\frac{U''}{U-U_s}<0$ or $0<\frac{U''}{U-U_s}$ in the flow,
where $U_s$ is the velocity at the inflection point, and $\mu_1$
is the eigenvalue of Poincar\'{e}'s problem. Second, this
criterion is generalized to barotropic geophysical flows in the
$\beta$ plane. Based on the stability criteria, the necessary
condition for wave-mean flow interaction is also obtained.

Then, the general stability criteria of two-dimensional inviscid
rotating flow with angular velocity $\Omega(r)$ are obtained
analytically. First, a necessary instability criterion for
centrifugal flows is derived as $\xi'(\Omega-\Omega_s)<0$ (or
$\xi'/(\Omega-\Omega_s)<0$) somewhere in the flow field, where
$\xi'$ is the vortictiy of profile and $\Omega_s$ is the angular
velocity at the inflection point $\xi'=0$. Second, a criterion for
stability is found as
$-(\mu_1+1/r_2)<f(r)=\frac{\xi'}{\Omega-\Omega_s}<0$, where
$\mu_1$ is an eigenvalue. The new criteria are the analogues of
the criteria for parallel flows, which are special cases of
Arnol'd's nonlinear criteria. Specifically, Pedley's cirterion is
proved to be an special case of Rayleigh's criterion. Moreover,
the criteria for parallel flows can also be derived from those for
the rotating flows. The analogy between rotation and
stratification in inviscid flow is also addressed. These results
extend the previous theorems and would intrigue future research on
the mechanism of hydrodynamic instability.

Besides, the essence of shear instability is fully revealed within
the linear context. The mechanism of shear instability is explored
by combining the mechanisms of both the Kelvin-Helmholtz
instability (K-H instability) and resonance of waves. The shear
instability requires both a concentrated vortex (with speed of
$U_s$) in the flow and resonant waves to interact with the
concentrated vortex. Physically, the standing waves (with phase
speed $c_r=U_s$) can interact with the concentrated vortex, so
they can trigger instability via K-H instability in the flows.
While the travelling waves (with $c_r\neq U_s$) have no
interaction with the concentrated vortex, so that they can not
trigger instability in the flows. The resonance of waves are
totally within the linear context.

In consequence, the above criteria would be helpful for
understanding the wave-mean flow interaction, especially the
Rossby wave-mean flow interaction in barotropic flows. According
to the stable criteria, the necessary condition for wave-mean flow
interaction can be obtained. And why the disturbed waves can't
take energy from the mean flow in the stable flow is revealed. If
the flow is stable, there is no wave-mean flow interaction at all.
This explains why the disturbed waves can't take energy from the
mean flow in the stable flow.

In the second part, we report the numerical simulations of the
partial-penetrating flow in horizontal convection within a squire
cavity tank at high Rayleigh numbers $10^7<Ra<10^{10}$. The
partial-penetrating flow was first reported in the experiment by
Wang and Huang (2005), which is though of an important material to
understand ocean circulation energy budget. The fast established
but slowly steadied flow is simulated, where a shallow and closed
circulation cell is obtained numerically as partial-penetrating
flow for the first time, which is consistent with the experiment.
As the partial-penetrating flow is shallow, it is seldom affected
by the bottom boundary. The depth of partial-penetrating
circulation satisfies minus 1/5-power law of Rayleigh number. The
larger the Rayleigh number is, the shallow the partial-penetrating
flow is. An objective definition of partial-penetrating is given
based on this power law. Then, further investigation points out
that the Prandtl number governs the partial-penetrating flows. As
$\Pran\geq 6$, there are always partial-penetrating flows. While
$\Pran\leq 4$, the flows tend to be full-penetrating.

Then, the horizontal convection at high Rayleigh number (Ra) in a
rectangle cavity with aspect ratio of $1:10$ is numerically
simulated. According to the results within the regime of $10^4
<Ra<10^{11}$, three continues regimes are obtained: linear regime
($10^4<Ra<10^6$), transition regime ($10^6<Ra<10^8$) and 1/5-power
law regime ($10^8<Ra<10^{11}$). For the flow strength, a 1/3-power
law of Ra is fitted when Ra is not high enough ($10^7<Ra<10^8$).
However, a 1/5-power law is obtained as Ra is high enough
($10^8<Ra<10^{11}$). The 1/5-power law confirms Rossby's analysis
and implies that 1/3-power law of Ra for Nusselt number by Siggers
et al. is over estimation.

Finally, the critical Rayleigh number for unset of the horizontal
convection is also addressed. The flow is found to be unsteady at
high Rayleigh numbers. There is a Hopf bifurcation of $Ra$ from
steady solutions to periodic solutions, and the critical Rayleigh
number $Ra_c$ is obtained as $Ra_c=5.5377\times 10^8$ for the
middle plume forcing at $Pr=1$, which is much larger than the
formerly obtained value. Besides, the unstable perturbations are
always generated from the central jet, which implies that the
onset of instability is due to velocity shear (shear instability)
other than thermally dynamics (thermal instability). Finally,
Paparella and Young's (2000) second hypotheses about the
destabilization of the flow is numerically proved, i.e. the middle
plume forcing can lead to a destabilization of the flow.

The report was supported by the National Foundation of Natural
Science (No. 40705027) and the National Science Foundation for
Post-doctoral Scientists of China (No. 20070410213).

\clearpage

\thispagestyle{empty} 

\makeatletter
    \renewcommand{\@dotsep}{0.8}
\makeatother

\setcounter{tocdepth}{2} \tableofcontents
\listoffigures 

\mainmatter  

\chapter{General Stability Criteria}\label{ch:GenStaCri}
\section{Two-Dimensional Problem}
\subsection{Introduction} The instability due to shear in the flow
is one of the fundamental and the most attracting problems in many
fields, such as fluid dynamics, astrophysical fluid dynamics,
oceanography, meteorology, etc. More generally, shear instability
is also referred as barotropic instability in geophysical flows,
where the gravitational and buoyancy effects are ignored. Shear
instability has been intensively investigated, which is to the
greatly helpful understanding of other instability mechanisms in
complex shear flows. Rayleigh investigated the growth of linear
disturbances by means of normal mode expansion, which leads to
Rayleigh's equation \cite[]{Rayleigh1880}. Using this equation,
Rayleigh first proved a necessary criterion for instability, i.e.,
Inflection Point Theorem, which is also called Rayleigh-Kuo
theorem \cite[e.g.][]{CriminaleBook2003} for Kuo's generalization
to barotropic geophysical flows in the $\beta$ plane
\cite[]{KuoHL1949}. Then, Fjortoft found a stronger necessary
criterion for instability \cite{Fjortoft1950}. Besides, Tollmien
gave a heuristic result that the criteria are also sufficient for
instability if the velocity profiles are the symmetrical or
monotonic \cite{Tollmien1935}. These criteria are well known and
have been widely used in various applications
\cite[e.g.][]{Drazin1981,Huerre1998,CriminaleBook2003}.

On the other hand,  Arnol'd considered the shear instability in a
totally different way
\cite[][]{Arnold1965a,Arnold1965b,ArnoldBook1998}. He studied the
conservation law of the inviscid flow via Euler's equations and
found two nonlinear stability theorems by means of variational
principle. Arnol'd's first stability theorem corresponds to Fj\o
rtoft's criterion \cite[]{Drazin1981,Dowling1995}. However,
Arnol'd's second nonlinear stability theorem, has no such
corresponding linear criterion. Though Arnol'd's second nonlinear
theorem is more useful in the geophysical flows
\cite[]{Dowling1995}, is seldom known by the scientists in other
fields. Dowling suggested that Arnol'd's idea should need to be
added to the general fluid-dynamics curriculum \cite{Dowling1995}.
Yet his suggestion has not been followed until now
\cite[e.g.][]{Drazin1981,Huerre1998,CriminaleBook2003,VallisAOFD2006},
since the proofs of Arnol'd's theorems are very advanced and
complex in mathematics.

The aim of this section is to find the elementary proofs for
Arnol'd's theorems, which could be used to teach undergraduate
students. As the variational method is too advanced and complex
for undergraduate students. The new proofs are obtained in a
totally different way, where the linear stability problem is
considered by using Rayleigh's equation.

\subsection{Stable criterion}

\begin{figure}
  \centerline{\includegraphics[width=6cm]{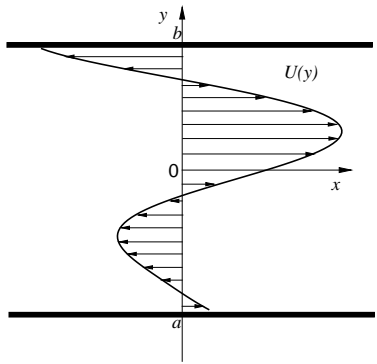}}
\caption{ Sketch of parallel flow.}
\label{Fig:stable_sketch_vorticity_profile}
\end{figure}

To find the criteria, Rayleigh's equation for an inviscid parallel
flow is employed
\cite[]{Rayleigh1880,Drazin1981,Huerre1998,SchmidBook2000,CriminaleBook2003},
which is the vorticity equation of the disturbance
\cite[]{Drazin1981,Huerre1998}. For a parallel flow with mean
velocity $U(y)$ in Fig.\ref{Fig:stable_sketch_vorticity_profile},
the vorticity is conserved along pathlines. The amplitude of
disturbed flow streamfunction, namely $\phi$, satisfies
 \begin{equation}
 (\phi''-k^2 \phi)-\frac{U''}{U-c}\phi=0,
 \label{Eq:stable_parallelflow_RayleighEq}
 \end{equation}
where $k$ is the nonnegative real wavenumber and $c=c_r+ic_i$ is
the complex phase speed and double prime $''$ denotes $d^2/dy^2$.
The real part of complex phase speed $c_r$ is the wave phase
speed, and $\omega_i=k c_i$ is the growth rate of the wave. This
equation is to be solved subject to homogeneous boundary
conditions
\begin{equation}
\phi=0 \,\, at\,\, y=a,b.
\label{Eq:stable_parallelflow_RayleighBc}
\end{equation}
There are three main categories of boundaries: (i) open channels
with both $a$ and $b$ being finite, (ii) boundary layers with
either $a$ or $b$ being infinite, and (iii) free shear flows with
both $a$ and $b$ being infinite.


It is obvious that the criterion for stability is $\omega_i=0$
($c_i=0$), for that the complex conjugate quantities $\phi^*$ and
$c^*$ are also a physical solution of
Eq.(\ref{Eq:stable_parallelflow_RayleighEq}) and
Eq.(\ref{Eq:stable_parallelflow_RayleighBc}).

 Multiplying
Eq.(\ref{Eq:stable_parallelflow_RayleighEq}) by the complex
conjugate $\phi^{*}$  and integrating over the domain $a\leq y
\leq b$, we get the following equations
\begin{equation}
\displaystyle\int_{a}^{b}
[(\|\phi'\|^2+k^2\|\phi\|^2)+\frac{U''(U-c_r)}{\|U-c\|^2}\|\phi\|^2]\,
dy=0,
\label{Eq:stable_parallelflow_Rayleigh_Int_Rea}
 \end{equation}
and
\begin{equation}
\displaystyle c_i\int_{a}^{b}
\frac{U''}{\|U-c\|^2}\|\phi\|^2\,dy=0.
\label{Eq:stable_parallelflow_Rayleigh_Int_Img}
 \end{equation}
Rayleigh  used only
Eq.(\ref{Eq:stable_parallelflow_Rayleigh_Int_Img}) to prove his
theorem, i.e., a necessary condition for instability is
$U''(y_s)=0$, where $y_s$ is the inflection point and $U_s=U(y_s)$
is the velocity at $y_s$. Fj\o rtoft noted that
Eq.(\ref{Eq:stable_parallelflow_Rayleigh_Int_Rea}) should also be
satisfied, then he obtained his necessary criterion. To find a
more restrictive criterion, we shall investigate the conditions
for $c_i=0$. Unlike the former investigations, we consider this
problem in a totally different way: if the velocity profile is
stable ($c_i=0$), then the hypothesis $c_i\neq0$ should result in
contradictions in some cases. Following this, some more
restrictive criteria can be obtained.

To find a stronger criterion, we need to estimate the ratio of
$\int_{a}^{b} \|\phi'\|^2 dy$ to $\int_{a}^{b} \|\phi\|^2 dy$.
This is known as Poincar\'{e}'s problem:
\begin{equation}
\int_{a}^{b}\|\phi'\|^2 dy=\mu\int_{a}^{b}\|\phi\|^2 dy,
\label{Eq:stable_parallelflow_Poincare}
\end{equation}
where the eigenvalue $\mu$ is  positive definite for any $\phi
\neq 0$. The smallest eigenvalue value, namely $\mu_1$, can be
estimated as $\mu_1>(\frac{\pi}{b-a})^2$, like what Tollmien have
done \cite{Tollmien1935} .

Then using Poincar\'{e}'s relation
Eq.(\ref{Eq:stable_parallelflow_Poincare}), a new stability
criterion may be found: the parallel flow is stable if
$-\mu_1<\frac{U''}{U-U_s}<0$ everywhere.

To get this criterion, we introduce an auxiliary function
$f(y)=\frac{U''}{U-U_s}$, where $f(y)$ is finite at the inflection
point. We will prove the criterion by two steps. First, we will
prove proposition 1: if the velocity profile is subject to
$-\mu_1<f(y)<0$, then $c_r\neq U_s$.

Proof: Since $-\mu_1<f(y)<0$, then
\begin{equation}
  -\mu_1<\frac{U''}{U-U_s}=\frac{U''(U-U_s)}{(U-U_s)^2}\leq\frac{U''(U-U_s)}{(U-U_s)^2+c_i^2}.
\label{Eq:stable_parallelflow_Rayleigh_inequ}
\end{equation}
Substitution of $c_r=U_s$ and
Eq.(\ref{Eq:stable_parallelflow_Rayleigh_inequ}) into
Eq.(\ref{Eq:stable_parallelflow_Rayleigh_Int_Rea}) results in
\begin{equation}
\displaystyle\int_a^b
[\|\phi'\|^2+k^2\|\phi\|^2+\frac{U''(U-U_s)}{\|U-c\|^2}\|\phi\|^2]\,
dy > 0.
\end{equation}
This contradicts
Eq.(\ref{Eq:stable_parallelflow_Rayleigh_Int_Rea}). So proposition
1 is proved.

Then, we will prove proposition 2: if $-\mu_1<f(y)<0$ and $c_r\neq
U_s$, there must be $c_i^2=0$.

Proof: If $c_i^2\neq0$, then multiplying
Eq.(\ref{Eq:stable_parallelflow_Rayleigh_Int_Img}) by
$(c_r-U_t)/c_i$, where the arbitrary real constant $U_t$ does not
depend on $y$, and adding the result to
Eq.(\ref{Eq:stable_parallelflow_Rayleigh_Int_Rea}), yields
\begin{equation}
\displaystyle\int_a^b
[(\|\phi'\|^2+k^2\|\phi\|^2)+\frac{U''(U-U_t)}{\|U-c\|^2}\|\phi\|^2]\,
dy=0.
\label{Eq:stable_parallelflow_Sun_Int} \end{equation}
But the above Eq.(\ref{Eq:stable_parallelflow_Sun_Int}) does not
hold for some special $U_t$. For example, if $U_t=2c_r-U_s$, then
there is $(U-U_s)(U-U_t)<\|U-c\|^2$, and
 \begin{equation}
\frac{U''(U-U_t)}{\|U-c\|^2}=
f(y)\frac{(U-U_s)(U-U_t)}{\|U-c\|^2}>-\mu_1.
\label{Eq:stable_parallelflow_Sun_Ust}
 \end{equation}
This yields
\begin{equation} \int_a^b
\{\|\phi'\|^2+[k^2+\frac{U''(U-U_t)}{\|U-c\|^2}]\|\phi\|^2\} dy>0,
\end{equation}
which also contradicts Eq.(\ref{Eq:stable_parallelflow_Sun_Int}).
So proposition 2 is also proved.

Using ``proposition 1: if $-\mu_1<f(y)<0$ then $c_r\neq U_s$" and
``proposition 2: if $-\mu_1<f(y)<0$ and $c_r\neq U_s$ then $c_i =
0$", we find a stability criterion. If the velocity profile
satisfies $-\mu_1<\frac{U''}{U-U_s}<0$  everywhere, the parallel
flow is stable. Moreover, the above proof is still valid for
$0<f(y)$, which is equivalent to Fj\o rtoft's criterion. Thus we
have the following theorem.

Theorem 1: If the velocity profile satisfies either
$-\mu_1<\frac{U''}{U-U_s}<0$ or $0<\frac{U''}{U-U_s}$, the flow is
stable.

This criterion covers Rayleigh's and Fj\o rtoft's criteria. And
the proofs here are very simple and easy to understand comparing
to Arnol'd's proofs. As mentioned above, we have investigated the
stable criterion via Rayleigh's equation, while Arnol'd
\cite{Arnold1969} considered the hydrodynamic stability in a
totally different way. Is there any relationship between these
proofs? Two points are outlined here. First, this criterion is
essentially the same as Arnol'd's second stability theorem and is
more restrictive than Fj\o rtoft's criterion. Second, the proofs
here are similar to Arnol'd's variational principle method. For
the arbitrary real number $U_t$, which is like a Lagrange
multiplier in variational principle method, plays a key role in
the proofs.

\subsection{Discussion }

One may note that the criterion above is something different from
Fj\o rtoft's criterion. Why are the functions of $U''/(U-U_s)$
used in Arnol'd's theorems and present theorems, unlike
$U''(U-U_s)$ in Fj\o rtoft's theorem? This is due to the property
of Rayleigh's equation. It can be seen from
Eq.(\ref{Eq:stable_parallelflow_RayleighEq}) that the stability of
profile $U(y)$ is not only Galilean invariant, but also
independent from the the magnitude of $U(y)$ due to the linearity.
So the stability of $U(y)$ is the same as that of $AU(y)+B$, where
$A$ and $B$ are any arbitrary nonzero real numbers. As the value
of $U''(U-U_s)$ is only Galilean invariant but not magnitude free,
it satisfies only part of the Rayleigh's equation's properties. On
the other hand the value of $U''/(U-U_s)$ satisfies both
conditions, this is the reason why the criteria in both Arnol'd's
theorems and present theorems are the functions of $U''/(U-U_s)$.
Since the stability of inviscid parallel flow depends only on the
velocity profile's geometry shape, namely $f(y)$, and the
magnitude of the velocity profile can be free, then the
instability of inviscid parallel flow could be called "geometry
shape instability" of the velocity profile. This distinguishes
from the viscous instability associated with the magnitude of the
velocity profile.

\begin{figure}
  \centerline{\includegraphics[width=6cm]{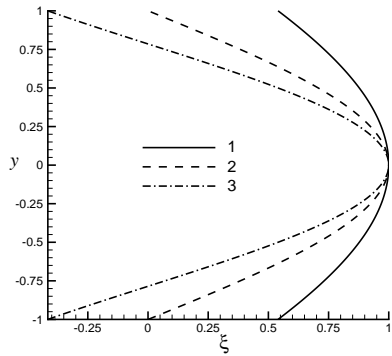}}
\caption{Vorticity profiles within the interval $-1\leq y\leq 1$.
Profile 2 ( $\xi=\cos(\pi y/2)$, dashed) is neutrally stable,
while profile 1 ($\xi=\cos( y)$, solid) and profile 3
($\xi=\cos(2y)$, dash doted) are stable and unstable,
respectively. } \label{Fig:vorticity_profile}
\end{figure}

Moreover, the above theorem is essentially associated with
vorticity distribution in the flow field. As known from Fj\o
rtoft's criterion, the necessary condition for instability is that
the base vorticity profile $\xi=-U'$ has a local maximum. Note
that $U''/(U-U_s)\approx \xi_s''/\xi_s$ near the inflection point,
where $\xi_s$ is the vorticity at the inflection point, that means
that the base vorticity $\xi$ must be convex enough near the local
maximum for instability, i.e., the vorticity should be
concentrated somewhere in the flow for instability. Otherwise, the
flow is stable if the vorticity distribution is smooth enough near
the inflection point at $y_s$. A simple example can be obtained by
following Tollmien's way \cite[]{Tollmien1935}.
Fig.\ref{Fig:vorticity_profile} depicts three vorticity profiles
within the interval $-1\leq y\leq 1$, which have local maximal at
$y=0$. Profile 2 ($U=-2\sin(\pi y/2)/\pi$) is neutrally stable,
while profile 1 ($U=-\sin(y)$) and profile 3 ($U=-\sin(2y)/2$) are
stable and unstable, respectively.

\begin{figure}
\centerline{\includegraphics[width=6cm]{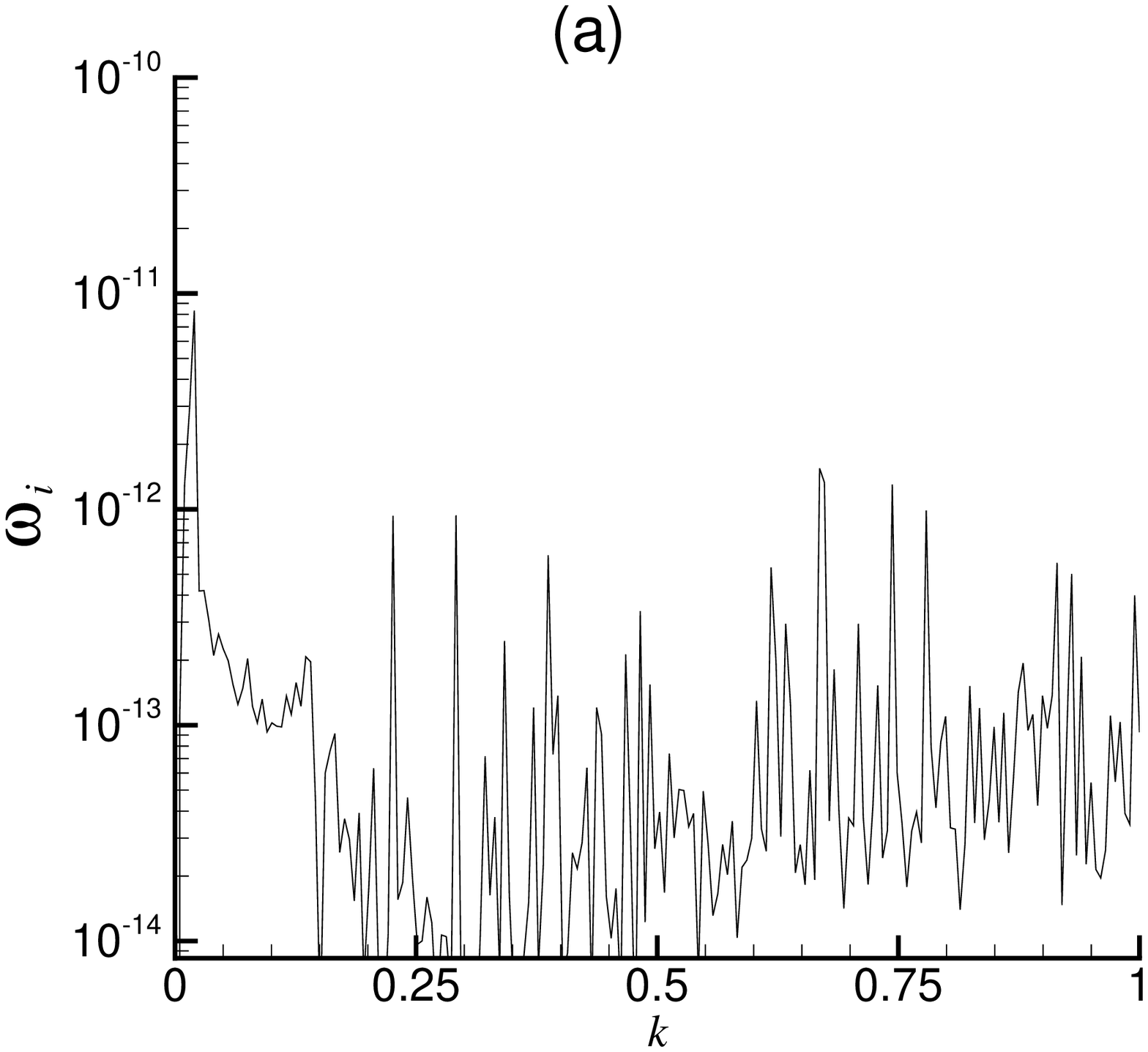}
  \includegraphics[width=6cm]{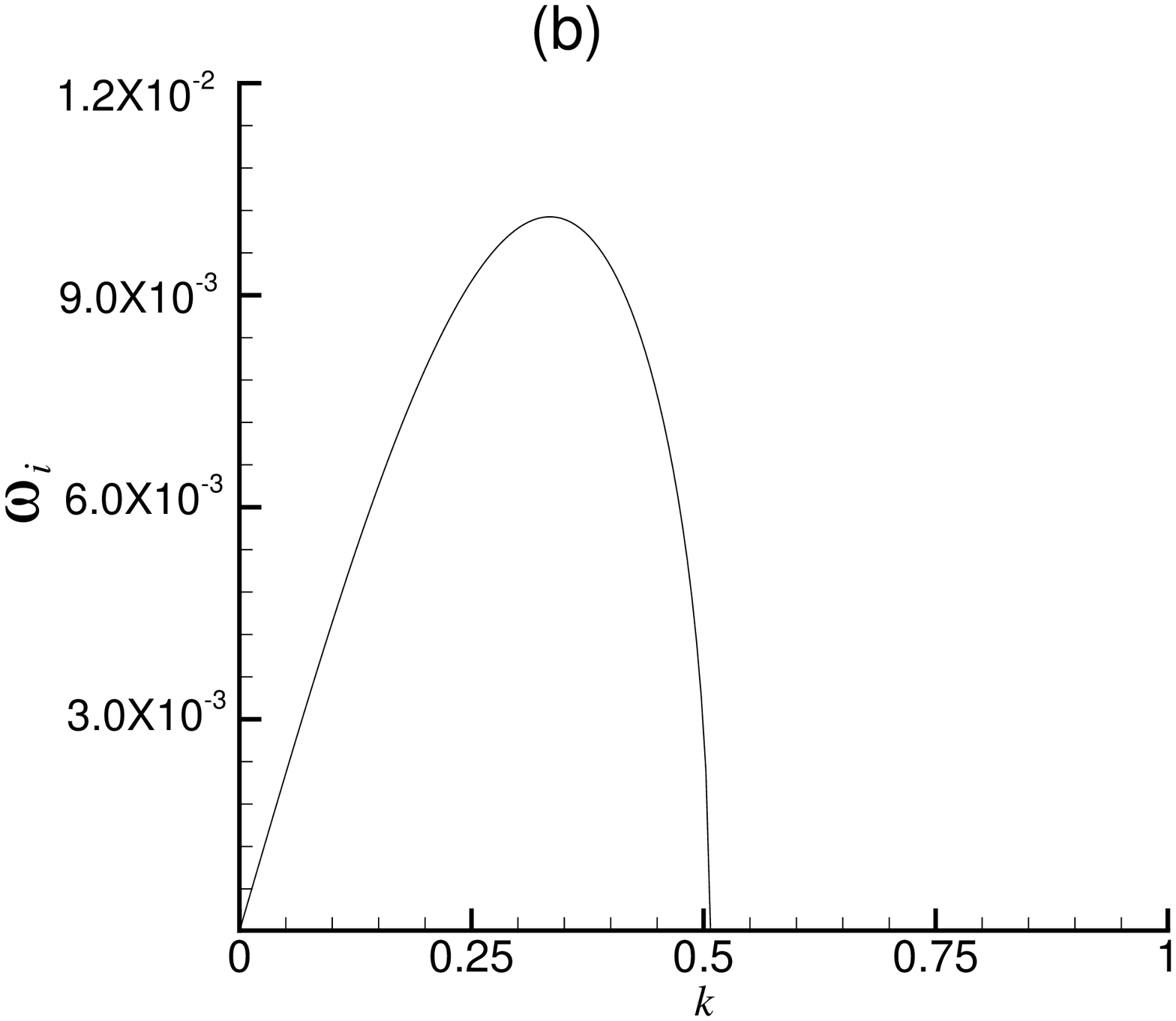}}
\caption{Growth rate $\omega_i$ as an function of wavenumber $k$,
(a) for $U_1=\tanh(1.1 y)$, (b) for $U_2=\tanh(1.3 y)$, both
within the interval $-1\leq y\leq 1$. } \label{Fig:Growth}
\end{figure}

To show the advantage of the criteria obtained above, we consider
the stability of velocity profile $U=\tanh(\alpha y)$ within the
interval $-1\leq y\leq 1$, where $\alpha$ is a constant. This
velocity profile is an classical model of mixing layer, and has
been investigated by many researchers (see
\cite{Huerre1998,SchmidBook2000,CriminaleBook2003} and references
therein). Since $U''(U-U_s)=-2\alpha^2\tanh^2(\alpha
y)/\cosh^2(\alpha y) <0$ for $-1\leq y\leq 1$, it might be
unstable for any $\alpha$ according to both Rayleigh's and Fj\o
rtoft's criteria. But it can be derived from Theorem 1 that the
flow is stable for $\alpha^2<\pi^2/8 $. For example, we choose
$\alpha_1=1.1$ and $\alpha_2=1.3$ for velocity profiles $U_1(y)$
and $U_2(y)$. The growth rate of the profiles can be obtained by
Chebyshev spectral collocation method \cite[]{SchmidBook2000} with
100 collocation points, as shown in Fig.\ref{Fig:Growth}. It is
obvious that $c_i=0$ for $U_1$ and $c_i>0$ for $U_2$, which agrees
well with the criteria obtained above. This is also a
counterexample that Fj\o rtoft's criterion is not sufficient for
instability. So this new criterion for stability is more useful in
real applications.

Then, recalling the proof of theorem 1, we will find that the
following Rayleigh's quotient $I(f)$ plays a key role in
determining the stability of parallel flows.
\begin{equation}
I(f)=\min_{\phi} \frac{\int_{a}^{b}
[\,\|\phi'\|^2+f(y)\|\phi\|^2\,]\, dy}{\int_{a}^{b} \|\phi\|^2}
\label{Eq:stable_paralleflow_sun_Energy}
\end{equation}
Note that the proof of theorem 1 is still valid in the case of
$I(f)>0$. We have such a result: parallel flows are stable if
$I(f)>0$. Although this criterion is more restrictive than that in
theorem 1, it is inconvenient for the real applications due to the
unknown Rayleigh's quotient $I(f)$. Theorem 1 is more convenient
for the real applications in different research fields.

\begin{figure}
  \centerline{\includegraphics[width=6cm]{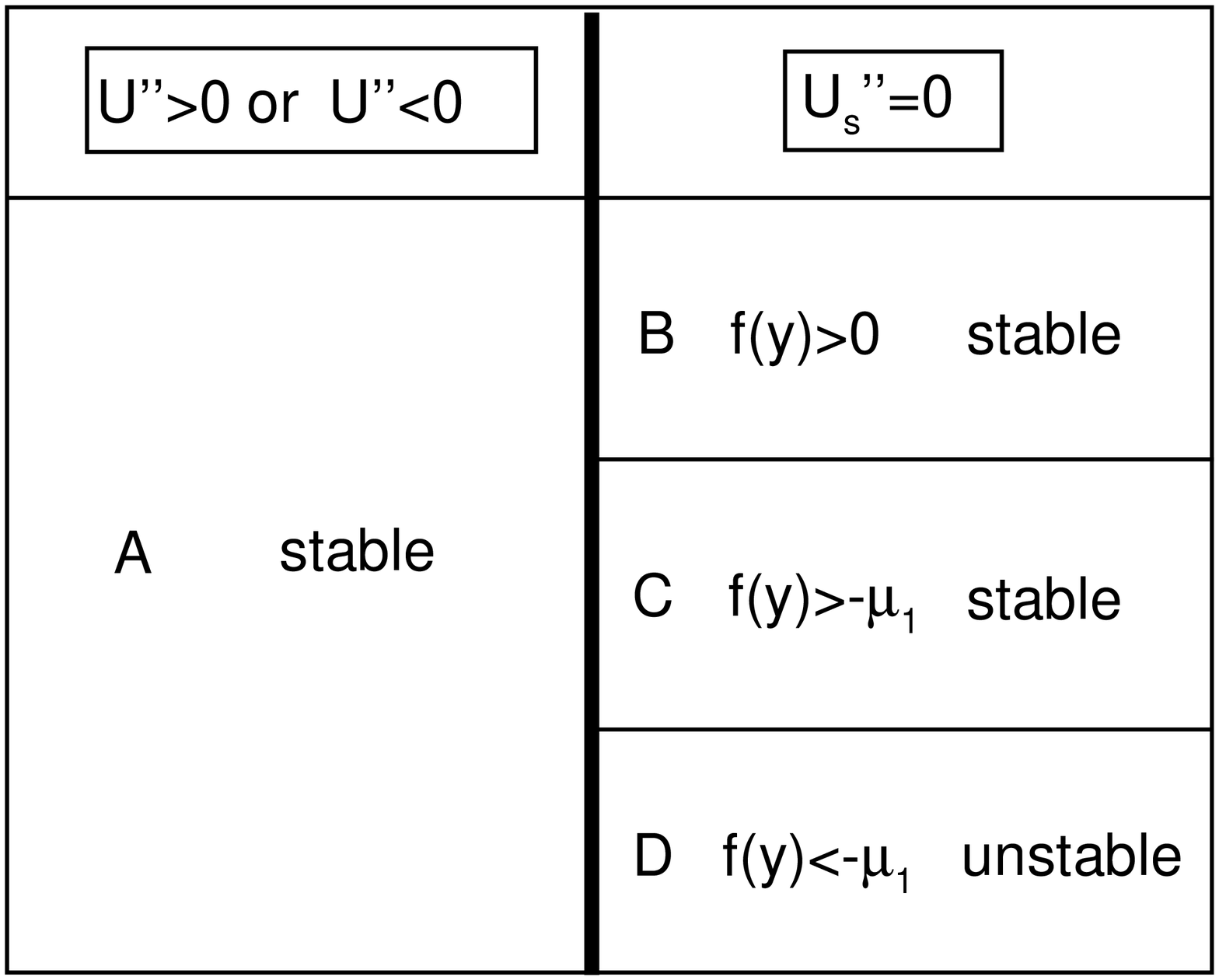}}
\caption{Diagram of stable velocity profiles.}
\label{Fig:parallel_stable_sun_shear_diagram}
\end{figure}
The present stability criteria give a affirmative answer to the
questions at the beginning, i.e., there are some stable flows if
$U''(U-U_s)<0$. Based on the former criteria, the velocity
profiles can be categorized as follows: (\romannumeral1) without
inflection point (Reyleigh's criterion), (\romannumeral2) $f(y)>0$
(Fj{\o}rtoft's criterion), and (\romannumeral3) $-\mu_1<f(y)<0$
(present criterion), which are shown in
Fig.\ref{Fig:parallel_stable_sun_shear_diagram}. Then the parallel
flow might be unstable only for $f(y)<-\mu_1$ and $f(y)$ changing
sign within the interval. However, if $f(y)$ changes sign
somewhere within the interval $[a,b]$, then the parallel flow is
stable. For that $f(y)$ changing sign implies $U'''_s=0$ but
$U''''_s\neq0$, so $U''$ does not change sign near the inflection
point. Thus $c_i$ must vanish in
Eq.(\ref{Eq:stable_parallelflow_Rayleigh_Int_Img}), i.e., the
parallel flow is stable for $f(y)$ changing sign within the
interval. In this way, the parallel flow might be unstable only
for $f(y)<\mu_1$ somewhere, which will intrigue further studies on
this problem. In fact, there are still stable flows if
$\mu_1<f(y)$ is violated.

Finally, the stable criterion for the parallel inviscid flows can
be applied to the barotropic geophysical flows in a differentially
rotating system. Considering the barotropic plane flows in a
rotating frame, which are the approximations of barotropic
geophysical flows
\cite[]{KuoHL1949,Dowling1995,CriminaleBook2003},
Eq.(\ref{Eq:stable_parallelflow_RayleighEq}) changes to
 \begin{equation}
 (\phi''-k^2 \phi)-\frac{U''-\beta}{U-c}\phi=0,
 \label{Eq:stable_parallelflow_BarotropicEq}
 \end{equation}
where $\beta$ is the gradient of the Coriolis parameter with
respect to $y$. Eq.(\ref{Eq:stable_parallelflow_BarotropicEq}) is
a generalized Rayleigh's equation, and there is a generalized
stability criterion for these flows.

Theorem 2: The flow is stable, if the velocity profile  satisfies
either $-\mu_1<\frac{U''-\beta}{U-U_s}<0$ or
$0<\frac{U''-\beta}{U-U_s}$, where $U_s$ is the velocity at the
point $U''(y_s)=\beta$.

The above criteria would be helpful for understanding the
wave-mean flow interaction, especially the Rossby wave-mean flow
interaction in barotropic flows. According to the stable criteria,
the necessary condition for wave-mean flow interaction can be
obtained. And why the disturbed waves can't take energy from the
mean flow in the stable flow is revealed. If the flow is stable,
there is no wave-mean flow interaction at all.

First, when the velocity profile has no inflection point, then the
speeds of barotropic waves $c_r-U\neq 0$. According to
Eq.(\ref{Eq:stable_parallelflow_Rayleigh_Int_Rea}), $U-c_r>0$
holds for $U''<0$, i.e., the barotropic waves are always
west-propagation relative to the mean flow $U$ if the vortex
gradient is positive. This extends the west-propagation theory of
Rossby waves. And $U-c_r<0$ holds for $U''>0$, i.e., the
barotropic waves are always east-propagation relative to the mean
flow $U$ if the vortex gradient is negative. Second, when the
velocity profile has an inflection point $U''_s=0$, the speed of
most favorite wave which might have interaction with the mean flow
should be $c_r=U_s$. However, the wave speed $c_r\neq U_s$ holds
for $-\mu_1<\frac{U''}{U-U_s}<0$, as has been pointed out in
proposition 1. In short, the waves have no interaction with the
mean flows whatever there is an inflection point or not. This
conclusion can be easily generated to the flows in $\beta$ plane.
Thus, the waves have no interaction with the stable flows. This is
the reason why the disturbed waves can't take energy from the
stably mean flow.

On the other hand, \cite{Howard1961} pointed out that $c_r$ of a
unstable wave must lie between the minimum and the maximum values
of the mean velocity profile. Thus, the unstable wave is
stationary relative to the mean flow, and can take energy from the
mean flow. So there are wave-mean flow interactions in unstable
flows.

\subsection{Conclusion}

In summary, the general stability criteria are obtained for
inviscid parallel flow. Both the criteria and the proofs are
remarkably simple and easy to understand, comparing to Arnol'd's
nonlinear theorems. The new criteria extend the former theorems
proved by Rayleigh, Tollmien and Fj\o rtoft. This may shed light
on the flow control and investigation of the vortex dynamics.
Based on the stability criteria, the reason why the disturbed
waves can't take energy from the stably mean flow is explained
that there is no wave-mean flow interaction at all.

\section{Longwave Instability}

\subsection{Introduction}
Shear instability, caused by the velocity shear, is one of most
important instabilities in the flows. Although the mechanism of
shear instability is not full revealed yet, it is applied to
explain the instability of mixing layer, jets in pipe, wakes
behind cylinder, etc. Some simple models were employed to study
the shear instability, including Kelvin-Helmholtz (K-H) model,
piecewise linear velocity profile \cite{Rayleigh1894} and
continued velocity profiles \cite{Rayleigh1880}, etc. To reveal
the mechanism of shear instability, the stability of perturbation
waves should be understood. In K-H model, the growth rate of the
disturbance wave is proportion to the product of wavenumber and
velocity shear, thus the shortwaves are more unstable than
longwaves in K-H model. However, Rayleigh \cite{Rayleigh1894}
found that the piecewise linear profile is linearly unstable only
in a finite range of wavenumbers $0\leq k<k_c$, which means the
shortwaves are stable in this case
\cite{Rayleigh1894,Chandrasekhar1961,Huerre1998,CriminaleBook2003}.
A contradiction \cite{Huerre1998,CriminaleBook2003} emerges as
shortwave and longwave dominate the shear instability in K-H model
and piecewise linear profiles, respectively. This was explained
either as viscous effect must be considered for shortwave
\cite{CriminaleBook2003} or as longwaves do not "feel" the finite
thickness of the layer \cite{Huerre1998,CriminaleBook2003}.

To dispel the contradiction, we should find out which dominates
the shear instability, shortwave or longwave? It is from the
numerical simulation (see
\cite{Huerre1998,SchmidBook2000,CriminaleBook2003} and references
therein) and some theoretical analysis (e.g. Tollmien
\cite{Tollmien1935} and Lin \cite{LinCCBook1955}) that longwave
might be more unstable than shortwave. They also proved the
long-wave instability subject to the velocity profile $U$ is
either symmetric or monotone. We will investigate this problem
following the way by Sun \cite{SunL2006a}.

Moreover, a prior estimation of growth rate, which may be obtained
from the investigation, is useful for unstable flows. For example,
Howard's semicircle theorem \cite{Howard1961} has been used to
validate the numerical calculations \cite{SchmidBook2000}. Another
useful result was found by H{\o}iland \cite{Hoiland1953} and
Howard \cite{Howard1961} that the growth rate $\omega_i$  must
less equate to half of the maxim of vorticity, i.e., $\omega_i\leq
|U'|_{\max}/2$. But this estimation is too cursory to use in
applications. For example, $U'$ is always great than zero even the
velocity profile has no inflection point. So this estimation is
trivial for these cases. Here we will show a refinement estimation
of growth rate, which can be applied to general velocity profiles
in parallel flows.

The motivation of this short letter is to investigate these
problems within the context of inviscid parallel flow. The aim
here is to find out some general characters for unstable waves.

\subsection{Model and Analysis}

To this purpose, long-wave instability in shear flows is
investigated via Rayleigh's equation
\cite{Rayleigh1880,Chandrasekhar1961,Huerre1998,CriminaleBook2003}.
For a parallel flow with mean velocity $U(y)$, where $y$ is the
cross-stream coordinate. The streamfunction of the disturbance
expands as series of waves (normal modes) with real wavenumber $k$
and complex frequency $\omega=\omega_r+i\omega_i$, where
$\omega_i$ relates to the grow rate of the waves. The flow is
unstable if and only if $\omega_i>0$. We study the stability of
the disturbances by investigating the growth rate of the waves,
this method is known as normal mode method. The amplitude of
waves, namely $\phi$, holds
 \begin{equation}
 (\phi''-k^2 \phi)-\frac{U''}{U-c}\phi=0,
 \label{Eq:stable_longwave_RayleighEq}
 \end{equation}
where $c=\omega/k=c_r+ic_i$ is the complex phase speed. The real
part of complex phase speed $c_r=\omega_r/k$ is the wave phase
speed. This equation is to be solved subject to homogeneous
boundary conditions
\begin{equation}
\phi=0 \,\, at\,\, y=a,b. \label{Eq:stable_longwave_RayleighBc}
\end{equation}

 From Rayleigh's equation, we get the
following equations:
\begin{equation}
\displaystyle\int_{a}^{b}
[(\|\phi'\|^2+k^2\|\phi\|^2)+\frac{U''(U-c_r)}{\|U-c\|^2}\|\phi\|^2]\,
dy=0,
\label{Eq:stable_longwave_Rayleigh_Int_Rea}
 \end{equation}
and
\begin{equation}
\displaystyle c_i\int_{a}^{b}
\frac{U''}{\|U-c\|^2}\|\phi\|^2\,dy=0.
\label{Eq:stable_longwave_Rayleigh_Int_Img}
 \end{equation}

Before the further discussion, we need estimate the rate of
$\int_{a}^{b} \|\phi'\|^2 dy$ to $\int_{a}^{b} \|\phi\|^2 dy$, as
Sun did \cite{SunL2006a}. This is known as Poincar\'{e}'s problem:
\begin{equation}
\int_{a}^{b}\|\phi'\|^2 dy=\mu\int_{a}^{b}\|\phi\|^2 dy,
\label{Eq:stable_longwave_Poincare}
\end{equation}
where the eigenvalue $\mu$ is  positive definition for $\phi \neq
0$. The smallest eigenvalue value, namely $\mu_1$, can be
estimated as $\mu_1>(\frac{\pi}{b-a})^2$. And an auxiliary
function $f(y)=\frac{U''}{U-U_s}$ is also introduced, where $f(y)$
is finite at inflection point.

With the preparation above, we have such consequence. If
$-f(y)<Q<\infty$, where $Q$ is a positive constant, then the
disturbances with shortwaves $k>k_c$ are always stable, where
$k_c$ is a critical wavenumber subject to $k_c^2=Q-\mu_1$. We will
prove the consequence by two steps. At first, we prove proposition
1: if $c_r=U_s$, the disturbances with shortwaves $k>k_c$ are
always stable.

Proof: Since $U''=f(y)(U-U_s)$ and $c_r=U_s$, this yields to
\begin{equation}
 \frac{U''(U-U_s)}{\|U-c\|^2}>f(y)\frac{U''}{U-U_s}>-Q,
\end{equation}
and
\begin{equation}
\begin{array}{rl} \displaystyle\int_a^b
[(\|\phi'\|^2+k^2\|\phi\|^2)+\frac{U''(U-U_s)}{\|U-c\|^2}\|\phi\|^2]\,
dy &\geq \\
\displaystyle\int_a^b [(\mu_1+k_c^2+\frac{U''(U-U_s)}{\|U-c\|^2})
\|\phi\|^2]&>0.
\end{array}
\end{equation}
This contradicts Eq.(\ref{Eq:stable_longwave_Rayleigh_Int_Rea}).
So proposition 1 is proved.

Then, we prove proposition 2: if $c_r\neq U_s$ , there must be
$c_i^2=0$ with $k>k_c$.

Proof: Otherwise if $c_i^2\neq0$, so according to
Eq.(\ref{Eq:stable_longwave_Rayleigh_Int_Rea}) and
Eq.(\ref{Eq:stable_longwave_Rayleigh_Int_Img}), for any arbitrary
real number $U_t$  which does not depend on $y$, it holds
\begin{equation}
\displaystyle\int_a^b
[(\|\phi'\|^2+k^2\|\phi\|^2)+\frac{U''(U-U_t)}{\|U-c\|^2}\|\phi\|^2]\,
dy=0.
\label{Eq:stable_longwave_Sun_Int} \end{equation}
But the above Eq.(\ref{Eq:stable_longwave_Sun_Int}) can not be
hold for some special $U_t$. For example, let $U_t=2c_r-U_s$, then
there is $(U-U_s)(U-U_t)<\|U-c\|^2$, and
 \begin{equation}
\frac{U''(U-U_t)}{\|U-c\|^2}=
f(y)\frac{(U-U_s)(U-U_t)}{\|U-c\|^2}>-Q.
\label{Eq:stable_longwave_Sun_Ust}
 \end{equation}
For $k>k_c$, this yields to
\begin{equation} \int_a^b
\{\|\phi'\|^2+[k^2+\frac{U''(U-U_t)}{\|U-c\|^2}]\|\phi\|^2\} dy>0,
\end{equation}
which also contradicts Eq.(\ref{Eq:stable_longwave_Sun_Int}). So
proposition 2 is also proved. These two propositions are natural
generalization of stabile criterion proved by Sun
\cite{SunL2006a}.

From the above two propositions, we can draw a conclusion that the
disturbances with shortwaves $k>k_c$ are always stable. This means
that the shear instability in flows must be long-wave instability.
Furthermore, the short-wave stability means that without any
viscous effect, the shortwaves can also be damped by shear flow.
This mechanism is unlike the viscous mechanism that the viscosity
has a damping effect on especially the shortwaves. It implies that
the shear flow itself can damp the shortwaves. This is very
general and important conclusion, which explains why the
instabilities found in shear flows are mostly long-wave
instabilities.

\subsection{Discussion}

This result is also very important for numerical calculation,
which means shortwaves can be truncated in the calculations
without changes the stability of shear flow. So the growth rates
of longwaves have enough information for judging the stability of
shear flow. On the other hand, the truncation of longwaves would
probably change the instability of the shear flow. So the
streamwise length scale must be longer enough to have longwaves
for the numerical simulations in shear flows, such as plane
parallel flow and pipe flow. Otherwise the instability of shear
flow would be damped without long-wave perturbations.

Then the growth rate of unstable waves can be estimated here by
following the former investigation. Similar to the assumption
above, $-f(y)<Q=p^2\mu_1<\infty$, where $1<p<2$, we will show that
the growth rate is subject to $\omega_i \leq (p-1)\sqrt{ \mu_1}
|U_{\max}-U_{\min}|$.

Proof: It is from Eq.(\ref{Eq:stable_longwave_Rayleigh_Int_Rea})
that gives
\begin{equation}
\int_{a}^{b} k^2\|\phi\|^2 dy=\int_{a}^{b}
-[\|\phi'\|^2+\frac{U''(U-c_r)}{\|U-c\|^2}\|\phi\|^2] dy.
\label{Eq:stable_longwave_Sun_Int_k2}
\end{equation}
Substituting Eq.(\ref{Eq:stable_longwave_Poincare}) to
Eq.(\ref{Eq:stable_longwave_Sun_Int_k2}) and recalling that
$\mu_1<\mu$, this yields
\begin{equation}
\int_{a}^{b} k^2\|\phi\|^2\, dy\leq \int_{a}^{b}
-[\frac{U''(U-c_r)}{\|U-c\|^2}+\mu_1]\|\phi\|^2\, dy.
\label{Eq:stable_longwave_k2Ineq}
\end{equation}
Multiplying above inequality (\ref{Eq:stable_longwave_k2Ineq}) by
$c_i^2$, we get
\begin{equation}
\omega_i^2\int_{a}^{b} \|\phi\|^2\, dy\leq \int_{a}^{b}
h(y)\|\phi\|^2\, dy,
\end{equation}
where
\begin{equation}
h(y)=-[\frac{U''(U-c_r)}{\|U-c\|^2}+\mu_1]c_i^2.
\end{equation}
Suppose the maxim of $h(y)$ is $P^2$, then the growth rate is
subject to
\begin{equation}
  \omega_i^2 \leq P^2 .
\end{equation}
This follows
\begin{equation}
h(y)\leq -[\frac{U''(U-c_r)}{(U-c_r)^2+c_i^2}+\mu_1]c_i^2.
\label{Eq:stable_longwave_hfunc1}
\end{equation}
Substitution of $f(y)$ into Eq.(\ref{Eq:stable_longwave_hfunc1}),
gives
\begin{equation}
h(y)\leq
\mu_1[\frac{(p^2-1)(U-c_r)^2-c_i^2}{(U-c_r)^2+c_i^2}]c_i^2.
\label{Eq:stable_longwave_hfunc2}
\end{equation}
When
\begin{equation}
c_i^2=(p-1)(U-c_r)^2,
 \label{Eq:stable_longwave_cimax}
\end{equation}
the right hand of Eq.(\ref{Eq:stable_longwave_hfunc2}) get its
largest value
\begin{equation}
P^2= (p-1)^2 \mu_1 (U-c_r)^2.
 \label{Eq:stable_longwave_omegaimax}
\end{equation}
Then the growth rate must be subject to
\begin{equation}
  \omega_i \leq (p-1) \sqrt{\mu_1} |U_{\max}-U_{\min}|,
\end{equation}
where $U_{\min}$ and $U_{\max}$ are minimum and maximum of $U(y)$,
respectively. And the wavenumber $k_{\max}$ corresponding to the
largest growth rate is
\begin{equation}
  k_{\max}=\sqrt{(p-1)\mu_1}.
  \label{Eq:stable_longwave_growth_max_k}
\end{equation}
So the result is proved. One should note that the fast growth rate
$\omega_i$ is only an approximation, but not a precision one, so
as to wavenumber $k_{\max}$.

Comparing with the previous one by H{\o}iland and Howard, the
above estimation includes not only a more precision estimation
about the the growth rate of unstable waves but also the regime of
unstable wavenumbers. This would be much helpful for validation in
numerical calculations.

%

As well known, the instability due to velocity shear is always
associated to Kelvin-Helmholtz instability (K-H instability), in
which the disturbances of all wavelengths are amplified. According
to K-H instability, the shorter the wavelength is, the faster the
perturbation wave amplifies. As the above investigation points out
that the shortwaves are always more unstable than longwaves in
this case, the contradiction at the beginning can be dispelled.

An physical explanation \cite{Huerre1998,CriminaleBook2003} to the
contradiction is that the K-H instability model has no intrinsic
length scale, while Rayleigh's model has width of shear layer as
length scale of waves. This can be noted from that Rayleigh's case
reduces to the Kelvein-Helmholtz vortex sheet model in the
long-wave limit $k\ll 1$ \cite{Huerre1998,CriminaleBook2003},
which is dramatically explained as longwave do not "feel" the
finite thickness of the shear layer \cite{Huerre1998}. Here we
will show that this explanation can be extend to shear flows. It
is from Eq.(\ref{Eq:stable_longwave_growth_max_k}) that the
fastest growing wavenumber $k_{\max}$ is proportion to
$\sqrt{\mu_1}$, then the corresponding wave length
$\lambda_{\max}$ is approximately $2(b-a)/\sqrt{p-1}$. So the
thinner the shear layer is, the lager the fastest growing
wavenumber is. It can be seen that the asymptotic case of infinite
small shear layer leads to K-H instability. This is another
evidence that K-H instability is in essence of long-wave
instability. In this case, the K-H instability is an approximation
of shear instability when the wave length of perturbation is much
longer than the width of shear layer.

In summary, two general properties of shear instability are
obtained in the investigations. Firstly, the shortwaves are always
stable in the continued profile flows and the shear instability is
due to long-wave instability. This implies that the shear flow
itself can damp the shortwaves and that K-H instability is in
essence of long-wave instability in shear flow. The result can be
used to understand the phenomena in hydrodynamics instabilities.
Secondly, the growth rate of unstable flow can be estimated as
$\omega_i \leq (p-1) \sqrt{\mu_1} (U_{\max}-U_{\min})$. This
estimation extend the previous result obtained by H{\o}iland and
Howard. Both results are important in numerical applications. The
first one provides the estimation of unstable wavenumbers, and the
second one provides the estimation of growth rate of unstable
waves. These results may be useful on both numerical calculation
and stability analysis.

\section{Mechanism of Shear Instability}
\subsection{Introduction}
The hydrodynamic instability is a fundamental problem in many
fields, such as fluid dynamics, astrodynamics, oceanography,
meteorology, etc. There are many kinds of hydrodynamic
instabilities, e.g., shear instability due to velocity shear,
thermal instability due to heating, viscous instability due to
viscosity and centrifugal instability due to rotation, etc. Among
them, the shear instability (inviscid) is the most important and
the simplest one, which has been intensively explored (see
\cite{Drazin1981,Huerre1998,CriminaleBook2003} and references
therein). Both linear and nonlinear stabilities of shear flow have
been considered, and some important conclusions have also be
obtained from the investigations.

On the one hand, the nonlinear stability of shear flow has been
investigated via variational principles. Kelvin \cite{Kelvin1875}
and Arnol'd \cite{Arnold1965b,Arnold1969,ArnoldBook1998} have
developed variational principles for two-dimensional inviscid flow
\cite{Saffman1992}. They showed that the steady flows are the
stationary solutions of the energy $H$. And if the second
variation $\delta^2 H$ is definite, then the steady flow is
nonlinearly stable. Moreover, Arnol'd proved two nonlinear
stability criteria, and that the flow is linearly stable as
$\delta^2 H$ is positive definite
\cite{Arnold1969,Saffman1992,ArnoldBook1998,Vladimirov1999}.
However, Arnol'd's theorem is not convenient to use as $\delta^2
H$ is always indefinite in sign, except for two special cases (see
\cite{Vladimirov1999} and references therein). Due to the lack of
the explicit expressions in both $H$ and the stability criteria
the variational principle is inconvenient for real applications.

On the other hand, the linear stability of shear flow has also
been investigated via Rayleigh's equation. Within the linear
context, there are three important general stability criteria,
namely Rayleigh's criterion \cite{Rayleigh1880}, Fj{\o}rtoft's
criterion \cite{Fjortoft1950} and Sun's criterion
\cite{SunL2006a}. As all the criteria have explicit expressions,
they are more convenient in real applications, and are widely used
in many fields. Based on the previous investigations, Sun
\cite{SunL2006a} also pointed out that the flow is stable for
Rayleigh's quotient $I(f)>0$ (see
Eq.(\ref{Eq:stable_paralleflow_sun_Energy}) behind). However, the
sufficient criterion for instability still lacks.

To understand the shear instability, some mechanisms were
suggested. Among them, Kelvin-Helmholtz instability (K-H
instability) is always taken as a prototype, which is physically
explained as the instability of a sheet vortex
\cite{Batchelor1967}. An another mechanism of instability is due
to the resonance of waves
\cite{Craik1971,Butler1992,Baines1994,Staquet2002,CriminaleBook2003}.
Butler and Farrell \cite{Butler1992} clearly showed with numerical
simulations that the resonance introduces an algebraic growth term
into the temporal development of a disturbance. Baines and
Mitsudera \cite{Baines1994} also used broken-line profile velocity
as a prototype to explain the interaction of waves. Their
explanation is so brilliant that it can explain why the
instability occurs for a finite range of wavenumber and how the
waves amplify each other. However, both mechanisms are independent
of the basic flows, for that Kelvin-Helmholtz model deals only
with vortex and resonance mechanism only considers the waves. Thus
the relationships between those mechanisms and the shear flows are
still covered.

Overview the previous investigations, the essence of the shear
instability is remained to be elucidated. Yet both the explicit
expression and the meaning of $H$ have not been revealed before,
although it is very important in variational principles. The
connection between linear and nonlinear stability criteria should
be retrieved explicitly. The relationships between instability
mechanisms and the physical explanation for shear instability are
needed. The aim of this paper is to reveal the essence of the
shear instability by investigating the inviscid shear flows in a
channel. And this would lead to a more comprehensive understanding
on shear instabilities.

\subsection{Stable Criterion}
For the two-dimensional inviscid flows with the velocity of $U$,
the vorticity $\xi=\nabla\times U$ is conserved along pathlines
\cite{Batchelor1967,Saffman1992,Huerre1998}:
 \begin{equation}
\frac{d\xi}{dt}=\frac{\partial \xi}{\partial t}+(U\cdot
\nabla)\xi=0.
 \label{Eq:stable_paralleflow_Conserve_Vor}
 \end{equation}
Its linear disturbance reduces to Rayleigh's equation provided the
basic flow $U$ being parallel. Considers an shear flow with
parallel horizontal velocity $U(y)$ in a channel, as shown in
Fig.\ref{Fig:stable_parallelflow_mechanism}. The amplitude of
disturbed flow streamfunction $\psi$, namely $\phi$, satisfies
\cite{Drazin1981,Huerre1998,CriminaleBook2003} :
 \begin{equation}
 (\phi''-k^2 \phi)-\frac{U''}{U-c}\phi=0,
 \label{Eq:stable_essence_RayleighEq}
 \end{equation}
where $k$ is the nonnegative real wavenumber and $c=c_r+ic_i$ is
the complex phase speed and double prime $''$ denotes the second
derivative with respect to $y$. The real part $c_r$ is the phase
speed of wave, and $c_i\neq 0$ denotes instability. This equation
is to be solved subject to homogeneous boundary conditions
$\phi=0$ at $y=a,b$.

Considers that the velocity profile $U(y)$ has an inflection point
$y_s$ at which $U''_s=U''(y_s)=0$ and $U_s=U(y_s)$. As Sun
\cite{SunL2006a} has pointed out, the following Rayleigh's
quotient $I(f)>0$ implies the flow is stable.
\begin{equation}
I(f)=\min_{\phi} \frac{\int_{a}^{b}
[\,\|\phi'\|^2+f(y)\|\phi\|^2\,]\, dy}{\int_{a}^{b} \|\phi\|^2\,
dy } \label{Eq:stable_essence_sun_Energy}
\end{equation}
where $f(y)=\frac{U''}{U-U_s}$. While if $I(f)<0$, then there is a
neutral stable model with $k_N^2=-I(f)$ and $c_r=U_s$. Moreover,
there are unstable modes with $c_r=U_s$ and $c_i\neq 0$ if
$I(f)<0$ and $0\leq k<k_N$. This can also be proved by following
the way by Tollmien \cite{Tollmien1935}, Friedrichs
\cite{Friedrichs1942,Drazin1966,Drazin1981} and Lin
\cite{LinCCBook1955}. Thus $I(f)=0$ means the flow is neutrally
stable. And there is only one neutral mode with $k=0$ and
$c_r=U_s$ in the flow. These conclusions can be summarized as a
new theorem.

Tollmien-Fridrichs-Lin theorem: The flows are neutrally stable, if
$I(f)=0$. The flows are stable and unstable for $I(f)> 0$ and
$I(f)< 0$, respectively.

From the above proof, it is obvious that the shortwaves (e.g.
$k\gg k_N$) are always more stable than longwaves (e.g. $k\ll
k_N$) in the inviscid shear flows \cite{Drazin1966,SunL2006b}. So
the shear instability is due to long-wave instability, and the
disturbances of shortwaves can be damped by the shear itself
without any viscosity \cite{SunL2006b}.

As mentioned above, the linear stability criterion can be derived
from the nonlinear one \cite{Arnold1969}. Moreover, the nonlinear
stability criterion can also be obtained from the linear one. To
illuminate this, the nonlinear criterion is retrieved
explicitly via the above theorem, 
which is briefly proved as follows.

Similar to Arnol'd's definition, the general energy $H$ here is
defined as
 \begin{equation}
  H=\frac{1}{2}\int_a^b [\frac{1}{2}\|\nabla\Psi\|^2+h(\Psi)] dy
\label{Eq:stable_essence_Energy_kinetic}
 \end{equation}
where $\Psi(y)$ is the streamfunction of the flow with
$\frac{\partial \Psi}{\partial y}=U(y)-U_s$, and $h(\Psi)$ is a
function of $\Psi$. The variation of $\delta H=0$ gives
 \begin{equation}
   \triangle \Psi= h'(\Psi).
\label{Eq:stable_essence_Energy_varEk}
 \end{equation}
So Arnol'd's variational principle is retrieved. And the function
$h=\|\nabla\Psi\|^2/2$ can also be obtained by solving
Eq.(\ref{Eq:stable_essence_Energy_varEk}). Then, the second variation
$\delta^2 H$ holds
 \begin{equation}
  \delta^2H=\frac{1}{4}\int_a^b [\|\nabla\psi\|^2+h''(\Psi)\psi^2]
  dy,
\label{Eq:stable_essence_Energy_2ndvar}
 \end{equation}
where $\psi$ denotes the variation $\delta\Psi$. Noting that
$h''(\Psi)$ remains unknown, Arnol'd's nonlinear criteria have not
be extensively used. Fortunately, we can obtain the explicit
expression of $h''(\Psi)$ here via the investigation on the linear
stability criterion. For $\frac{\partial \Psi}{\partial y}=U-U_s$,
 \begin{equation}
h''(\Psi)=\frac{dh'}{d\Psi}=\frac{dh'}{dy}\frac{dy}{d\Psi}=\frac{U''}{(U-U_s)}=f(y).
\label{Eq:stable_essence_confunc_2ndvar}
 \end{equation}
As $h''(\Psi)$ is solved explicitly, $\delta^2 H$ has an explicit
expression, which is greatly helpful for real applications.

Let the streamfunction of the perturbation expands as travelling
waves $\psi(x,y,t)=\phi(y)e^{i(kx-\omega t)}$, where $\omega$ is
the frequency. The averages of $\|\nabla\psi\|^2$ and $\|\psi\|^2$
are $(\|\phi'\|^2+k^2\|\phi\|^2)/2$ and $\|\phi\|^2/2$ along the
flow direction $x$, respectively. So
Eq.(\ref{Eq:stable_essence_Energy_2ndvar}) reduces to
 \begin{equation}
  \delta^2H=\frac{1}{8}\int_a^b [\|\phi'\|^2+k^2\|\phi\|^2+f(y)\|\phi\|^2]
  dy.
\label{Eq:stable_essence_Energy_2ndvar2}
 \end{equation}
The sign of $\delta^2H$ is then associated with $I(f)$ in
Eq.(\ref{Eq:stable_essence_sun_Energy}). If $I(f)<0$, the second
variation $\delta^2H$ can be both negative and positive, i.e., the
stationary solution is a saddle point. And $I(f)>0$ implies
$\delta^2H$ is positive definite and vice versa. So the stable
flow has the minimum value of the total kinetic energy $H$. The
physical meaning of $H$ can also be revealed, as the explicit
expressions of $H$, $\frac{\partial \Psi}{\partial y}=U-U_s$ and
$h''(\Psi)=f(y)$ have been obtained.

First, according to the expressions, the velocity $U$ in vorticity
conservation law Eq.(\ref{Eq:stable_paralleflow_Conserve_Vor}) can
be decomposed to two parts: the rotational flow $U-U_s$ and the
irrotational advection flow $U_s$. The vorticity $\xi$ in
Eq.(\ref{Eq:stable_paralleflow_Conserve_Vor}) depends only on
$U-U_s$, and $U_s$ is only advection velocity. Then $U-U_s$ and
$U_s$ are associated with the dynamics and kinetics of the flow,
respectively. Eq.(\ref{Eq:stable_paralleflow_Conserve_Vor})
physically shows that the vorticity field is advected by $U_s$,
which can also known from the conservation of vorticity in the
inviscid flows. A similar example is the dynamics of vortex in the
wake behind cylinder, where the vortices dominate the dynamics of
the flow and they are advected by mean flow (see Fig.2 in
\cite{Ponta2004}). The decomposition of velocity may be useful in
vortex dynamics, for that our investigation clearly shows that the
dynamics of the flow is dominated by vorticity distribution.

Then the physical meaning of $H$ can also be understood from the
above investigation. It is not $U$ but $U-U_s$ that is associated
with the general energy $H$, so $H=\frac{1}{2}\int_a^b (U-U_s)^2
dy$ is not the total kinetic energy but the kinetic energy of flow
with vorticity. Thus the stable steady states are always
minimizing the kinetic energy of the flow associated with
vorticity. This is also the reason why the flow with maximum
vorticity might be unstable, as Fj\o rtoft's criterion shows. We
would like to restate it as a theorem and to name it after Kelvin
\cite{Kelvin1875} and Arnol'd \cite{Arnold1969} for their
contributions on this field \cite{Saffman1992}.

Kelvin-Arnol'd theorem: the stable flow minimizes the kinetic
energy of flow associated with vorticity.

Both Tollmien-Fridrichs-Lin theorem and Kelvin-Arnol'd theorem are
equivalent to the following simple principle \cite{SunL2005a}: The
flow is stable, if and only if all the disturbances with $c_r=U_s$
are neutrally stable.

\subsection{Physical explanation}

We have obtained the sufficient and necessary conditions for
instability, then the physical mechanism of instability can be
understood from them. According to Fj\o rtoft's and Sun's criteria
\cite{SunL2006a}, the necessary conditions for instability require
that the base vorticity must be concentrated enough (e.g. sheet
vortex). We call it "concentrated vortex" for latter convenience.
The following investigation will reveal that the essence of shear
instability is due to the interaction between the "concentrated
vortex" and the corresponding resonant waves.

As mentioned above, the concentrated vortex is a general model of
sheet vortex in the Kelvin-Helmholtz model, for that the sheet
vortex can be recovered as the concentrated vortex
$\xi(y_s)\rightarrow\infty$ (see \cite{VallisAOFD2006} for a
comprehensive discussion about the Kelvin-Helmholtz model and
continued shear profiles). Then how the shear flow becomes
unstable, if there is a concentrated vortex? As the sufficient
condition for instability is $I(f)<0$, the normal modes in the
regime of $0\leq k< k_N$ with $c_r=U_s$ are unstable. These modes
are stationary or standing waves, comparing to the velocity at
inflection point. So that the shear instability is due to the
disturbance of concentrated vortex by the standing waves with
$c_r=U_s$. In this case, the resonance mechanism is valid. The
interaction waves propagate at the same speed with the
concentrated vortex, so that they are locked together and
amplified by simple advection \cite{Baines1994}. In short, the
disturbances on concentrated vortex is amplified like that in
Kelvin-Helmholtz model.

This instability mechanism combines both K-H instability and
resonance mechanism. Physically, the standing waves (with
$c_r=U_s$) can interact with the concentrated vortex, so they can
trigger instability in the flows. While the travelling waves (with
$c_r\neq U_s$) have no interaction with the concentrated vortex,
so that they can not trigger instability in the flows. This is the
mechanism of shear instability. As pointed out above, the inviscid
shear instability is due to long-wave instability. If the
longwaves are unstable, they can obtain the energy from background
flows. Thus, the energy within small scales transfer to and
concentrate on large scales. In a word, the shear instability
itself provides a mechanism to inverse energy cascade and to
maintain the large structures or coherent structures in the
complex flows.

\begin{figure}
\centerline{\includegraphics[width=6cm]{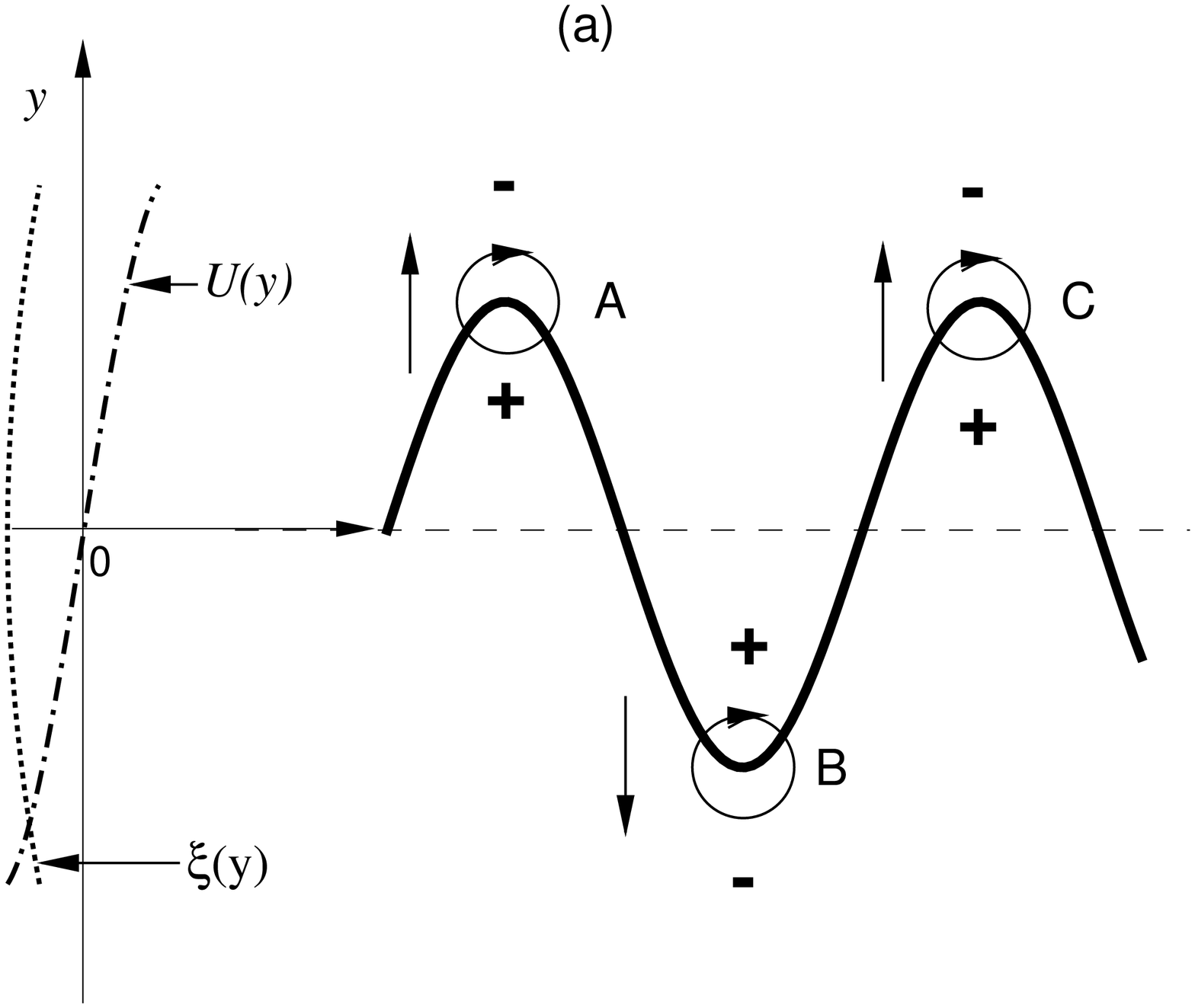}
  \includegraphics[width=6cm]{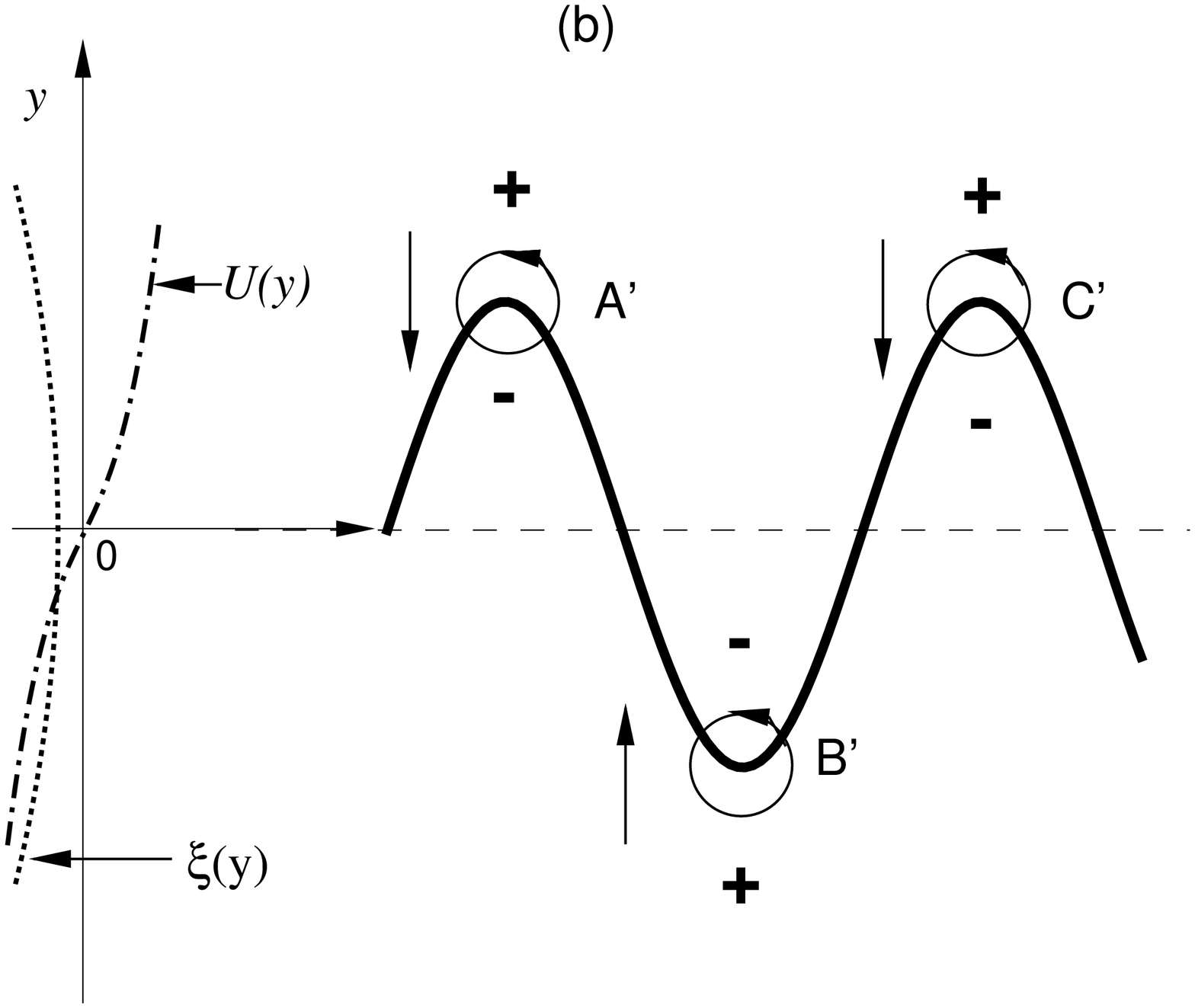}}
\caption{Sketch of shear instability: physical interpretation.
Left parts depict the profiles of velocity $U(y)$ and vorticity
$\xi(y)$, right ones depict the disturbance of vorticities. The
unstable veloctiy profile $U(y)$ has a local maximum in vorticity
$\xi(y)$ (a). If the vortices (A, B and C) disturbed from their
original positions (dashed line) to new places (solid curve), they
will be taken away from their original positions due to pressure
difference. The stable veloctiy profile $U(y)$ has a local minimum
in vorticity $\xi(y)$ (b). The disturbed vortices (A', B' and C')
will be brought back to their original positions due to pressure
difference. } \label{Fig:stable_parallelflow_mechanism}
\end{figure}

To illuminate the mechanism of shear instability, a physical
diagram is also presented here by following the way of
interpreting the K-H instability \cite{Batchelor1967}.
Fig.\ref{Fig:stable_parallelflow_mechanism} sketches the mechanism
of shear instability in terms of wave disturbances of vortices.
The mean velocity profile $U(y)$ has an inflection point at
$y_s=0$ with $U_s=0$, and the corresponding vorticity is
$\xi(y)=-U'(y)$. There is a local maximum at $y_s$ in the unstable
vorticity profile in Fig.\ref{Fig:stable_parallelflow_mechanism}a.
According to Eq.(\ref{Eq:stable_paralleflow_Conserve_Vor}), the
vorticity is conserved in the inviscid flows. If the vortices at
the local maximum (A, B and C) are sinusoidally disturbed from
their original positions (dashed line) to new places (solid
curve), they have negative vorticities with respect to the
undisturbed ones. The vortices will induce cyclone flows around
them in consequence. The flows around the vortices become faster
(slower) in the upper (lower) of vortex A referred to the basic
flow $U(y)$. The pressures at upper and lower decrease (indicated
by - signs) and increase (indicated by + signs) according to
Bernoulli's theorem, respectively. Then vortex A gets a upward
acceleration due to the disturbed pressure difference as the
uparrow shows. This tends to take the vortex away from its
original position, so the flow is unstable. On the other hand,
Fig.\ref{Fig:stable_parallelflow_mechanism}b depicts the
disturbances in a stable velocity profile, where a local minimum
is in the vorticity profile. The disturbed vortices have positive
vorticities with respect to the undisturbed ones. The vortices
will induce anticyclone flows around them in consequence. So
vortex A' get a downward acceleration due to the disturbed
pressure difference as the downarrow shows. This tends to bring
the vortex back from its original position, so the flow is stable.
In this interpretation, the advection of $U_s$ is independent of
the shear instability, only the flow field of $U-U_s$ and the
corresponding vorticity $\xi$ are the dominations. The unstable
disturbances in Fig. \ref{Fig:stable_parallelflow_mechanism} have
$c_r=U_s$, which consists with the above discussions. This
physically explains why the maximum and minimum vorticities have
different stable aspects.

Though Tollmien-Fridrichs-Lin theorem is a sufficient and
necessary condition for stability, the unknown $\phi$ in the
theorem restricts its application. So some simple criteria may be
more useful. For the parallel flows within interval $a\leq y\leq
b$, there are two simple criteria.

Corollary 1: The flow is stable for $f(y)>-(\frac{\pi}{b-a})^2$
\cite{SunL2005a,SunL2006a}.

Corollary 2: The flow is unstable for $f(y)<-(\frac{\pi}{b-a})^2$
\cite{SunL2005a}.

 In summary, the general stability and
instability criteria are retrieved for inviscid parallel flow
within linear context, which are associated with the nonlinear
stability criteria, i.e., minimizing the kinetic energy of flow.
Then the mechanism of shear instability is explained as the
resonance of standing waves with the concentrated vortex at
$c_r=U_s$. The physical process is also sketched by extending the
way of interpreting K-H instability. Finally, some useful criteria
are given. These results would lead to a more comprehensive
understanding on shear instabilities, especially for undergraduate
students.

\section{Rotating Flow}

\subsection{Introduction}
The instability of the rotating flows is one of the most
attractive problems in many fields, such as fluid dynamics,
astrophysical hydrodynamics, oceanography, meteorology, etc. Among
them, the simplest one is the instability of pure rotation flow
between coaxial cylinders, i.e., Rayleigh-Taylor problem, which
has intensively been explored \cite{Chandrasekhar1961,Drazin1981}.

Two kinds of instabilities in inviscid rotating flow have been
theoretically addressed in the literatures. One is centrifugal
instability, which was first investigated by Rayleigh
\cite{Rayleigh1880,Drazin1981}. He derived the circulation
criterion for the inviscid rotating flows that a necessary and
sufficient condition for stability to axisymmetric disturbances is
that the square of the circulation does not decrease anywhere.
This criterion is also be stated as the Rayleigh discriminant
$\Phi\geq 0$ (see
Eq.(\ref{Eq:stable_taylorflow_RayleighDiscriminant}) behind), and
is always be used in astrophysical hydrodynamics \cite{Majun2006}.
It is also generalized to non-axisymmetric flows
\cite{Paul2005jfm}. The other is known as instability due to
two-dimensional disturbances in rotating flows, which is similar
to the shear instability in parallel flow. We call this
instability as the shear instability in rotating flow hereafter.
For this instability, Rayleigh also obtained a criterion, i.e.,
inflection point theorem in inviscid rotating flows, which is the
analogue of the theorem in parallel flows \cite{Rayleigh1880}.
Following this way, Howard and Gupta \cite{Howard1962b} found a
stability criterion for two-dimensional disturbance in inviscid
rotating flow. However, the theoretical results remains scarce,
due to the complexity of rotating flow.

Comparing this to shear instability in parallel flows, the
criteria for parallel flows are much more abundant. Fj\o rtoft
\cite{Fjortoft1950} and Sun \cite{SunL2006a} proved some more
strict criteria. And the stability of two-dimensional in a
rotating frame was also addressed by Pedley \cite{Pedley1969},
which seems to be more complex than the stability problem in the
pure rotation flows.

Motivated, then, by the theoretical criteria for parallel flows
\cite{Fjortoft1950,SunL2006a}, our study focuses on the
instability due to shear in inviscid rotating flows. The aim of
this letter is to obtain such criteria for the inviscid rotating
flows, and the relationship between previous criteria is also
discussed. Thus other instabilities may be understood via the
investigation here.

\subsection{Howard-Gupta Equation}
For this purpose, Howard-Gupta equation (hereafter H-G equation)
\cite{Howard1962b} is employed. To obtain H-G equation, Euler's
equations
\cite{Chandrasekhar1961,Batchelor1967,Drazin1981,CriminaleBook2003}
for incompressible barotropic flow in cylindrical polar
coordinates $(r,\theta)$ are then given by

 \begin{equation}
 \frac{\partial u_r}{\partial t} + u_r\frac{\partial u_r}{\partial
 r}+\frac{u_\theta}{r}\frac{\partial u_r}{\partial
 \theta}-\frac{u_\theta^2}{r}=-\frac{1}{\rho} \frac{\partial p}{\partial
 r},
 \label{Eq:stable_taylorflow_Radial}
 \end{equation}
and
 \begin{equation}
 \frac{\partial u_\theta}{\partial t} + u_r\frac{\partial u_\theta}{\partial
 r}+\frac{u_\theta}{r}\frac{\partial u_\theta}{\partial
 \theta}+\frac{u_ru_\theta}{r}=-\frac{1}{\rho r} \frac{\partial p}{\partial
 \theta}.
 \label{Eq:stable_taylorflow_Angular}
 \end{equation}
Under the condition of incompressible barotropic flow, the
evolution equation for the vorticity can be obtained from
Eq.(\ref{Eq:stable_taylorflow_Radial}) and
Eq.(\ref{Eq:stable_taylorflow_Angular}),
 \begin{equation}
\frac{\partial \xi}{\partial t} + u_r\frac{\partial \xi}{\partial
 r}+\frac{u_\theta}{r}\frac{\partial \xi}{\partial
 \theta}=0,
 \label{Eq:stable_taylorflow_Vorticity}
 \end{equation}
where $\xi=\frac{1}{r}\frac{\partial }{\partial
r}(ru_\theta)-\frac{1}{r}\frac{\partial u_r}{\partial \theta}$ is
the vorticity of the background flow.
Eq.(\ref{Eq:stable_taylorflow_Vorticity}) can also be derived from
Fridman's vortex dynamics equation
\cite{Batchelor1967,Saffman1992}. And it admits a steady basic
solution,
 \begin{equation}
 u_r=0, u_\theta=V(r)=\Omega(r) r,
 \label{Eq:stable_taylorflow_Basicflow}
 \end{equation}
where $\Omega(r)$ is the mean angular velocity. And Rayleigh
discriminant is defined by
 \begin{equation}
 \Phi=\frac{1}{r^3}\frac{d}{dr}(\Omega r^2)^2.
 \label{Eq:stable_taylorflow_RayleighDiscriminant}
 \end{equation}

\subsection{Stable Criterion}

Then, consider the evolution of two-dimensional disturbances. The
disturbances $\psi'(r,\theta,t)$, which depend only on $r$,
$\theta$ and $t$, expand as series of waves,
 \begin{equation}
\psi'(r,\theta,t)=\phi(r)e^{i(n\theta-\omega t)},
 \label{Eq:stable_taylorflow_Disturbance}
 \end{equation}
where $\phi(r)$ is the amplitude of disturbance, $n$ is real
wavenumber and $\omega=\omega_r+i\omega_i$ is complex frequency.
Unlike the wavenumber in Rayleigh's equation for inviscid parallel
flows, the wavenumber $n$ here must be integer for the periodic
condition of $\theta$. The flow is unstable if and only if
$\omega_i>0$. In this way, the amplitude $\phi$ satisfies
 \begin{equation}
(n\Omega-\omega)[D^*D-\frac{n^2}{r^2}]\phi-\frac{n}{r}(D\xi)\phi=0,
 \label{Eq:stable_taylorflow_HowardGuptaEq}
 \end{equation}
where  $D=d/dr$, $D^*=d/dr+1/r$. This equation is known as H-G
equation and to be solved subject to homogeneous boundary
conditions
\begin{equation}
D\phi=0 \,\, at\,\, r=r_1,r_2.
\label{Eq:stable_taylorflow_HowardGuptaBc}
\end{equation}
%


By multiplying $\frac{r\phi^{*}}{\omega-\Omega n}$ to H-G equation
Eq.(\ref{Eq:stable_taylorflow_HowardGuptaEq}), where $\phi^{*}$ is
the complex conjugate of $\phi$, and integrating over the domain
$r_1\leq r \leq r_2$, we get the following equation
\begin{equation}
\displaystyle\int_{r_1}^{r_2}
r\{\phi^*(D^*D)\phi-[\frac{n^2}{r^2}+\frac{nD(\xi)}{r(n\Omega-\omega)}]\|\phi\|^2\}dr\,=0.
\label{Eq:stable_taylorflow_HowardGupta_Inta}
 \end{equation}
Then the integration gives
\begin{equation}
\displaystyle\int_{r_1}^{r_2}
r\{\|\phi'\|^2+[\frac{n^2}{r^2}+\frac{n(\Omega
n-\omega^*)\xi'}{r\|\Omega n-\omega
\|^2}]\|\phi\|^2\}dr\,=0,
\label{Eq:stable_taylorflow_HowardGupta_Intb}
 \end{equation}
where $\phi'=D\phi$, $\xi'=D(\xi)$ and $\omega^{*}$ is the complex
conjugate of $\omega$. Thus  the real part and image part are
\begin{equation}
\displaystyle\int_{r_1}^{r_2}
r\{\|\phi'\|^2+[\frac{n^2}{r^2}+\frac{(\Omega-c_r)\xi'}{r\|\Omega-c\|^2}]\|\phi\|^2\}dr=0,
\label{Eq:stable_taylorflow_HowardGupta_Int_Rea}
 \end{equation}
and
\begin{equation}
\displaystyle\int_{r_1}^{r_2} \frac{c_i \xi'}{\|\Omega-c\|^2}\|\phi\|^2dr\,=0,
\label{Eq:stable_taylorflow_HowardGupta_Int_Img}
 \end{equation}
where $c=\omega/n=c_r+ic_i$ is the complex angular phase speed.
Rayleigh used only
Eq.(\ref{Eq:stable_taylorflow_HowardGupta_Int_Img}) to prove his
theorem: The necessary condition for instability is that the
gradient of the basic vorticity $\xi'$ must change sign at least
once in the interval $r_1<r<r_2$. The point at $r=r_s$  is called
the inflection point with $\xi'_s=0$, at which the angular
velocity of $\Omega_s=\Omega(r_s)$. This theorem is the analogue
of Rayleigh's inflection point theorem for parallel flow
\cite{Rayleigh1880,Drazin1981}.

Similar to the proof of Fj\o rtoft theorem \cite{Fjortoft1950} in
the parallel flow, we can prove the following criterion.

Theorem 1: A necessary condition for instability is that
$\xi'(\Omega-\Omega_s)<0$ (or $\xi'/(\Omega-\Omega_s)<0$)
somewhere in the flow field.

The proof of Theorem 1 is trivial, and is omitted here. This
criterion is more restrictive than Rayleigh's. Moreover, some more
restrictive criteria may also be found, if we follow the way given
by Sun \cite{SunL2006a}. If the velocity profile $\Omega(r)$ is
stable ($c_i=0$), then the hypothesis $c_i\neq0$ should result in
contradictions in some cases. So that a more restrictive criterion
can be obtained.

To find the criterion, we need estimate the rate of
$\int_{r_1}^{r_2} r\|\phi'\|^2 dr$ to $\int_{r_1}^{r_2} \|\phi\|^2
dr$,
\begin{equation}
\int_{r_1}^{r_2}r\|\phi'\|^2 dr=\mu\int_{r_1}^{r_2}\|\phi\|^2 dr,
\label{Eq:stable_taylorflow_Poincare}
\end{equation}
where the eigenvalue $\mu$ is  positive definition for $\phi \neq
0$. According to boundary condition
Eq.(\ref{Eq:stable_taylorflow_HowardGuptaBc}), $\phi$ can expand
as Fourier series. So the smallest eigenvalue value, namely
$\mu_1$, can be estimated as $\mu_1>r_1\pi^2/(r_2-r_1)^2$.

Then there is a criterion for stability using relation
(\ref{Eq:stable_taylorflow_Poincare}), a new stability criterion
may be found: the flow is stable if
$-(\mu_1+1/r_2)<\frac{\xi'}{\Omega-\Omega_s}<0$ everywhere.

To get this criterion, we introduce an auxiliary function
$f(r)=\frac{\xi'}{\Omega-\Omega_s}$, where $f(r)$ is finite at
inflection point. We will prove the criterion by two steps. At
first, we prove proposition 1: if the velocity profile is subject
to $-(\mu_1+1/r_2)<f(r)<0$, then $c_r\neq \Omega_s$.

Proof: Since $-(\mu_1+1/r_2)<f(r)<0$, then
\begin{equation}
   -(\mu_1+1/r_2)<\frac{\xi'}{\Omega-\Omega_s}\leq\frac{\xi'(\Omega-\Omega_s)}{(\Omega-\Omega_s)^2+c_i^2},
\end{equation}
and if $c_r=\Omega_s$ and $1\leq n$, this yields to
\begin{equation}
\begin{array}{rl} \displaystyle\int_{r_1}^{r_2}
r\|\phi'\|^2+[\frac{n^2}{r}+\frac{\xi'(\Omega-\Omega_s)}{\|\Omega-c\|^2}]\|\phi\|^2\,
dr &\geq \\
\displaystyle\int_{r_1}^{r_2}
[(\mu_1+\frac{1}{r_2})+\frac{1}{r}+\frac{\xi'}{(\Omega-\Omega_s)}]
\|\phi\|^2dr &>0.

\end{array}
\end{equation}
This contradicts
Eq.(\ref{Eq:stable_taylorflow_HowardGupta_Int_Rea}). So
proposition 1 is proved.

Then, we prove proposition 2: if $-(\mu_1+1/r_2)<f(r)<0$ and
$c_r\neq \Omega_s$, there must be $c_i^2=0$.

Proof: If $c_i^2\neq0$, then multiplying
Eq.(\ref{Eq:stable_taylorflow_HowardGupta_Int_Img}) by
$(c_r-c_t)/c_i$, where the arbitrary real number $c_t$ does not
depend on $r$, and adding the result to
Eq.(\ref{Eq:stable_taylorflow_HowardGupta_Int_Rea}), it satisfies
\begin{equation}
\displaystyle\int_{r_1}^{r_2}
r\{\|\phi'\|^2+[\frac{n^2}{r^2}+\frac{\xi'(\Omega-c_t)}{r\|\Omega-c\|^2}]\|\phi\|^2\}\,
dr=0.
\label{Eq:stable_taylorflow_Sun_Int} \end{equation}
But the above Eq.(\ref{Eq:stable_taylorflow_Sun_Int}) can not be
hold for some special $c_t$. For example, let $c_t=2c_r-\Omega_s$,
then there is $(\Omega-\Omega_s)(\Omega-c_t)<\|\Omega-c\|^2$, and
 \begin{equation}
\frac{\xi'(\Omega-c_t)}{\|\Omega-c\|^2}=
f(r)\frac{(\Omega-\Omega_s)(\Omega-c_t)}{\|\Omega-c\|^2}>-(\mu_1+\frac{1}{r_2}).
\label{Eq:stable_taylorflow_Sun_Ust}
 \end{equation}
This yields
\begin{equation} \int_{r_1}^{r_2}
[r\|\phi'\|^2+(\frac{n^2}{r}+\frac{\xi'(\Omega-c_t)}{\|\Omega-c\|^2})\|\phi\|^2]
dr>0,
\end{equation}
which also contradicts Eq.(\ref{Eq:stable_taylorflow_Sun_Int}). So
the second proposition is also proved.

Using 'proposition 1: if $-(\mu_1+1/r_2)<f(r)<0$ then $c_r\neq
\Omega_s$' and 'proposition 2: if $-(\mu_1+1/r_2)<f(r)<0$ and
$c_r\neq \Omega_s$ then $c_i = 0$', we find a stability criterion.

Theorem 2: If the velocity profile satisfy $-(\mu_1+1/r_2)<f(r)<0$
everywhere in the flow, it is stable.

This criterion is the analogue of the theorem proved by Sun
\cite{SunL2006a}. Both theorem 1 and theorem 2 here are more
restrictive than Rayleigh's theorem for the inviscid rotating
flows. The theorems indicate the probability that a vorticity
profile with local maximum or minimum would be stable, if it
satisfies the stable criteria. Theorem 2 implies that the rotating
flow is stable, if the distribution of vorticity is relatively
smooth. As shown by Sun \cite{SunL2006a}, the instability of
inviscid parallel flows must have vortices concentrated enough.
This is also the shear instability in rotating flows. Since
several stable criteria for inviscid rotating flows have been
obtained, it is convenient to explore the relationship among them,
as discussed followed.

The criteria for rotating flow can be applied to parallel flows,
given narrow-gap approximation. First, Pedley's criterion is
covered by the centrifugal instability criteria. As mentioned
above, Pedley \cite{Pedley1969} considered the stability of
two-dimensional flows $U$ in a frame rotating with angular
velocity $\Omega$. A criterion is found that instability occurs
locally when $2\Omega(2\Omega-U')<0$, where $U'=dU/dr$ represents
radial shear of horizontal velocity. Pedley's criterion, which is
recovered by later researches \cite{Tritton1992,Cambon1997}, is in
essence the special case of Rayleigh's circulation criterion,
i.e., $\frac{d}{dr}(\Omega^2r^4)<0$ for instability. Here the
proof is briefly given. Considering the narrow-gap approximation
$r_2-r_1=d\ll r_1$ and the large radii approximation
$1/r_1\rightarrow 0$ in Rayleigh's circulation criterion, there is
$\Omega'r\approx -U'$. So $\Phi=d(\Omega^2r^4)/{dr}/r^3=2\Omega
(2\Omega-U')<0$, which is exactly Pedley's criterion. Thus
Pedley's criterion is covered by Rayleigh's circulation criterion.
Second, the stable criteria for parallel flows, such as Rayleigh's
theorem \cite{Rayleigh1880} and Fj\o rtfot's theorem
\cite{Fjortoft1950}, can be derived from those for rotating flows,
given the narrow-gap approximation $r_2-r_1=d\ll r_1$ and the
large radii approximation $1/r_1\rightarrow 0$. Following this
way, the results of the Taylor-Couette system can also be applied
to the plane Couette system \cite{Faisst2000}. The proof is
omitted here, as the approach is trivial too. As pointed out by
Sun \cite{SunL2006a}, all of the shear instability criteria for
two-dimensional flows are the special cases of Arnol'd's nonlinear
criteria \cite{Arnold1969}, which are much more complex yet not
widely used. In general, all the known stability criteria for
parallel flows (even in a rotating frame) can be derived from the
stability criteria for rotating flows.

\begin{figure}
 \centerline{
 \includegraphics[width=6cm]{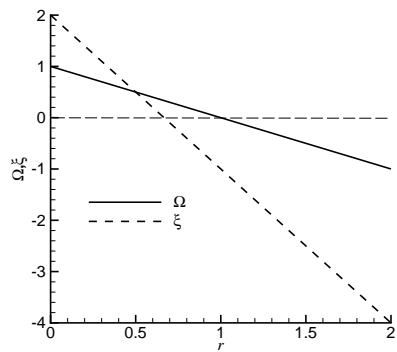}}
\caption{Angular velocity $\Omega$ (solid line) and corresponding
vorticity $\xi$ (dashed line) versus $r$. }
\label{Fig:Taylor_vorticity_profile}
\end{figure}

\subsection{Discussion}
As well known, these two kinds of instabilities are independent
with each other. Howard \cite{Howard1962a,Howard1962b, Drazin1981}
has given an example which is stable to axisymmetric disturbances
but unstable to two-dimensional (shear) disturbances. Here another
example is given that the flow is unstable to axisymmetric
disturbances but stable to two-dimensional disturbances according
the stability criteria above. To illuminate this, a simple example
is given as $\Omega(r)=1-r$ with the vorticity $\xi=2-3r$. As
shown in Fig.~\ref{Fig:Taylor_vorticity_profile}, both $\Omega$
(solid line) and $\xi$ (dashed line) are plotted within the
interval $r_1\leq r \leq r_2$. It is from Rayleigh's inflection
point criterion that the flow is always stable to two dimensional
disturbances. While the flow is unstable for the axisymmetric
disturbances, if $2/3<r_1<1$ or $2/3<r_2<1$. Then, a shear stable
flow may be unstable due to centrifugal instability. Considering
that there are three-dimensional disturbances, the flow might be
unstable, even it is centrifugal stable and shear stable. This
might be the reason why there is no general criterion known for
H-G equation when non-axisymmetric disturbances are considered
\cite{Howard1962b}.

And the relationship between rotation and stratification is also
noted. As "Rayleigh observed that there is an analogy between the
stability of rotating flows and the stability of a stratified
fluid at rest in a gravitational field" \cite{Drazin1981}, the
criteria between them can also be analogy with each other. For the
stratified flow with vertical velocity shear, Howard
\cite{Howard1961} obtained a stability criterion, i.e., the flow
is stable if $N^2>U'^2/4$, where $N$ is known as
Brunt-V\"{a}is\"{a}l\"{a} frequency, $U$ is horizontal velocity
and $U'=dU/dz$ represents vertical shear of horizontal velocity.
This criterion is then an analogue of the following criterion by
Howard and Gupta \cite{Howard1962b}: The rotating flow is stable
for axisymmetric disturbances if
Eq.(\ref{Eq:stable_taylorflow_HowardGupta_criterion}) holds,
\begin{equation}
\Phi\geq \frac{1}{4}(\frac{dW}{dr})^2,
\label{Eq:stable_taylorflow_HowardGupta_criterion}
\end{equation}
where $W$ is the axis velocity, and it vanishes in the pure
rotating flows, as in Rayleigh's criteria. Comparing the two
stability criteria, the analogues of $\Phi$ and $dW/dr$ are $N^2$
and $V'$, respectively. Physically the rotating establishes a
potential distribution along the ratio, which is denoted by
Rayleigh discriminant $\Phi$. Similarly, the stratification
establishes a potential distribution along the gravity, which is
denoted by Brunt-V\"{a}is\"{a}l\"{a} frequency. And the analogue
of the radial shear in rotating flow is right the vertical shear
in stratified flow. So the criteria between rotation and
stratification can also be analogy with each other in this way.

In short, the general stability criterion is obtained for inviscid
rotating flow. Then Pedley's cirterion is proved to be an special
case of Rayleigh's criterion. These results extend Rayleigh's
inflection point theorem for curved and rotating flows, and they
are analogues of the theorems proved by Fj\o rtoft and Sun for the
two-dimensional inviscid parallel flows.

\section{General Barotropic Wave}
Barotropic waves are widely existed in the flow with shear.
However, the general theory for barotropic waves still lacks
though the stability of linear waves (e.g. Rossby wave) are always
the most concerned. The theory of wave-mean flow interaction
points out that the flow is unstable, the disturbed waves can grow
by taking energy from the mean flow. However, why the disturbance
waves can't take energy from the mean flow in the stable flow is
not known. Nevertheless, the general criterion can be great
helpful to understand these problems. As Arnol'd's criteria deal
with the total energy, they can not provide such kind of
information. In this way, a simply linear criterion corresponding
to Arnol'd's second nonlinear stability theorem is necessary for
considering the stability of waves. It also provides a new way to
investigate shear instability. The aim of this short paper is to
find such a stability criterion from Rayleigh's equation in normal
mode way, through which the shear instability may be understood.

The above criteria would be helpful for understanding the
wave-mean flow interaction, especially the Rossby wave-mean flow
interaction in barotropic flows. According to the stable criteria,
the necessary condition for wave-mean flow interaction can be
obtained. And why the disturbed waves can't take energy from the
mean flow in the stable flow is revealed. If the flow is stable,
there is no wave-mean flow interaction at all.

First, when the velocity profile has no inflection point, then the
speeds of barotropic waves $c_r-U\neq 0$. According to
Eq.(\ref{Eq:stable_parallelflow_Rayleigh_Int_Rea}), $U-c_r>0$
holds for $U''<0$, i.e., the barotropic waves are always
west-propagation relative to the mean flow $U$ if the vortex
gradient is positive. This extends the west-propagation theory of
Rossby waves. And $U-c_r<0$ holds for $U''>0$, i.e., the
barotropic waves are always east-propagation relative to the mean
flow $U$ if the vortex gradient is negative. Second, when the
velocity profile has an inflection point $U''_s=0$, the speed of
most favorite wave which might have interaction with the mean flow
should be $c_r=U_s$. However, the wave speed $c_r\neq U_s$ holds
for $-\mu_1<\frac{U''}{U-U_s}<0$, as has been pointed out in
proposition 1. In short, the waves have no interaction with the
mean flows whatever there is an inflection point or not. This
conclusion can be easily generated to the flows in $\beta$ plane.
Thus, the waves have no interaction with the stable flows. This is
the reason why the disturbed waves can't take energy from the
stably mean flow.

On the other hand, \cite{Howard1961} pointed out that $c_r$ of a
unstable wave must lie between the minimum and the maximum values
of the mean velocity profile. Thus, the unstable wave is
stationary relative to the mean flow, and can take energy from the
mean flow. So there are wave-mean flow interactions in unstable
flows.

In summary, the general stability criterion is obtained for
inviscid both parallel and rotating flows. Then Pedley's cirterion
is proved to be an special case of Rayleigh's criterion. These
results extend Rayleigh's inflection point theorem for curved and
rotating flows, and they are analogues of the theorems proved by
Rayleigh, Tollmien and Fj\o rtoft for the two-dimensional inviscid
parallel flows. Besides, this would intrigue future research on
the mechanism of hydrodynamic instability.

\chapter{Horizontal Convection}\label{ch:HoriConv}

\section{Establishment of Horizontal Convection}

\subsection{Introduction}
The abyssal ocean circulation is thought of an important energy
conveyor belt, which has great impact to climate change
\cite[]{Wunsch2002,Rahmstorf2003,SunL2005}. As former
investigators noted the fact that density water sinks by surface
cooling at North Atlantic Ocean, this is so called density-driven
flow. Horizontal convection, in which the flow is uneven heated at
the horizontal surface, was taken as a model of such circulation.
However, a novel idea emerges when the ocean energy balance is
considered \cite[]{Munk1998,HuangRX1999,wunsch2004_ARFM}. The
horizontal convection become a key model to exam the theories. Two
classes of flows are often used. The first one, named as two-basin
forcing flows hereafter, has symmetric surface forcing like that
of South and North Atlantic Ocean basins. The multiple equilibria
and bifurcation phenomena of such flows are often discussed,
especially in numerical way
\cite[]{Quon1992,Dijkstra1997a,Fleury1997,Paparella2002}. The
second one, named as one-basin forcing flows hereafter, has
monotone forcing from one side to another like that of North
Atlantic Ocean basin. This kind of flow is more conveniently used
in the experiments
\cite[]{Rossby1965,Mullarney2004,WangWei2005}. 

Some general properties of horizontal convection have been
obtained by theoretic studies. Above all, \cite{Rossby1965} found
the 1/5-power laws of Rayleigh number $Ra$ for flow strength
$\Psi_{\max}$ and the Nusselt number $Nu$, which are consistent
with experiments and numerical simulations. This 1/5-power law of
$Ra$ is generally valid no mater what the flow is steady or
non-steady
\cite[e.g.][]{Rossby1965,Rossby1998,Quon1992,Mullarney2004,Siggers2004,WangWei2005}.
Thus it is useful to consider what the horizontal convection would
be as $Ra\rightarrow \infty$. \cite{Paparella2002} focused on the
energy dissipation under the conditions of the viscosity $\nu$ and
thermal diffusivity $\kappa$ being lowered to zero with Prandtl
number $\Pran$ fixed. They proved that the horizontal convection
is non-turbulent under certain definition of turbulence. Motivated
by above results, \cite{Siggers2004} claimed $Nu$ is bounded by
$Ra^{1/3}$ as $Ra\rightarrow \infty$.

On the other hand, numerical simulations and experiments were also
used to find details of the flows. \cite{Rossby1965,Rossby1998}
found by both experiment and numerical simulations that the flows
are steady and stable in one-basin circulation. Comparing to
Rossby's one-basin forcing, \cite{Quon1992} studied the multiple
equilibria under symmetric two-basin forcing numerically. They
also claimed that the flow is always symmetric if only theromal or
salt forcing at least for their parameter regimes. If both kinds
of forcing are considered, there will be symmetry breaking and
even unsteady flows
\cite[e.g.][]{Cessi1992,Quon1992,Dijkstra1997a,Fleury1997}. For
one-basin circulation, it was thought that horizontal convection
must be steady and stable even for $Ra\rightarrow \infty$,
according to Sandst\"{o}rm's theorem
\cite[e.g.][]{Sandstorm1904,HuangRX1999,Wunsch2000}.

But this scenario seems to be violated in recent investigations.
\cite{Paparella2002} studied the horizontal convection under the
symmetric two-basin forcing. The smaller the Prandtl number is,
the more unstable the flow is. As the Prandtl number increases to
10, the circulation tends to be a shallow cell. They also obtained
the thresholds for the transition from steady flows to unstable
and steady flows. These unsteady flows then also ware reported in
experiments \cite[]{Mullarney2004}. However, such unsteady
convective mixing and turbulent interior motion in the exmpriments
are nonturbulent according to the definition of turbulence
\cite[]{Paparella2002}. As the plume rises from bottom to top and
through the full depth of the tank, this kind of flows are
referred as full-penetrating flows. While \cite{WangWei2005}
showed in their experiments a totally different result. The motion
of the circulation, though being visible to the naked eye, is
vanishingly small. The convection cells, shallow and near the
heating surface, are steady and stable. This kind of flows are
quite different from those in \cite{Mullarney2004}, and they were
referred as partial-penetrating flows \cite[]{WangWei2005}.

Though lots of numerical simulations on the horizontal convection,
none of them have obtained such partial-penetrating flow and the
onset of such flow needs to reveal yet. The main purpose of this
paper is to investigate the partial-penetrating flow by numerical
simulation, thus resulting in a more comprehensive view on this
issue. The model and the scheme are sketched
in~\S\,\ref{sec:estab_modelscheme}, in which a benchmark solution
is included. The establishment of the circulation and the
partial-penetrating cell of horizontal convection are depicted
in~\S\,\ref{sec:estab_results}, with the onset of the
partial-penetrating flow being discussed.

\subsection{Model and Scheme}\label{sec:estab_modelscheme} We consider
the the horizontal convection flows within the two-dimensional
domain, and Boussinesq approximation is assumed to these flows.
The horizontal (y) and vertical (z) regimes are $0\leq y \leq L$
and $0\leq z\leq D$, respectively. Similar to
\cite{Paparella2002}, the depth $D$ is taken as reference length
scale and $A=D/L$ denotes the aspect ratio. We use $A=1$ in
present work, which is consistent with the experiments by
\cite{WangWei2005}. Taking account of nondivergence of velocity
filed in Boussinesq approximation, the lagrangian streamfunction
$\Psi$ and the corresponding vorticity $\omega$ are introduced.
The velocity $\overrightarrow{\mathrm{u}}=(v,w)$, where horizontal
velocity $v=\frac{\parti \Psi}{\parti z}$ and vertical velocity
$w=-\frac{\parti \Psi}{\parti y}$, respectively. The governing
equations
\cite[]{Quon1992,Dijkstra1997a,Fleury1997,Paparella2002,Siggers2004}
in vorticity-streamfunction formulation are

\begin{subeqnarray} \frac{\partial T}{\partial t} + J(\Psi,T) &=&
(\frac{\parti^2 T }{\parti y^2}+\frac{\parti^2
T }{\parti z^2})\\
\frac{\partial \omega}{\partial t} + J(\Psi,\omega) &=-& \Pran
(\nabla^2 \omega+  \Ra \frac{\parti
T}{\parti y})\\
 \nabla^2  \Psi&=&-\omega
 \label{Eq:thermo_ctl_Horizontal}
 \end{subeqnarray}
where $J(\Psi,\phi)=\frac{\parti \Psi}{\parti y}\frac{\parti
\phi}{\parti z}-\frac{\parti \phi}{\parti y}\frac{\parti
\Psi}{\parti z}$ denotes the nonlinear advection term. There are
two important dimensionless parameter in
Eq.(\ref{Eq:thermo_ctl_Horizontal}), i.e. Rayleigh number
$\Ra=\alpha_T \Delta T gL^3/(\kappa \nu)$ and Prandtl number
$\Pran=\nu/\kappa$, where $g$, $\alpha_T$, $\Delta T$, $L$,
$\kappa$ and $\nu$ are gravity acceleration, thermal expansion
coefficient, surface temperature difference, length of horizontal
domain, thermal diffusivity and kinematic viscosity, respectively.
The surface buoyancy forcing is $T=\sin(\frac{\pi}{2}y)$, so that
only one-basin flows instead of symmetric two-basin flows can be
obtained.

There are two important quantity describing the circulation, i.e.
the non-dimensional streamfunction maximum and the non-dimensional
heat flux. The non-dimensional streamfunction maximum
$\Psi_{\max}=\Psi^*_{\max}/\nu$, where $\Psi^*_{\max}$ is the
maximum of the dimensional streamfunction. For the non-dimensional
heat flux is defined as $f_T=\partial T/\partial z$ at the heated
surface. 

The above Eq.(\ref{Eq:thermo_ctl_Horizontal}) is solved with
Arakawa scheme \cite[e.g.][]{Arakawa1966,Olandi2000} and
non-uniform grids. Comparing to the other schemes, Arakawa scheme
is more accuracy but more expensive, and it has also been applied
to horizontal convection flows at high Rayleigh number
\cite[]{SunL2006_jhd1}.

\begin{table}
\center
\def~{\hphantom{0}}
\begin{tabular}{lcccccc}
author & $Ra$ & $\Psi_{mid}$ & $\Psi_{\max}$ & $u_{\max}$($y$) & $v_{\max}$($x$) & $Nu$ \\
Present & $10^6$ & $16.430$ & $16.863$ & 64.47(0.85) & 219.17 (0.038) &$8.828$ \\
\cite{Lequere1991} & $10^6$ & $16.386$ & $16.811$ & 64.83(0.85) & 220.56 (0.038) &$8.822$ \\
\cite{TianZF2003} & $10^6$ & $16.386$ & $16.811$ & 64.83(0.85) & 220.57 (0.038) &$8.825$ \\
Present & $10^7$ & $29.586$ & $30.426$ & 146.05(0.886) & 687.17 (0.022) &$16.535$ \\
\cite{Lequere1991} & $10^7$ & $29.361$ & $30.165$ & 148.59(0.879) & 699.17 (0.021) &$16.523$ \\
\cite{TianZF2003} & $10^7$ & $29.356$ & $30.155$ & 148.57(0.879) & 699.17 (0.021) &$16.511$ \\

\end{tabular}
\caption{Comparison of the bench mark solutions from Le
Qu\'{e}r\'{e} and \cite{TianZF2003}, $\Pran=0.71$. $\Psi_{mid}$,
$\Psi_{\max}$ are the values in the midpoint and the maximum of
streamfunction, respectively. $u_{\max}$ and $v_{\max}$ are the
maximum of horizontal and vertical velocity component on the
vertical midplane $x=0.5$ and $y=0.5$, together with its position
$y$ and $x$, respectively. And $Nu$ is average Nusselt number at
the heated wall. The resolution is $80\times80$ meshes for present
results.} \label{Table:NatConv_Benchmark}
\end{table}

To validate the scheme, we calculate the nature convection problem
with the resolution of $80\times80$ meshes. Table \ref{Table:NatConv_Benchmark} compares the present
solutions with bench mark solutions from \cite{Lequere1991}, who
solved the problems using a pseudo-spectral Chebyshev algorithm
and \cite{TianZF2003}, who used fourth-order compact finite
difference scheme. It is obvious that the our scheme is very
accuracy for such flows.

\subsection{Results}\label{sec:estab_results}

\subsubsection{the establishment of circulation}\label{sec:results-estab}

Similar to the experiments by \cite{WangWei2005}, we use $A=1$ for
the numeric simulations, the circulations are obtained for
$\Ra>10^7$. The dimensionless time $t=t^*/\tau$, where $t^*$ and
$\tau=D^2/\kappa$ are the dimensional and the unit scaling times,
respectively. A typical value of $t=1$ is about 80 hours in the
dimensional time, given $D=20\, cm$ and $\kappa=1.4 \times
10^{-3}\, cm^2/s$, which are approximate to the values used in the
experiments by \cite{Mullarney2004,WangWei2005}. According to the
flow pattern, the establishment of the circulation can be divided
into three stages: (1) startup of circulation, (2) damp of
secondary circulation, (3) amplification of primary circulation.
The former two stages are relatively short, while the third one is
very long. At the end of last stage, the circulation is fully
established.

\begin{figure}
\centerline{\includegraphics[width=14cm]{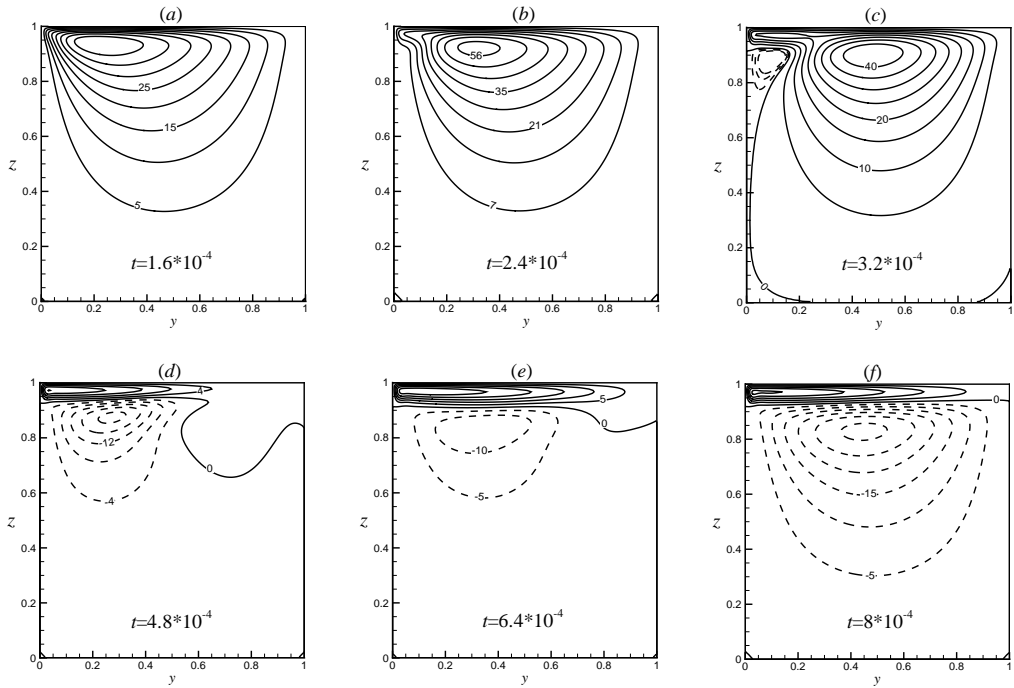}}
  \caption{Snapshots of the flow fields (streamfunction $\Psi$),
  with solid counter curves for $\Psi>0$ and dashed counter lines for $\Psi<0$.}
\label{Fig:HConv_Pr8H1_Psi_Ra5E8_dyn1}
\end{figure}
\begin{figure}
\centerline{\includegraphics[width=14cm]{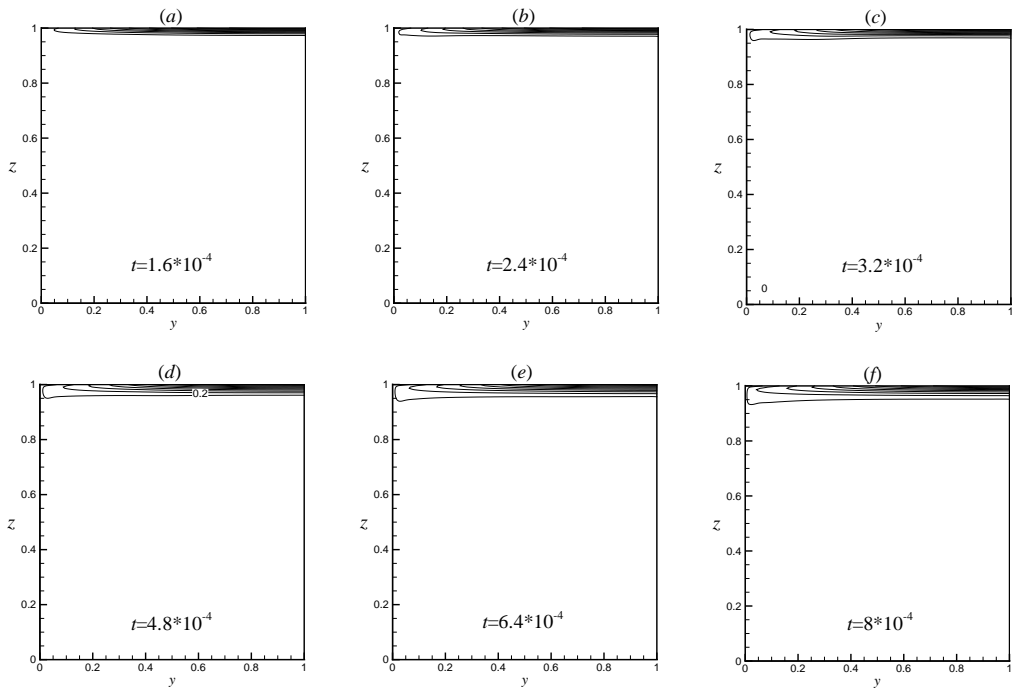}}
  \caption{Snapshots of the temperature fields corresponding to these in Fig.\ref{Fig:HConv_Pr8H1_Psi_Ra5E8_dyn1},
  with counter intervals being 0.1.}
\label{Fig:HConv_Pr8H1_Tem_Ra5E8_dyn1}
\end{figure}

To illuminate this, the circulation of $\Pran=8$ and
$Ra=5\times10^8$ is taken as an example. During the first stage,
the circulation is established as soon as the surface forcing is
superposed (Fig.\ref{Fig:HConv_Pr8H1_Psi_Ra5E8_dyn1}\textit{a}).
The gradient of horizontal buoyancy drives the water like a lid,
so that a very energetic primary circulation ($\Psi_{\max}>56$)
generates (Fig.\ref{Fig:HConv_Pr8H1_Psi_Ra5E8_dyn1}\textit{b}),
which was also observed by the experimentalist
\cite[]{WangWei2005}. But this process is much more faster, and
soon the secondary circulation emerges below the primary
circulation at $t=3.2\times10^{-4}$
(Fig.\ref{Fig:HConv_Pr8H1_Psi_Ra5E8_dyn1}\textit{c}). The
secondary circulation becomes stronger and stronger, and the
primary circulation becomes weaker and weaker. At the end of the
stage, the primary circulation becomes partial-penetrating, the
secondary circulation becomes full-penetrating
(Fig.\ref{Fig:HConv_Pr8H1_Psi_Ra5E8_dyn1}\textit{f}). Meanwhile,
the heat conducts from top to bottom along the side walls
(Fig.\ref{Fig:HConv_Pr8H1_Tem_Ra5E8_dyn1}).

\begin{figure}
\centerline{\includegraphics[width=14cm]{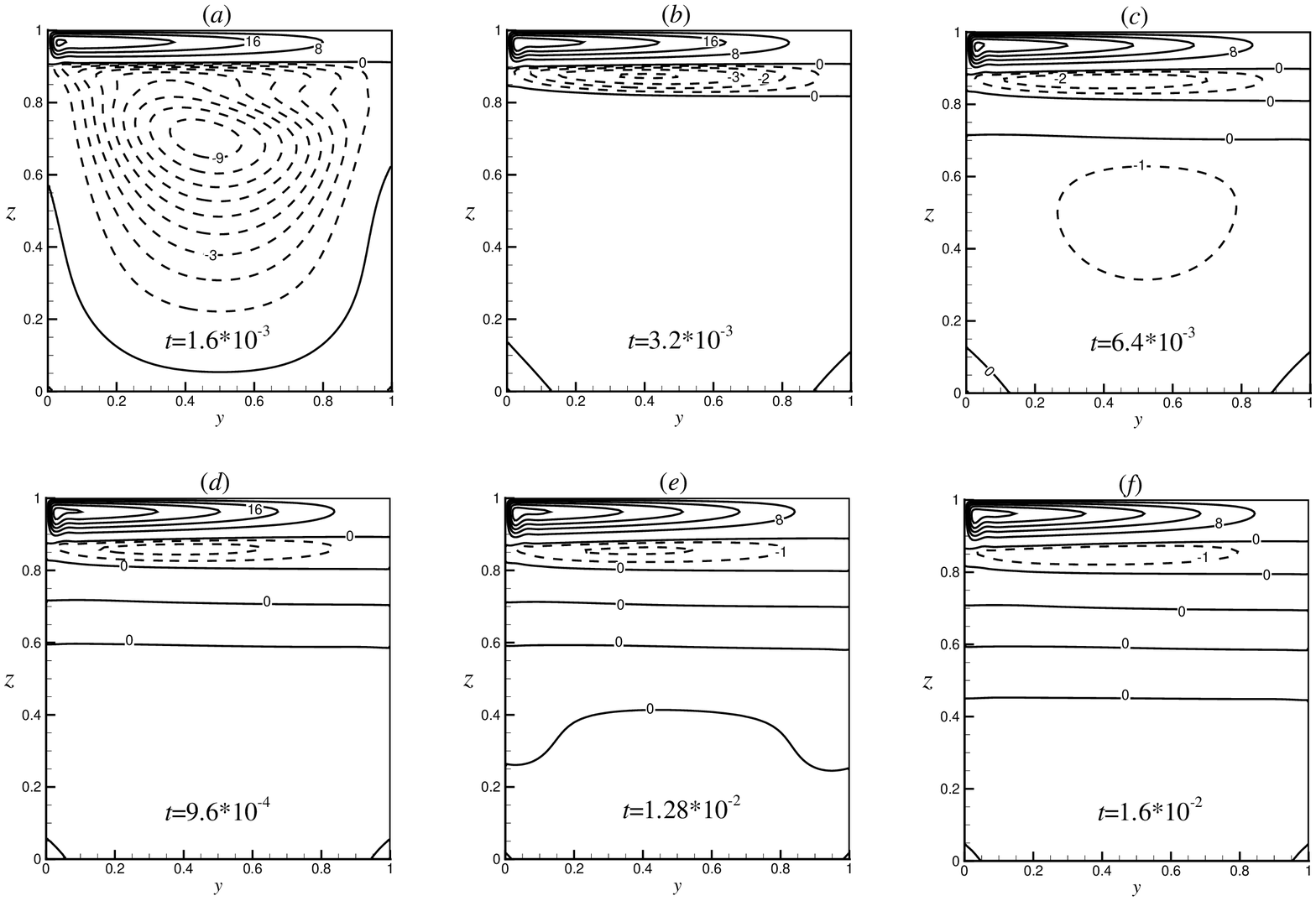}}
  \caption{Snapshots of the flow fields (streamfunction
  $\Psi$), with solid counter curves for $\Psi>0$ and dashed counter lines for
  $\Psi<0$.
  The counter intervals are 8 and -1 for $\Psi>0$ and $\Psi<0$, respectively.}
\label{Fig:HConv_Pr8H1_Psi_Ra5E8_dyn2}
\end{figure}
\begin{figure}
\centerline{\includegraphics[width=14cm]{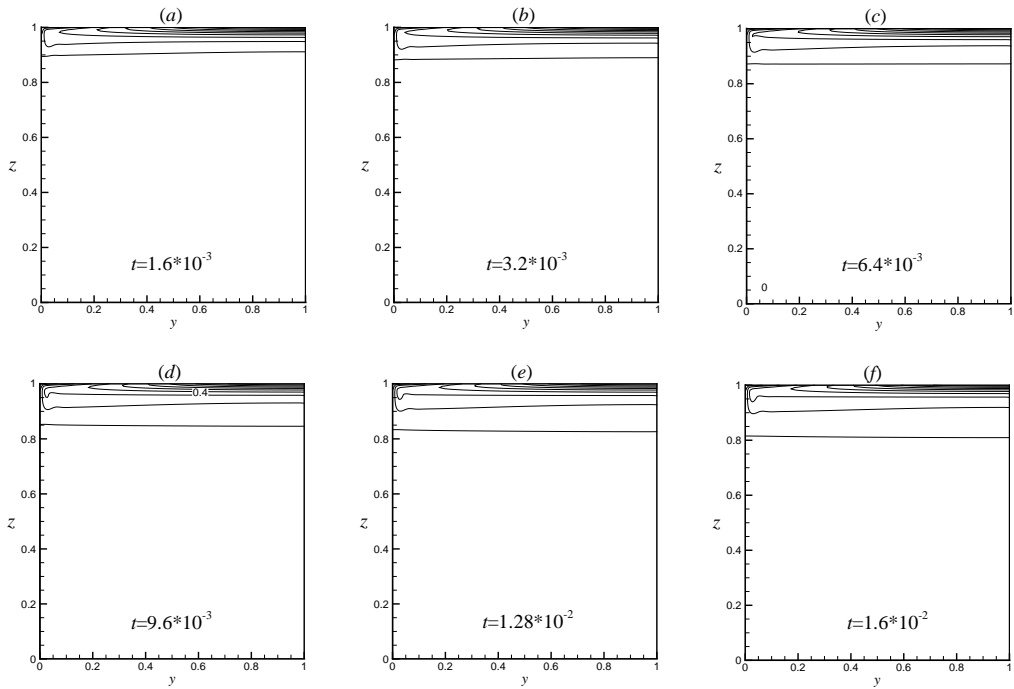}}
  \caption{Snapshots of the temperature fields corresponding to these in Fig.\ref{Fig:HConv_Pr8H1_Psi_Ra5E8_dyn2},
    with counter intervals being 0.1.}
\label{Fig:HConv_Pr8H1_Tem_Ra5E8_dyn2}
\end{figure}

Comparing to the first stage, the second one is longer, during
which the secondary circulation damps and breaks into several
weaker circulations (Fig.\ref{Fig:HConv_Pr8H1_Psi_Ra5E8_dyn2}). In
consequence, the smaller circulations emerges from the bottom one
after another. And the flow field fulfills such
partial-penetrating circulations, as observed by
\cite{WangWei2005}. The damp of the secondary circulation is a
puzzle for the experimentalists, as the corresponding temperature
field was not well measured then. Here the temperature fields
(Fig.\ref{Fig:HConv_Pr8H1_Tem_Ra5E8_dyn2}) explains the reason why
such full-penetrating circulation can't be maintained. The lack of
driven forcing and the viscous friction are the reasons. As in
Fig.\ref{Fig:HConv_Pr8H1_Tem_Ra5E8_dyn2}, there is vanishingly
small buoyancy gradient in the secondary circulation, so that it
damps down due to viscosity and boundary friction, especially the
circulation near the bottom. Meanwhile, the primary circulation
covers the convection region, where the horizontal buoyancy
gradient is remarkably large.

\begin{figure}
\centerline{\includegraphics[width=14cm]{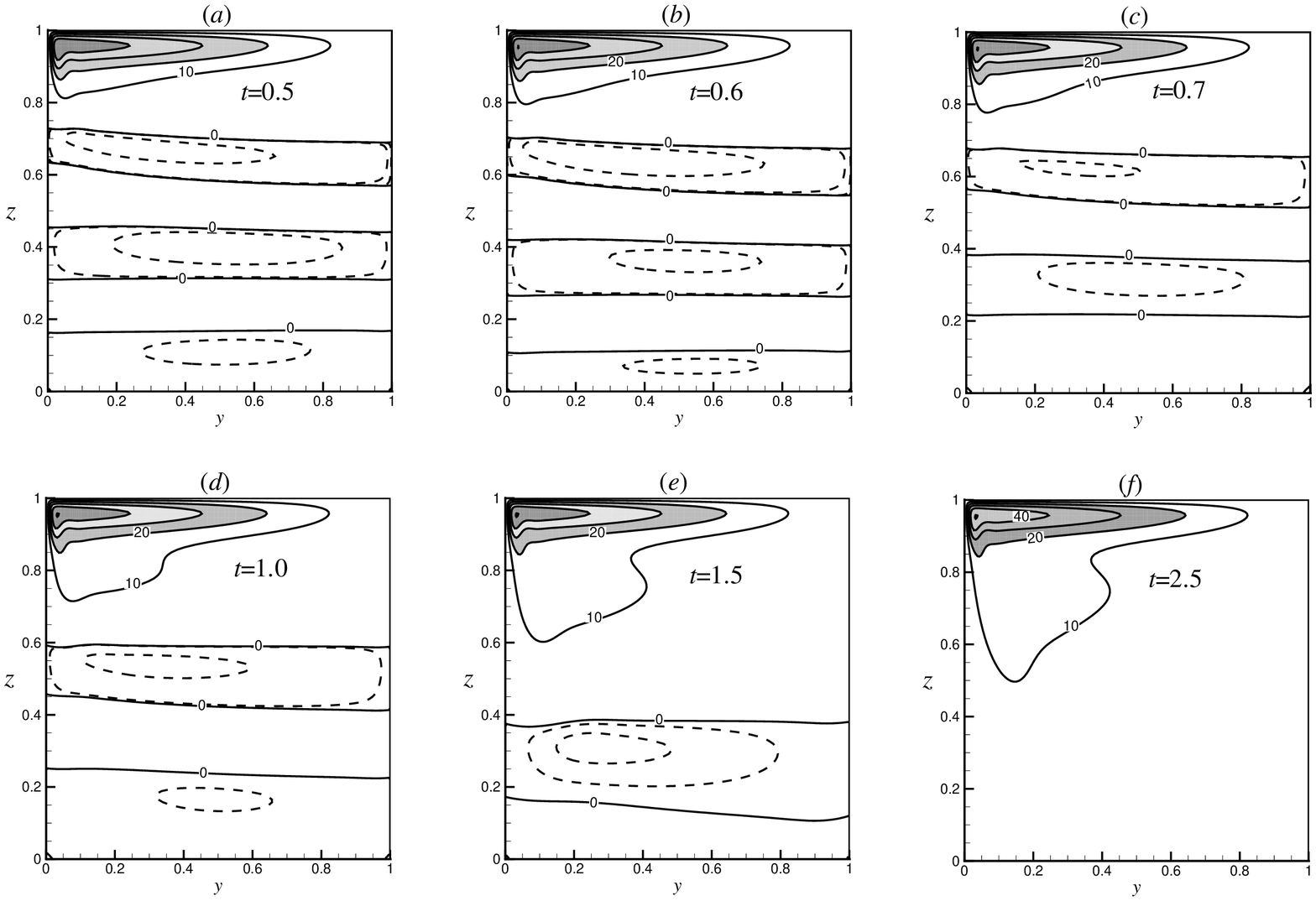}}
  \caption{Snapshots of the flow fields (streamfunction
  $\Psi$) for $\Pran=8$ and $Ra=5\times 10^8$, with solid counter curves for $\Psi>0$ and dashed counter lines for
  $\Psi<0$.
  The counter intervals are 10 for $\Psi>0$.}
\label{Fig:HConv_Pr8H1_Psi_Ra5E8_dyn3}
\end{figure}
\begin{figure}
\centerline{\includegraphics[width=14cm]{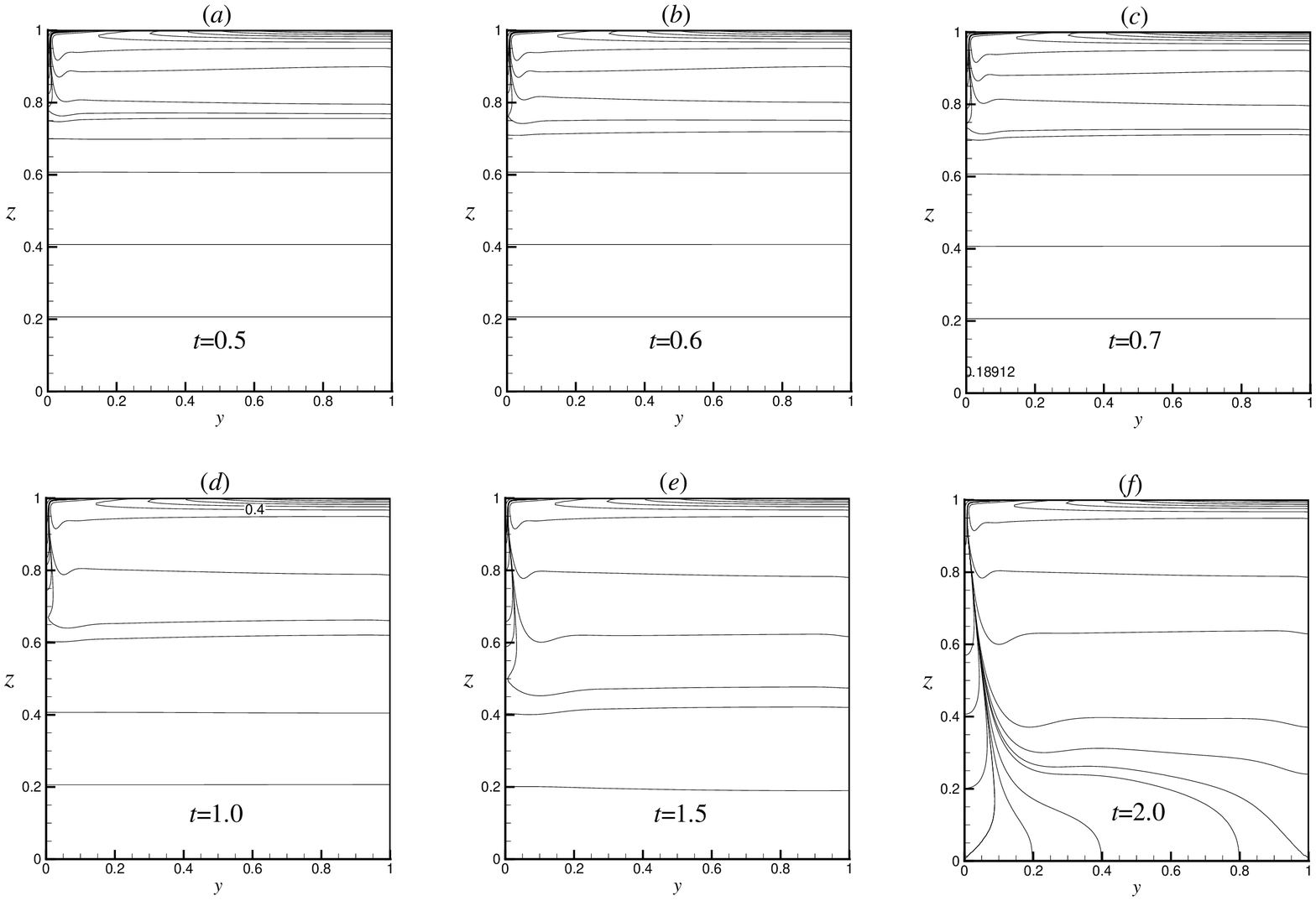}}
  \caption{Snapshots of the temperature fields corresponding to these in Fig.\ref{Fig:HConv_Pr8H1_Psi_Ra5E8_dyn3}.}
\label{Fig:HConv_Pr8H1_Tem_Ra5E8_dyn3}
\end{figure}

The last stage is a long and slowly process of approaching to
quasi-equilibrium state. The primary circulation is amplified in
this stage, and the secondary circulations disappear from the
bottom, like their emerging process
(Fig.\ref{Fig:HConv_Pr8H1_Tem_Ra5E8_dyn3}). It is notable that the
primary circulation (e.g. the shadowed shallow circulation cell in
Fig.\ref{Fig:HConv_Pr8H1_Tem_Ra5E8_dyn3}) seldom changes during
this stage, which implies that the flow near surface approaches to
quasi-equilibrium state relatively faster. While the process to
quasi-equilibrium state is very slowly near the bottom. Comparing
the flow field and temperature field, it is clear that the flow
field can't be steady until thermal conduction is balanced.

The above stages have different time scales. The startup stage is
the most fast stage, during with the circulations are established
within $t=10^{-3}$ (several minutes in laboratorial time). The
second stage and the last state are time scales of $t=10^{-2}$
(several hours) and $t=2.5$ (several days), respectively.

In the following simulations, we use the total kinetic energy
$E_k$ and $\Psi_{\max}$ as indexes to oversee the evolution of the
circulations, where $E_k=\oint (v^2+w^2)/2\, dy \,dz$. It is
notable that the time approaching to quasi-equilibrium state is
very long (about one week) in our numerical simulations, but is is
remarkably shorter in the experiments
\cite[]{Mullarney2004,WangWei2005}. According to the parameters
used by \cite{WangWei2005}, not all of the experiments take enough
time, so that the circulations are not fully established in some
cases. This is more serious in the experiments by
\cite{Mullarney2004}, where the experiment time of 30 hours is
much less than $t_e$ of 200 hours.

\begin{figure}
\centerline{\includegraphics[width=10cm]{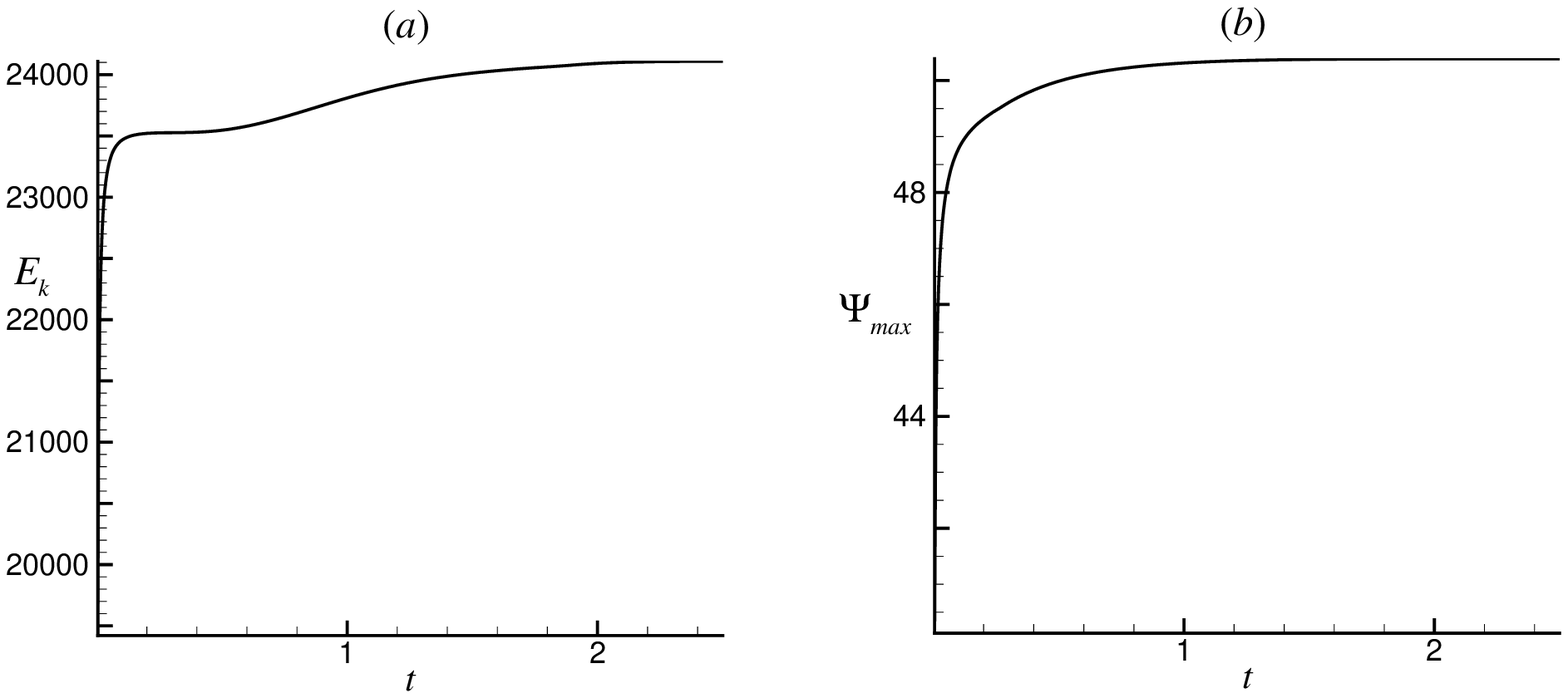}}
  \caption{The time evolution of the flow field for total kinetic energy $E_k$
  (\textit{a}) and maximum flow $\Psi_{\max}$(\textit{b}) at $Pr=8$ and $Ra=5\times10^8$.}
\label{Fig:HConv_Pr8H1_EkPsi_Ra5E8}
\end{figure}

The establishment time $t_e$ of the circulation is defined as the
time when $d\Psi/dt<0.001\Psi_{\max}$ holds for whole field. It is
found that the partial-penetrating circulation is established very
fast, but it takes a very long time for the flow to approach to
equilibrium state. Fig.\ref{Fig:HConv_Pr8H1_EkPsi_Ra5E8} displays
the time evolutions of total kinetic energy $E_k$ and
$\Psi_{\max}$ with time $t$. The circulation is established very
fast, and both $E_k$ and $\Psi_{\max}$ reach their $95\%$ of the
equilibrium values within $t=0.05$ (4 hours). In addition,
Fig.\ref{Fig:HConv_Pr8H1_EkPsi_Ra5E8}\textit{b} agrees well with
the experiments (e.g. Fig.2 in \cite{WangWei2005}), and this fast
establishment of the circulation was also noted by
\cite{Mullarney2004}. However the total time to get the
quasi-equilibrium state is relatively longer, it takes at least
$t_e=2.5$ (200 hours) for the flow being steady in present
simulation.

Moreover, the numerical simulation shows that $t_e=2.5$ is widely
valid, at least for all the cases within the regime of $1\leq
\Pran \leq 10$ and $10^7 \leq Ra \leq 10^{10}$. It also implies
that the establishment of the circulation is governed by
conduction process along the side walls, as $t_e$ is only
determined by $D$ and $\kappa$. The conclusion of $t_e=2.5$ is
also potentially useful for both numerical simulations and
laboratorial experiments.

The fast established but slowly steadied flow is due to the
horizontal evolution of buoyancy (or temperature). It is well
known that the horizontal gradient of buoyancy drives the
circulation, which can also be known from
Eq.(\ref{Eq:thermo_ctl_Horizontal}\textit{b}). Thus the fast
establishment of the temperature gradient near the surface makes
establishment of the circulation very fast. To illuminate this,
the temperature fields of the circulation are shown in
Fig.\ref{Fig:HConv_Pr8H1_Tem_Ra5E8_dyn3}. The horizontal
temperature gradient is relatively large near the top surface, but
it is relatively to small to be vanished near the bottom.
Comparing Fig.\ref{Fig:HConv_Pr8H1_Psi_Ra5E8_dyn3}\textit{a} with
Fig.\ref{Fig:HConv_Pr8H1_Tem_Ra5E8_dyn3}\textit{a}, the main
circulation is right within the zone where temperature gradient is
remarkably large, so as to the figures at other times. As the time
goes on, the main circulation becomes deeper and deeper with the
downward propagation of buoyancy gradient. So the slowly heat
conduction from the top surface to the bottom makes the
establishment of the circulation near bottom to be a long time
process. Hence, the depth $D$ of the tank and the thermal
diffusivity $\kappa$ both determinate the establishment time $t_e$
of the circulation. And it is from
Fig.\ref{Fig:HConv_Pr8H1_EkPsi_Ra5E8} that $t_e$ is 2.5 for the
circulation of $\Pran=8$ and $Ra=5\times 10^8$.

\subsubsection{power laws}

\begin{figure}
\centerline{
  \includegraphics[width=8cm]{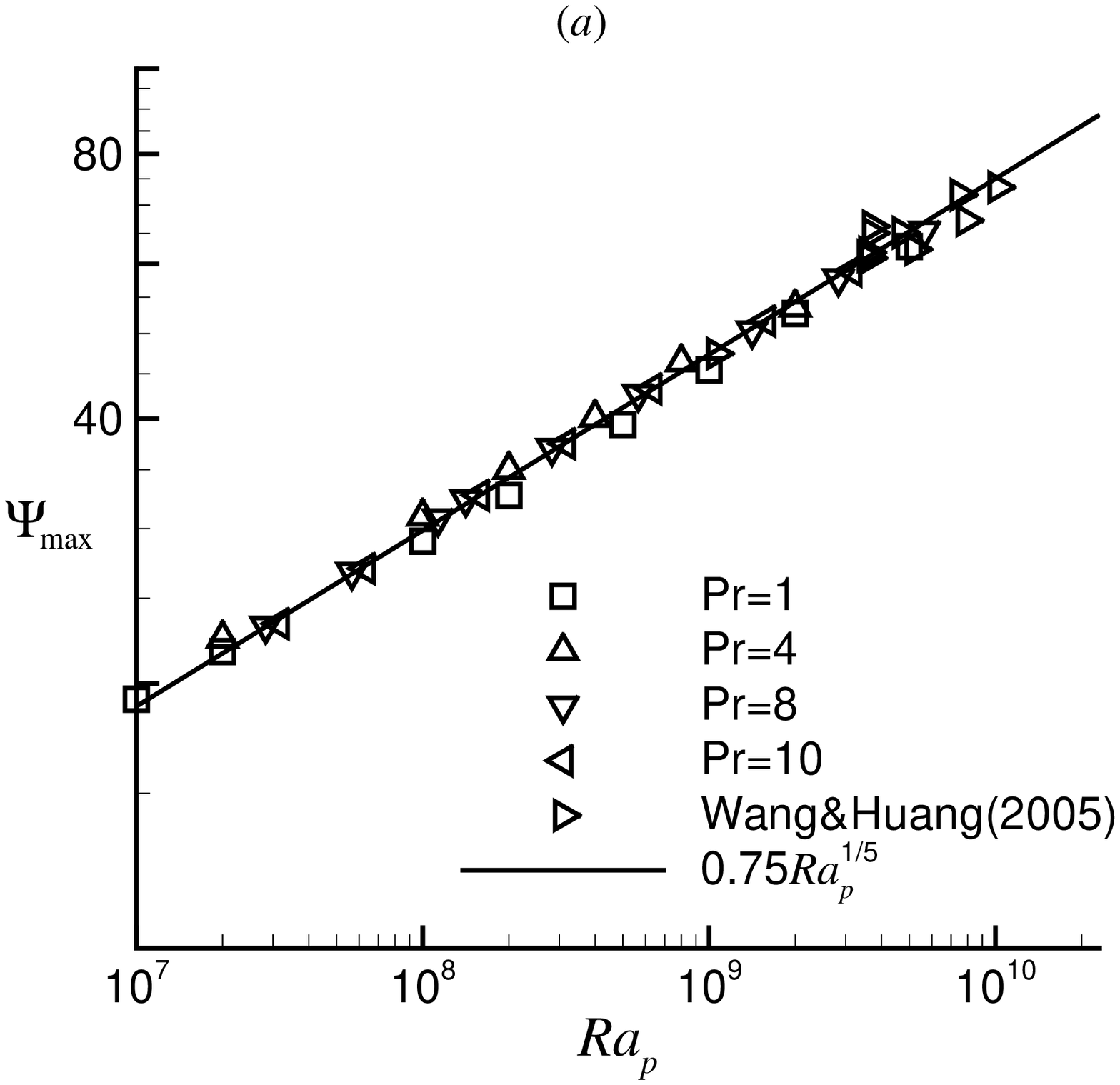}
  \includegraphics[width=8cm]{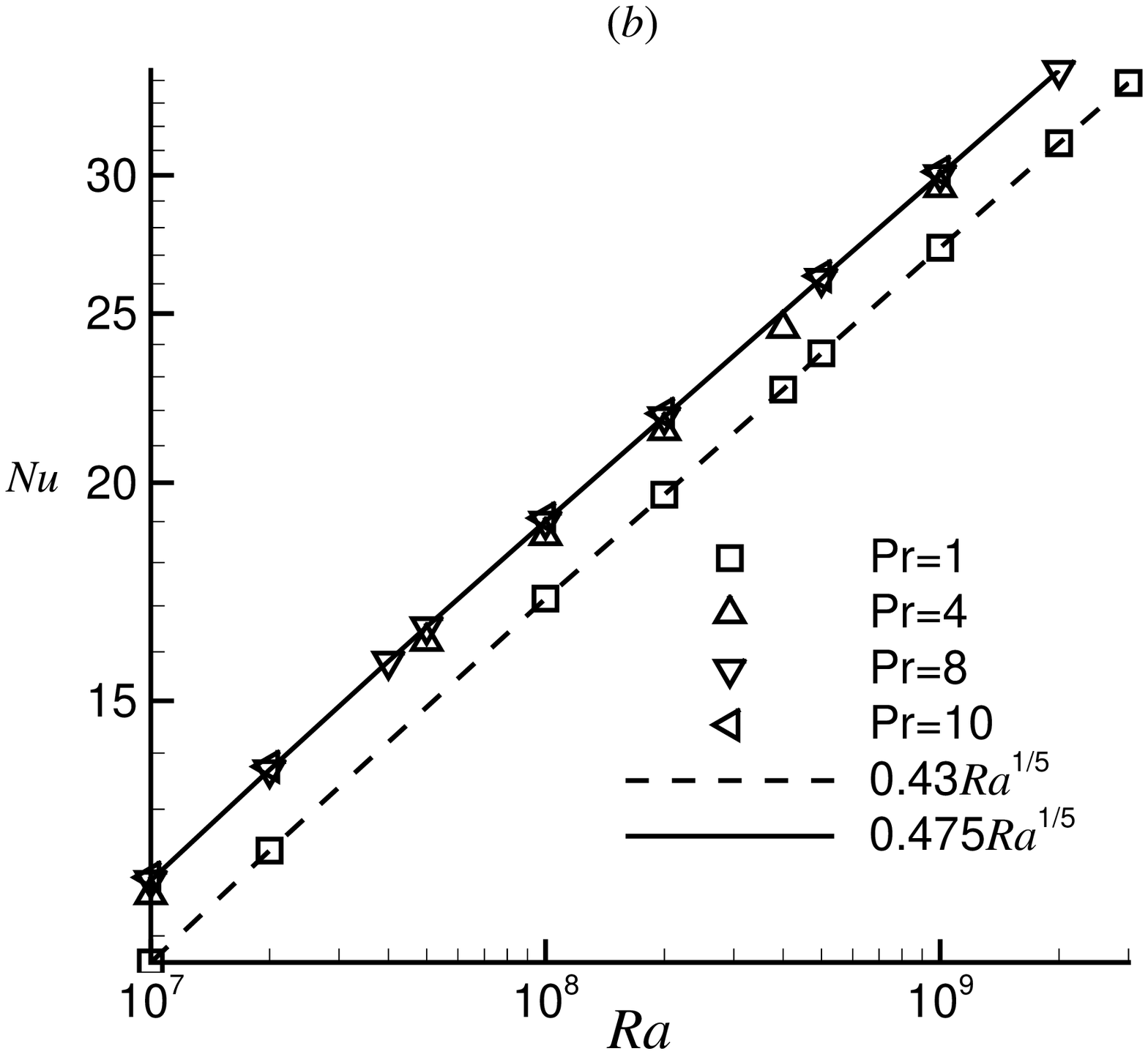}
} \caption{The flow streamfunction $\Psi_{\max}$ (\textit{a}) and
the heat flux $Nu$ (\textit{b}) vs. Rayleigh number, where
$Ra_p=RaPr^{1/2}$ in (\textit{a}).}
\label{Fig:HConv_Bc1101_PsiNu_Ra}
\end{figure}

As the parameters in different ocean global circulation models
(OGCM) varies in wide regime
\cite[e.g.][]{wunsch2004_ARFM,DijkstraB2005}, it is very useful to
know how the sensitivity of the results to the parameters. This is
also useful when the experimental results are extrapolated to real
ocean circulations \cite[]{Mullarney2004}.

Then, the power laws at different Prandtl numbers (e.g. $\Pran=1$,
$\Pran=4$, $\Pran=8$ and $\Pran=10$) are calculated to investigate
the sensitivities of $Nu$ and $\Psi_{\max}$ to $\Pran$.
Fig.\ref{Fig:HConv_Bc1101_PsiNu_Ra}\textit{a} and
Fig.\ref{Fig:HConv_Bc1101_PsiNu_Ra}\textit{b} show the 1/5-power
laws of $\Psi_{\max}$ and $Nu$, respectively. Noting that
$\Psi_{\max}$ instead of $\Psi_{\max}$ and $Ra_p=RaPr^{1/2}$
instead of $Ra$ are in
Fig.\ref{Fig:HConv_Bc1101_PsiNu_Ra}\textit{a}, it implies that the
flow is dominated by thermal diffusivity $\kappa$. The larger
Prandtl number is, the stronger the flow is. All the direct
numeric simulation (DNS) data at $\Pran=1$, $\Pran=4$, $\Pran=8$
and $\Pran=10$ lie in the line of this power law,
$\Psi_{\max}=0.75Ra_p^{1/5}$, where $Ra_p=Ra\Pran^{1/2}$. On the
other hand, \cite{WangWei2005} fitted  their experiments data into
two different 1/5-power laws. It is amazing that more than half of
the experimental data (14/25) by \cite{WangWei2005}, of which the
reduced acceleration due to gravity $g'>2.5 \, cm/s^2$, lie in
this line (right triangles in
Fig.\ref{Fig:HConv_Bc1101_PsiNu_Ra}\textit{a}). Taking account
that the aspect ratio in DNS is not exact the same as in their
experiments, these data are consistent well with each others.
Meanwhile, the Nusselt number $Nu$ in DNS lie in two slightly
displaced parallel lines, one for $\Pran=1$ and the other for
$\Pran=4$, $\Pran=8$ and $\Pran=10$. It seems that $\Psi_{\max}$
is more sensitive to Prandtl number than Nusselt number does.

It is notable that 1/5-power law in
Fig.\ref{Fig:HConv_Bc1101_PsiNu_Ra}\textit{b} is something
different with that of \cite{Siggers2004}, where $Nu$ is very
sensitive to $\Pran$ ($0.2<\Pran<4$). The fact that the power law
of $Nu$ is little sensitive to $\Pran$ as $\Pran>4$ in
Fig.\ref{Fig:HConv_Bc1101_PsiNu_Ra}\textit{b} is consistent with
the results of \cite{Rossby1998}. Here a simple explanation is
presented for these results.

Considering the steady state solution of
Eq.(\ref{Eq:thermo_ctl_Horizontal}) as $Ra\rightarrow \infty$, we
obtain the following equation by taking $\Psi=\Psi^*/\kappa$
\cite[]{Rossby1965,Quon1992}.
\begin{subeqnarray} J(T,\Psi) &=&
(\frac{\parti^2 T }{\parti y^2}+\frac{\parti^2
T }{\parti z^2})\\
Pr^{-1} J(\nabla^2\Psi,\Psi) &=& (\frac{\parti^4 \Psi }{\parti
y^4}+\frac{\parti^4 \Psi }{\parti z^4})+  \Ra \frac{\parti
T}{\parti y}
 \label{Eq:thermo_ctl_Horizontal_Gill}
 \end{subeqnarray}
When $\Pran\gg 1$, then the left convection term $Pr^{-1}
J(\nabla^2\Psi,\Psi)$ in
Eq.(\ref{Eq:thermo_ctl_Horizontal_Gill}\textit{b}) can be ignored,
and the buoyancy forcing term $Ra \frac{\partial T}{\partial y}$
balances the viscous term $\nabla^4 \Psi$. In this case, the
governing equations is independent of Prandtl number $\Pran$,
which is due to \cite{Rossby1965}. This is the reason why $Nu$ and
$\Psi_{\max}$ are little sensitive to $\Pran$ for $\Pran
>10$ in the numerical simulations by \cite{Rossby1998}. When $\Pran$ is order of 1 or even less than 1,
the convection term can not be ignored and $\Pran$ plays a role in
this case, as obtained by \cite{Siggers2004}. As $\Pran \ll 1$,
the buoyancy forcing should be balanced by strong convection at
high $Ra$ numbers, which can be seen from
Eq.(\ref{Eq:thermo_ctl_Horizontal}\textit{b}). This is also can
explain the numerical results by \cite{Paparella2002}, where the
flow becomes unsteady and has strong eddy diffusion at
$\Pran=0.1$.

\section{Partial-Penetrating Flow}

\subsection{Introduction}
Horizontal convection, in which the water is unevenly heated at
the horizontal surface, was taken as a model of abyssal ocean
circulation. The circulation, driven by density gradient and
referred as thermohaline circulation (THC) at North Atlantic
Ocean, is thought of an important energy conveyor belt and has
great impact to climate change
\cite[]{Wunsch2002,Rahmstorf2003,SunL2005}. The horizontal
convection become an important model to discuss the ocean energy
balance \cite[]{Munk1998,HuangRX1999,wunsch2004_ARFM}. Unlike the
Rayleigh-B\'{e}nard convection, the horizontal convection can be
set to motion by any small temperature gradient. Moreover, the
horizontal convection yields 1/5-power laws of $Ra$, comparing
with the 1/4-power laws in the Rayleigh-B\'{e}nard convection.

The 1/5-power laws of $Ra$ for flow strength (streamfunction
maximum $\Psi_{\max}$) and the heat flux (Nusselt number $Nu$),
first found by \cite{Rossby1965}, were later approved by both
experiments \cite[e.g.][]{Rossby1965,Mullarney2004,WangWei2005}
and numerical simulations
\cite[e.g.][]{Beardsley1973,Rossby1998,Quon1992,Siggers2004}.
According to the scaling analysis \cite[]{Rossby1965,Quon1992},
there is a boundary layer near the surface, which is inverse
proportion to 1/5 power of $Ra$. Both the flow strength and the
heat flux are dominated by the scale of boundary-layer. Although
this 1/5-power law of $Ra$ is obtained for steady flow, it is
still valid even for unsteady flow \cite[e.g.][]{Mullarney2004}.

The unsteady flow in horizontal convection was first found by
numerically \cite[]{Paparella2002}, then was observed in the
experiment \cite[]{Mullarney2004}. This unsteady flow is proved to
be non-turbulent for that the energy dissipation turns to zero
under the conditions of the viscosity $\nu$ and thermal
diffusivity $\kappa$ being lowered to zero with Prandtl number
$\Pran=\nu/\kappa$ being fixed \cite[]{Paparella2002}. Motivated
by above investigations, \cite{Siggers2004} tried to find the
bounded of $Nu$ as $Ra\rightarrow \infty$. However, their
conclusion of $Nu\leq c Ra^{1/3}$ (for some constant $c$ ) seems
not tight enough as all the numerical simulations yield 1/5-power
law.

In a recent experiment, a new flow configuration referred as
"parti-penetrating flow" was reported \cite[]{WangWei2005}.
According to the measurement, the circulation cell is shallow and
and no longer occupies the whole length of the tank. Though lots
of numerical simulations on the horizontal convection, none of
them have obtained such partial-penetrating flow. The main reason
is that all of the former simulations used free slip condition on
the walls \cite[]{Quon1992,Rossby1998,Siggers2004,SunL2006_jhd1},
the energetic circulation turns to be full-penetrating. While the
laboratory experiments always require no slip on the walls
\cite[]{WangWei2005}, the viscous drag slows down the vigorous
circulation. Noting that the power law fitted from the
experimental data is somehow coarse, this can be improved by
numerical simulations. Moreover, the partial-penetrating flow and
the onset of such flow needs to reveal too.

The main purpose of this paper is to investigate the
partial-penetrating flow and to find a more accurate power law by
numerical simulation, thus resulting in a more comprehensive view
on this issue. The model and the scheme are sketched
in~\S\,\ref{sec:partial_modelscheme}, in which a benchmark
solution is included. The establishment of the circulation and the
partial-penetrating cell of horizontal convection are depicted
in~\S\,\ref{sec:partial_results}, with the onset of the
partial-penetrating flow being discussed. Finally~\S\,\ref
{sec:conclusion} concludes.

\subsection{Model and Scheme}\label{sec:partial_modelscheme}
We consider the the horizontal convection flows within the
two-dimensional domain, and the Boussinesq approximation is
assumed to these flows. The  horizontal (y) and vertical (z)
regimes are $0\leq y \leq L$ and $0\leq z\leq D$, respectively.
Similar to \cite{Rossby1965}, the depth $L$ is taken as reference
length scale and $A=D/L$ denotes the aspect ratio. We use $A=1$ in
present work as \cite{Rossby1965} did, while $A=0.675$ is used in
the experiments by \cite{WangWei2005}. Taking account of
nondivergence of velocity field in Boussinesq approximation, the
lagrangian streamfunction $\Psi$ and the corresponding vorticity
$\omega$ are introduced. The velocity
$\overrightarrow{\mathrm{u}}=(v,w)$, where horizontal velocity
$v=\frac{\parti \Psi}{\parti z}$ and vertical velocity
$w=-\frac{\parti \Psi}{\parti y}$, respectively. The governing
equations
\cite[]{Quon1992,Dijkstra1997a,Fleury1997,Paparella2002,Siggers2004}
in vorticity-streamfunction formulation are

\begin{subeqnarray} \frac{\partial T}{\partial t} + J(\Psi,T) &=&
(\frac{\parti^2 T }{\parti y^2}+\frac{\parti^2
T }{\parti z^2})\\
\frac{\partial \omega}{\partial t} + J(\Psi,\omega) &=-& \Pran
(\nabla^2 \omega+  \Ra \frac{\parti
T}{\parti y})\\
 \nabla^2  \Psi&=&-\omega
 \label{Eq:thermo_ctl_Horizontal_parti}
 \end{subeqnarray}
where $J(\Psi,\phi)=\frac{\parti \Psi}{\parti y}\frac{\parti
\phi}{\parti z}-\frac{\parti \phi}{\parti y}\frac{\parti
\Psi}{\parti z}$ denotes the nonlinear advection term. There are
two important dimensionless parameter in
Eq.(\ref{Eq:thermo_ctl_Horizontal_parti}), i.e. Rayleigh number
$\Ra=\alpha_T \Delta T gL^3/(\kappa \nu)$ and Prandtl number
$\Pran=\nu/\kappa$, where $g$, $\alpha_T$, $\Delta T$, $L$,
$\kappa$ and $\nu$ are gravity acceleration, thermal expansion
coefficient, surface temperature difference, length of horizontal
domain, thermal diffusivity and kinematic viscosity, respectively.
The boundary condition is the same with the experiment: the
surface buoyancy forcing is $T=\sin(\frac{\pi}{2}y)$, and no slip
boundary condition is applied to walls except for surface.

There are two important quantity describing the circulation, i.e.
the non-dimensional streamfunction maximum and the non-dimensional
heat flux. The non-dimensional streamfunction maximum
$\Psi_{\max}=\Psi^*_{\max}/\nu$, where $\Psi^*_{\max}$ is the
maximum of the dimensional streamfunction. The non-dimensional
heat flux is defined as $f_T=\partial T/\partial z$ on the heated
surface. Nusselt number $Nu$, which is defined here as the maximum
of $\partial T/\partial z$ on the top surface. This definition of
$Nu$ is something different from the others
\cite[e.g.][]{Mullarney2004,Siggers2004}.

The above Eq.(\ref{Eq:thermo_ctl_Horizontal_parti}) is solved with
finite different method in non-uniform grids. Crank-Nicholson
scheme and Arakawa scheme \cite[e.g.][]{Arakawa1966,Olandi2000}
are applied to discretize the linear and nonlinear terms,
respectively. Comparing to the other schemes, Arakawa scheme is
more accuracy but more expensive, and it has also been applied to
horizontal convection flows at high Rayleigh number
\cite[]{SunL2006_jhd1}.

To validate the scheme, we calculate the nature convection problem
with the resolution of $80\times80$ meshes. Table
\ref{Table:NatConv_Benchmark} compares the present solutions with
bench mark solutions from \cite{Lequere1991}, who solved the
problems using a pseudo-spectral Chebyshev algorithm and
\cite{TianZF2003}, who used fourth-order compact finite difference
scheme. It is obvious that the our scheme is very accuracy for
such flows.

\subsection{Results}\label{sec:partial_results}
\subsubsection{spatial resolution }
First, we investigate grid dependency of the solutions to ensure
that the numerical simulations are valid. The boundary condition
is the same with the experiment: the surface buoyancy forcing is
$T=\sin(\frac{\pi}{2}y)$, and no slip boundary condition is
applied to walls except for surface. To this purpose, a case of
$Ra=2\times10^8$ and $\Pran=1$ is calculated with grids of three
different resolution, i.e. the horizontal number of meshes $N=40$,
$N=64$ and $N=80$. We find that the resolution of grids must be
fine enough, otherwise some unphysical time-depend solutions would
be obtained.

\begin{figure}
\centerline{
  \includegraphics[width=12cm]{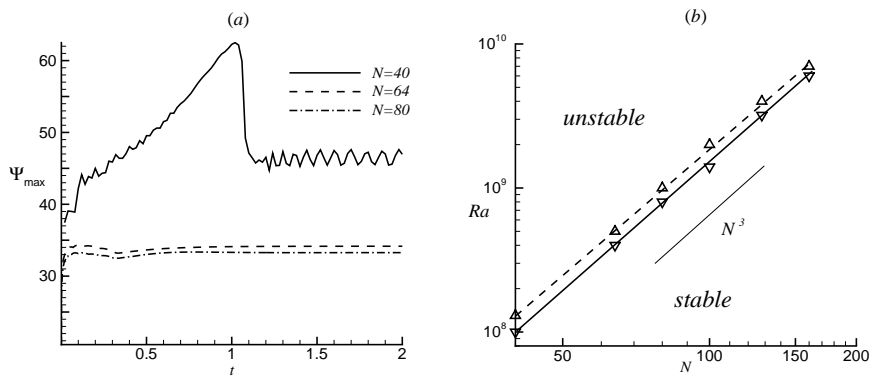} }
\caption{(\textit{a}) The maximum of streamfunction $\Psi_{\max}$
vs time $t$ for $Ra=2\times10^8$. The solid, dashed and dash-doted
curves are solutions with $N=40$, $N=64$ and $N=80$, respectively.
(\textit{b}) The stable and unstable regime on the plot of
Rayleigh number $Ra$ vs $N$. } \label{Fig:HConv_Psi_nodes}
\end{figure}

Fig.\ref{Fig:HConv_Psi_nodes}\textit{a} depicts the time evolution
of the maximum $\Psi_{\max}$. The solutions tend to be steady as
time $t>1$ for $N=64$ and $N=80$. While it becomes time-dependent
for $N=40$. It implies that some unphysical time-dependent
solutions might be obtained if the spatial resolution  is not fine
enough. To exclude the unphysical time-dependent solutions, the
numerical simulations must be obtained with sufficient spatial
resolution which depends on the Rayleigh number $Ra$. As
Fig.\ref{Fig:HConv_Psi_nodes}\textit{b} shows, the minimal number
of horizontal meshes $N$ is to obtain correct results directly
proportion to $\Ra^{1/3}$. Taking account of $\Ra\propto L^3$,
this means $N \propto L$: the longer $L$ is, the larger N is. To
obtain the physical solutions, $N$ must be within the stable
regime in Fig.\ref{Fig:HConv_Psi_nodes}\textit{b}. According to
our calculations, the flow is still steady and stable for
$Ra\leq10^{10}$.

It is from Fig.\ref{Fig:HConv_Psi_nodes}\textit{b} that $\Delta
y=L/N=C_R (\kappa \nu)^{1/3}/(\alpha_T \Delta T g)^{1/3}$, where
$\Delta y$ and $C_R=10$ are the mesh size in $y$ direction and the
coefficient, respectively. The smaller $\kappa$ and $\nu$ are, the
smaller the mesh should be. For the molecular kinematic viscosity
$\nu=1.5 \times 10^{-2}\, cm^2/s$ and thermal diffusivity
$\kappa=1.3 \times 10^{-3}\, cm^2/s$ in the case of ``run 16" by
\cite{WangWei2005}, the mesh $\Delta y$ should be $2.1\, mm$,
which is smaller than Kolmogorov scale
$\eta=(\nu^3/\epsilon)^{1/4}=5.8\, mm$, where $\epsilon=2 \times
10^{-4} cm^2/s$ is dissipation rate in the field
\cite[]{WangWei2005}. So this implies that the mesh should be fine
enough to resolute Kolmogorov scale eddies.

The resolution requirement of $\Delta y$ implies that the bound of
$L$ in numerical simulations is about laboratory scale if
molecular viscosity and diffusivity are used. And eddy viscosity
and diffusivity are required, when the length of ocean scale is
considered.

\subsubsection{the establishment of circulation}

Then the circulations are obtained numerically for $\Ra>10^7$.
Similar to \cite{Paparella2002}, the dimensionless time
$t=t^*/\tau$, where $t^*$ and $\tau=D^2/\kappa$ are the
dimensional and the unit scaling times, respectively. A typical
value of $t=1$ is about 80 hours in the dimensional time, given
$D=20\, cm$ and $\kappa=1.4 \times 10^{-3}\, cm^2/s$, which are
approximate to the values used in the experiments by
\cite{Mullarney2004,WangWei2005}. In the
following simulations, we use the total kinetic energy $E_k$ and
$\Psi_{\max}$ as indexes to oversee the evolution of the
circulations, where $E_k=\oint (v^2+w^2)/2\, dy \,dz$. It is found
that the flow is established very fast, but it takes a very long
time for the flow to approach to equilibrium state.

\begin{figure}
\centerline{\includegraphics[width=12cm]{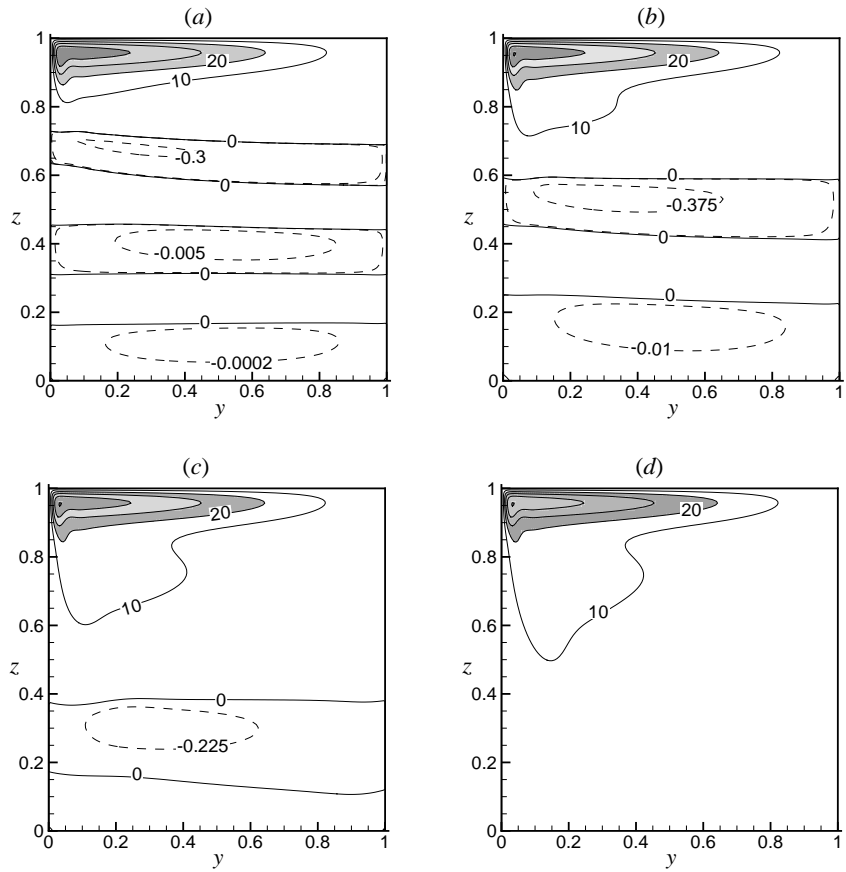}}
  \caption{The flow fields (streamfunction $\Psi$) at four different time steps: $t=0.5$ (\textit{a}),
  $t=1.0$ (\textit{b}), $t=1.5$ (\textit{c}) and $t=2.5$
 (\textit{d}) at Pr=8 and $Ra=5\times10^8$, solid curves for clockwise flow ($\Psi>0$) and dashed curves for anticlockwise flow ($\Psi<0$), respectively.
 The partial-penetrating cells ($\Psi>20$) are shadowed and the counter intervals are 10 for $\Psi>0$ in each figures.}
\label{Fig:HConv_Pr8H1_Psi_Ra5E8}
\end{figure}

To illuminate this, the circulation of $\Pran=8$ and
$Ra=5\times10^8$ is taken as an example.
Fig.\ref{Fig:HConv_Pr8H1_EkPsi_Ra5E8} displays the time evolutions
of total kinetic energy $E_k$ and $\Psi_{\max}$ with time $t$. The
circulation is established very fast, and both $E_k$ and
$\Psi_{\max}$ reach their $95\%$ of the equilibrium values within
$t=0.05$ (4 hours). In addition,
Fig.\ref{Fig:HConv_Pr8H1_EkPsi_Ra5E8}\textit{b} agrees well with
the experiments (e.g. Fig.2 in \cite{WangWei2005}), and this fast
establishment of the circulation was also noted by
\cite{Mullarney2004}. However the total time to get the
quasi-equilibrium state is relatively longer, it takes at least
$t_e=2.5$ (200 hours) for the flow being steady in present
simulation. During this process, the flow near the surface is
fully established and approaches to equilibrium state within
$t=1$, as Fig.\ref{Fig:HConv_Pr8H1_Psi_Ra5E8}\textit{a} and
Fig.\ref{Fig:HConv_Pr8H1_Psi_Ra5E8}\textit{b} show. Meanwhile,
there are several very weakly clockwise and anticlockwise
circulation cells below the primary circulation cell, which are
secondary flows due to the drag by upper primary circulation.
These secondary cells were also observed in the experiments by
\cite{WangWei2005}. As the time goes on, the primary circulation
cell becomes stronger and larger, and the secondary circulations
are weaker and smaller. So that the counter line of $\Psi=10$
becomes deeper and deeper as shown from
Fig.\ref{Fig:HConv_Pr8H1_Psi_Ra5E8}\textit{a} to
Fig.\ref{Fig:HConv_Pr8H1_Psi_Ra5E8}\textit{d}. Finally, the
primary cell fulfills the whole tank at $t=2.5$
(Fig.\ref{Fig:HConv_Pr8H1_Psi_Ra5E8}\textit{d}).

Then, the flows within $10^7\leq Ra \leq 10^{10}$ are calculated.
Fig.\ref{Fig:HConv_Bc1101_PsiNu_Ra}\textit{a} shows the 1/5-power
law of $Ra$ for $\Psi_{\max}$ at different Prandtl numbers (e.g.
$\Pran=1$, $\Pran=4$, $\Pran=8$ and $\Pran=10$). Noting that
$Ra_p=RaPr^{1/2}$ instead of $Ra$ are used in
Fig.\ref{Fig:HConv_Bc1101_PsiNu_Ra}\textit{a}, it implies that the
flow is dominated by thermal diffusivity $\kappa$. The larger
Prandtl number is, the stronger the flow is. All the direct
numeric simulation (DNS) data at $\Pran=1$, $\Pran=4$, $\Pran=8$
and $\Pran=10$ yield,
\begin{equation}
\Psi_{\max}=0.75Ra_p^{1/5}=0.75Ra^{1/5}Pr^{1/10}
 \label{Eq:Hcon_Psi_Rap}
 \end{equation}
On the other hand, \cite{WangWei2005} fitted  their experiments
data into two different 1/5-power laws: bigger one for the reduced
acceleration due to gravity $g'>2.5 \, cm/s^2$, the smaller one
for $g'<2.5 \, cm/s^2$. Equation (\ref{Eq:Hcon_Psi_Rap}) is
similar to but more accurate than the bigger one obtained by
\cite{WangWei2005}. It is notable that all the experimental data
lie around this line (right triangles in
Fig.\ref{Fig:HConv_Bc1101_PsiNu_Ra}\textit{a}), as the Prandtl
number is considered. Taking account that the aspect ratio in DNS
is not exact the same as that in their experiments, these data are
consistent well with each others.
Fig.\ref{Fig:HConv_Bc1101_PsiNu_Ra}\textit{b} shows the 1/5-power
law for $Nu$. This is consistent with the scaling analysis, i.e.
the thermal boundary-layer is inverse proportion to $Ra^{1/5}$. It
is obvious that $Nu$ is less sensitive to $Pr$ than that of
$\Psi_{\max}$, so that $Nu$ seldom changes for $\Pran>4$.

\subsubsection{the partial-penetrating cell}\label{sec:results-pr8}

It is notable that the main circulation near the surface is seldom
changed during the establishment process, especially the close
circulation cell shadowed in Fig.\ref{Fig:HConv_Pr8H1_Psi_Ra5E8}.
As this cell is shallow and the circulation is only near the
surface, it is referred as the "partial-penetrating cell" after
\cite{WangWei2005}. For example, the shadowed cell height is about
1/6 of the total depth in Fig.\ref{Fig:HConv_Pr8H1_Psi_Ra5E8}, so
that the whole cell is within the boundary-layer near the surface.
This kind of flow also exist in other Rayleigh numbers, where the
shallow and close cells like that in
Fig.\ref{Fig:HConv_Pr8H1_Psi_Ra5E8}.

An objective definition of "partial-penetrating cell" is more
convenient for further discussion. In present investigation, the
penetrating depth of the cell $D_c$ is defined as the depth of a
close circulation cell, which contains $60\%$ of the total amount.
Here the concept of partial-penetrating cell has such meanings:
(1) above all, the close circulation cell is very shallow
comparing to its width, so that  $D_c$ is within the
boundary-layer near the surface, (2) consequently, the close cell
is seldom affected by the bottom boundary, (3) moreover, the flow
in close circulation cell is dominant of the main circulation,
e.g., the flow rate in the shallow cell is about $60\%$ of the
total amount (e.g. $\Psi_c=20$  in
Fig.\ref{Fig:HConv_Pr8H1_Psi_Ra5E8}). On the contrary, "the
full-penetrating circulation" is referred to the circulation takes
whole depth of the tank hereinafter.

\begin{figure}
\centerline{\includegraphics[width=12cm]{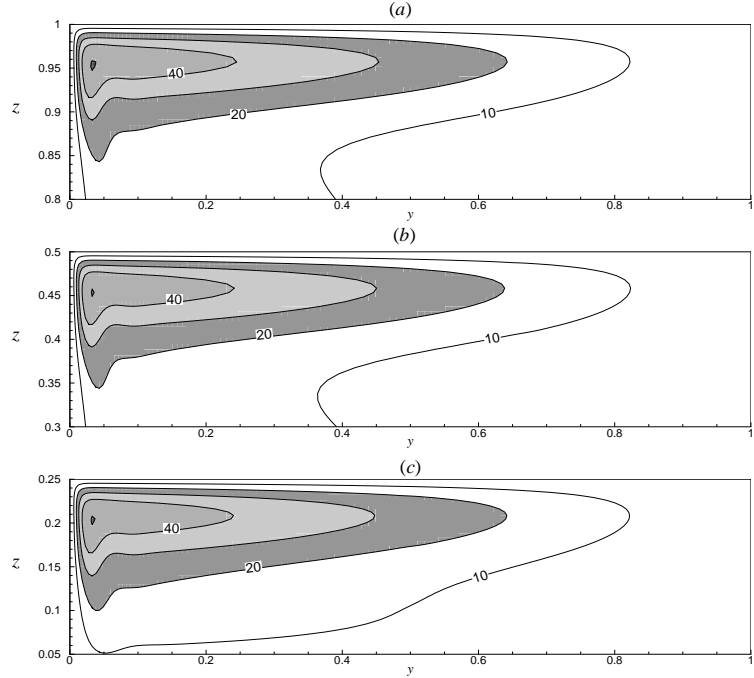}}
  \caption{The flow fields (streamfunction $\Psi$) near the forcing surface of three respective aspect ratios: $A=1$ (\textit{a}),
  $A=0.5$ (\textit{b}) and $A=0.25$ (\textit{c}) at Pr=8 and $Ra=5\times10^8$. The partial-penetrating cells
  are
 shadowed and the counter intervals are 10 in each figures.}
\label{Fig:HConv_Pr8Ra5E8_Psi_H}
\end{figure}

Using the above definition, the shadowed cell in
Fig.\ref{Fig:HConv_Pr8H1_Psi_Ra5E8} is the partial-penetrating
cell. First, it is obvious that the shadowed cell satisfies
conditions (1) and (3). Second, the close cell is seldom affected
by the bottom boundary. To illuminate this, we descend the depth
of the water tank (or equally descending aspect radio $A$).
Fig.\ref{Fig:HConv_Pr8Ra5E8_Psi_H} clear depicts that the
partial-penetrating part is little sensitive to bottom, if the
depth of the tank is large enough ($D_c<D<2D_c$ in
Fig.\ref{Fig:HConv_Pr8Ra5E8_Psi_H}\textit{c}). It may also be
noted that the flow in
Fig.\ref{Fig:HConv_Pr8Ra5E8_Psi_H}\textit{c} is very similar to
the flow field near the surface ($0.75\leq z\leq 1$) in
Fig.\ref{Fig:HConv_Pr8H1_Psi_Ra5E8}\textit{a}. For that at that
time in Fig.\ref{Fig:HConv_Pr8H1_Psi_Ra5E8}\textit{a}, the
boundary of main circulation $\Psi=0$ is about $z=0.75$, hence the
flow near the top is approximation of the case in
Fig.\ref{Fig:HConv_Pr8Ra5E8_Psi_H}\textit{c}. This also implies
that the shadowed cell in Fig.\ref{Fig:HConv_Pr8H1_Psi_Ra5E8} is
partial-penetrating.

 As the partial-penetrating cell is near the forcing
surface, it is quite independent of the flows in the middle and
the bottom. First, the partial-penetrating cell changes little
even when the full-penetrating circulation emerges in
Fig.\ref{Fig:HConv_Pr8H1_Psi_Ra5E8}\textit{c}, \textit{d}. Second,
In addition, Fig.\ref{Fig:HConv_Pr8Ra5E8_Psi_H} clear depicts that
the partial-penetrating part is little sensitive to bottom, if the
depth is large enough. Thus the partial-penetrating cell
($\Psi>\Psi_c$) is unchanged given the tank is deep enough. On the
contrary, the full-penetrating part is very sensitive to the
bottom, as the flow this part of flow approaches to the bottom.

\subsubsection{the constrain for partial-penetrating flow}

\begin{figure}
\centerline{\includegraphics[width=8cm]{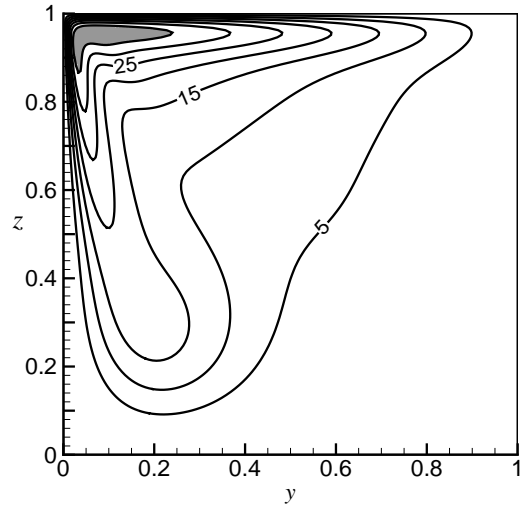}}
  \caption{The flow fields (streamfunction $\Psi$) for full-penetrating flow at Pr=1 and
  $Ra=5\times10^8$. The counter interval is 5, and $\Psi>35$ is shadowed. }
\label{Fig:HConv_Pr1H1_Psi_Ra5E8_128}
\end{figure}

\begin{figure}
\centerline{\includegraphics[width=12cm]{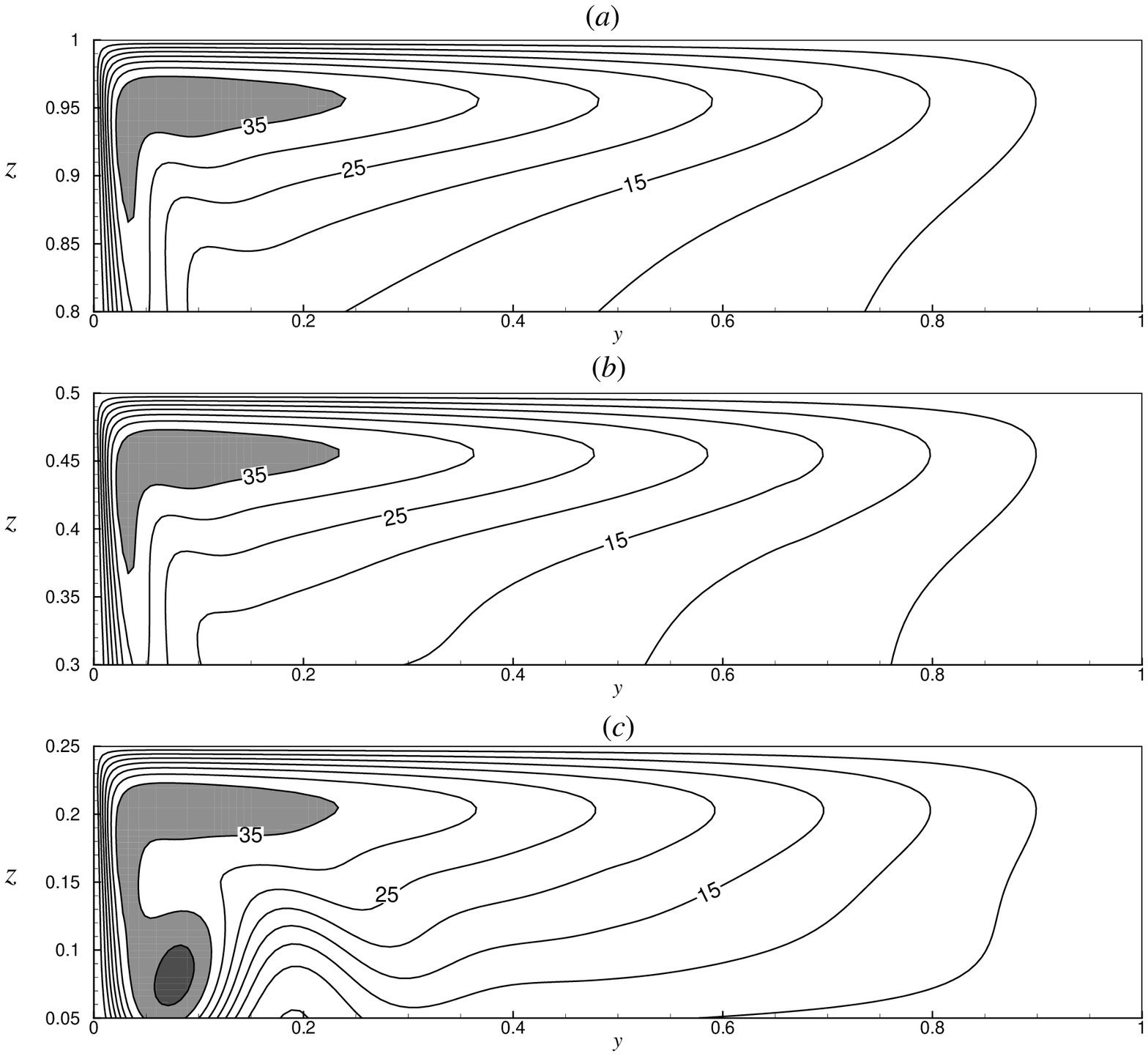}}
  \caption{The flow fields (streamfunction $\Psi$) near the forcing surface of three respective aspect ratios: $A=1$ (\textit{a}),
  $A=0.5$ (\textit{b}) and $A=0.25$ (\textit{c}) at Pr=1 and $Ra=5\times10^8$. The
  streamfunction $\Psi>35$
  are
 shadowed and the counter intervals are 5 in each figures.}
\label{Fig:HConv_Pr1Ra5E8_Psi_H}
\end{figure}

However, not all the flows have partial-penetrating cells like
that in Fig.\ref{Fig:HConv_Pr8H1_Psi_Ra5E8}, e.g. the flow at Pr=1
and $Ra=5\times10^8$ in Fig.\ref{Fig:HConv_Pr1H1_Psi_Ra5E8_128}.
It is obvious that the circulation of $\Pran=1$, which is mainly
full-penetrating, is more deep comparing to $\Pran=8$. So only the
close circulation cell for $\Psi>35$ is shallow enough, and it is
about $1/6$ of the total depth
(Fig.\ref{Fig:HConv_Pr1H1_Psi_Ra5E8_128}). While this shallow cell
is neither dominant in the whole circulation nor immune to the
changes of bottom boundary, which is shown in
Fig.\ref{Fig:HConv_Pr1Ra5E8_Psi_H}. Even this small shallow cell
is sensitive to the bottom boundary. Compare this with
Fig.\ref{Fig:HConv_Pr8Ra5E8_Psi_H}, the close circulation cell for
$\Psi>35$ is not like the partial-penetrating cell.

\begin{figure}
\centerline{\includegraphics[width=8cm]{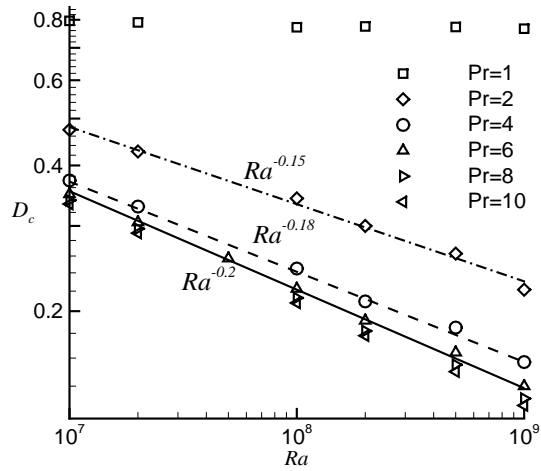}}
  \caption{$D_c$ vs. Ra. The solid, dashed and dash dotted lines are power laws of $Ra$ respectively
  for $D_c$ at $\Pran=6$,
  $\Pran=4$ and $\Pran=2$.}
\label{Fig:HConv_Bc1101_Dc_Ra}
\end{figure}

Then a new problem emerges. As there might be full-penetrating
flows, what's the constrain for the partial-penetrating flow? To
investigate this, $D_c$ is employed as an index of
partial-penetrating flow according to the above definition. If
$D_c$ satisfies $-1/5$ power law of $Ra$, then the flow is within
the boundary-layer and the flow is partial-penetrating, for that
the boundary-layer satisfies $-1/5$ power law of $Ra$
\cite[]{Rossby1965,Quon1992,Mullarney2004}. Else if $D_c$ does not
satisfy $-1/5$ power law of $Ra$, then the flow is
full-penetrating. Thus, according to the power law for $D_c$, the
flow can be referred as either partial-penetrating or
full-penetrating.

In fact, the critical parameter governing the exitance of
partial-penetrating cell is Prandtl number $\Pran$. The larger the
Prandtl number is, the more obvious the partial-penetrating cell
is. It is from Fig.\ref{Fig:HConv_Bc1101_Dc_Ra} that the smaller
$\Pran$ is, the bigger or deeper $D_c$ is. Only for the flows at
$\Pran \geq 6$, does $D_c$ satisfy the $-1/5$ power law of $Ra$.
$D_c$ seldom changes with $Ra$ even for the flows at $\Pran=1$.

It is notable that partial-penetrating flows exist for
$\Pran\geq6$, this can be understood from scale analysis
\cite[]{Rossby1965,Quon1992}. When $\Pran\gg 1$, then the left
convection term $Pr^{-1} J(\nabla^2\Psi,\Psi)$ in
Eq.(\ref{Eq:thermo_ctl_Horizontal_parti}\textit{b}) can be
ignored, and the buoyancy forcing term $Ra \frac{\partial
T}{\partial y}$ balances the viscous term $\nabla^4 \Psi$. In this
case, the governing equations is independent of $\Pran$ and the
boundary layer is inverse proportion to $Ra^{1/5}$. When $\Pran$
is order of 1 or even less than 1, the convection term can not be
ignored. So that the strong convection is full-penetrating as
depicted in Fig.\ref{Fig:HConv_Pr1H1_Psi_Ra5E8_128}.

In words, it is $\Pran$ that governs the flow pattern. When $\Pran
\geq 6$, the flow is partial-penetrating. While $\Pran \leq 4$,
the flow is full-penetrating.

\subsection{Conclusion}\label{sec:conclusion}
The experiments of horizontal convection in a rectangle cavity are
simulated numerically. The simulations agree well with the
experimental data, and a more extensive 1/5-power law of $Ra$ is
obtained by fitting the DNS data. The partial-penetrating flows
are revisited by numerical simulations. It is $\Pran$ that governs
the the existent of the partial-penetrating flow. When $\Pran \geq
6$, the flow is partial-penetrating. While $\Pran \leq 4$, the
flow is full-penetrating.

\section{Power Laws}

\subsection{Introduction}
The power laws in horizontal convection is one of the fundamental
for understanding the flows. It is reported from
\cite[]{Rossby1965} and \cite[]{WangWei2005} that the flow
strength (the maximum of stream function $\Psi_{max}$) is always
increasing with $Ra^{1/5}$. However, the flux of the flow can't be
measured very accuracy, and the parameters are bounded within
$Ra<10^{10}$ due to experimental inconvenientness. Moreover, the
numerical simulations of \cite{Paparella2002} ($0.1<Pr<10$),
\cite{Siggers2004} ($0.2<Pr<4$) can obtain both the flow strength
and the heat flux. But the Rayleigh number $Ra$ used in numerical
simulation are lower than the maximum value used in the
experiments. It is found that $\Psi_{max}\sim Ra^{1/4}$ to
$\Psi_{max}\sim Ra^{1/3}$, and $Nu\propto Ra^{1/5}$
\cite{Siggers2004}, which is consist with the results by
\cite[]{SunL2006_jhd1} at $Pr=10$. The flow strength depends
continuously to $Ra$ from $\Psi_{max}\propto Ra$ to
$\Psi_{max}\propto Ra^{1/3}$.

On the other hand, Paparella and Young (2002)
\cite[]{Paparella2002} have claimed that the flux trends to zero
as thermal diffusion trends to zero and $Pr$ is fixed. Besides,
SIggers et. al (2004) obtained that the maximum of heat flux $Nu$
is proper to $Ra^{1/3}$, which is not supported by their numerical
simulations. Then two possibilities are suggested. First, their
theory is not accuracy. Second, the numerical simulation is not
accuracy.

Motivated by the above problem, the power laws of horizontal
convection is investigated. Similar to the previous
investigations, we consider the horizontal convection flows within
the two-dimensional domain with aspect ratio $A=0.1$. In the
following parts, the flows at very high Rayleigh numbers are
simulated, and the power laws are obtained by fitting the
individual result.

\subsection{Results}
\begin{figure}
\centerline{
  \includegraphics[width=12cm]{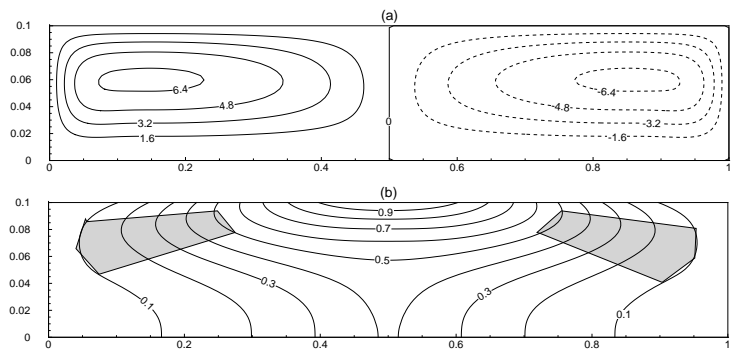}}
  \caption{The flow stream function (a) and temperature field (b) of $Ra=10^{7}$.
  It is steady and stable and symmetric with middle plume forcing, solid and dashed
  curves for positive and negative values, respectively.}
\label{Fig:HConv_Pr1H01_Ra1E7_PsiDen}
\end{figure}

When the Rayleigh number is relatively lower, the flow is weak and
the heat transportation is dominated by heat conduction
\cite{Quon1992,Siggers2004}. As the Rayleigh number increases, the
thermal convection enhances to balance the surface heat flux.
Fig.\ref{Fig:HConv_Pr1H01_Ra1E7_PsiDen} shows the stream function
and temperature field of $Ra=10^{7}$. There is a weak upwelling
flow in the center and two strong downwelling flow in the two
sidewalls, which are driven by central heating and sidewall
cooling. Accordingly, the temperature field deforms it horizontal
isothermal lines as the convection. There is a warm tongue
(shadowed in Fig.\ref{Fig:HConv_Pr1H01_Ra1E7_PsiDen}b), which is
due to the vigorous central flow near the top surface and the
central flow  near the bottom bringing the cold water to the
middle. For example, the isothermal line of $T=0.3$ extends to
cold water near top but but to warm water near bottom.

\begin{figure}
\centerline{
  \includegraphics[width=12cm]{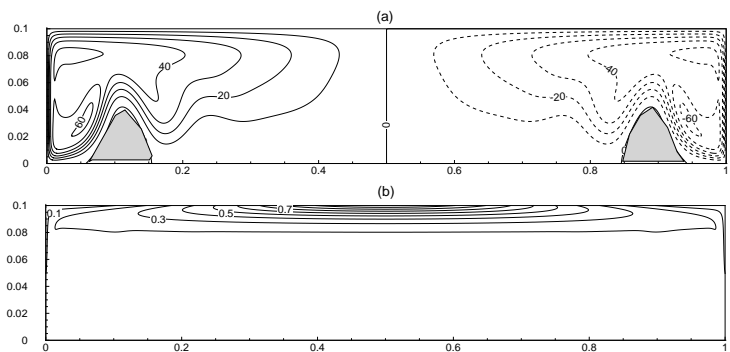}}
  \caption{The flow stream function (a) and temperature field (b) of $Ra=10^{10}$.
  It is steady and stable and symmetric with middle plume forcing, solid and dashed
  curves for positive and negative values, respectively.}
\label{Fig:HConv_Pr1H01_Ra1E10_PsiDen}
\end{figure}

When the Rayleigh number is higher, the flow field is more
complex. Fig.\ref{Fig:HConv_Pr1H01_Ra1E10_PsiDen} depicts the flow
filed and temperature field of $Ra=10^{10}$, which are something
different from the flow of $Ra=10^7$. First, the downwelling flow
departs into two separate parts: one descends the middle, the
other can descend to the bottom. Thus there are two vortex centers
in the main circulation as in
Fig.\ref{Fig:HConv_Pr1H01_Ra1E10_PsiDen}a. Besides, there is a
reverse secondary circulation along the bottom near the sidewalls
($1/10$ departs from sidewall), which have no mass interchange
with the main circulations (dead water). Consequently, they have
no contribution to the main thermal convection. Similar to
Fig.\ref{Fig:HConv_Pr1H01_Ra1E7_PsiDen}a, the isothermal lines are
concentrated  and horizontal in the center. Moreover, the
isothermal lines in Fig.\ref{Fig:HConv_Pr1H01_Ra1E10_PsiDen}b
shows there are a temperature boundary layer and a thermocline
near the surface. In total, the flow field are dominated by cold
water, where $T<0.2$ except for the boundary layer. This contracts
to the Fig.\ref{Fig:HConv_Pr1H01_Ra1E7_PsiDen}b.

Compare the temperature fields in
Fig.\ref{Fig:HConv_Pr1H01_Ra1E7_PsiDen}b and in
Fig.\ref{Fig:HConv_Pr1H01_Ra1E10_PsiDen}b, the temperature
boundary emerges near the top surface as the increase of Ra. The
temperature is homogenous in the middle and bottom, except for
near the sidewalls. The stable stratification also emerges during
this process. It is horizontal temperature gradient that driven
the surface circulation, which can be seen from the flow filed.
However, the bottom circulations are dominated by kinetic
adjustment, which has nothing to do with temperature distribution.

\subsection{Power laws}

The above investigation have obtained the flow pattern of two
different Rayleigh numbers. The general properties of the
horizontal flows are addressed here, especially for the
sensitivity of Ra. To this purpose, the power laws for flow
strength and heat flux are calculated.

\begin{figure}
\centerline{
  \includegraphics[width=8cm]{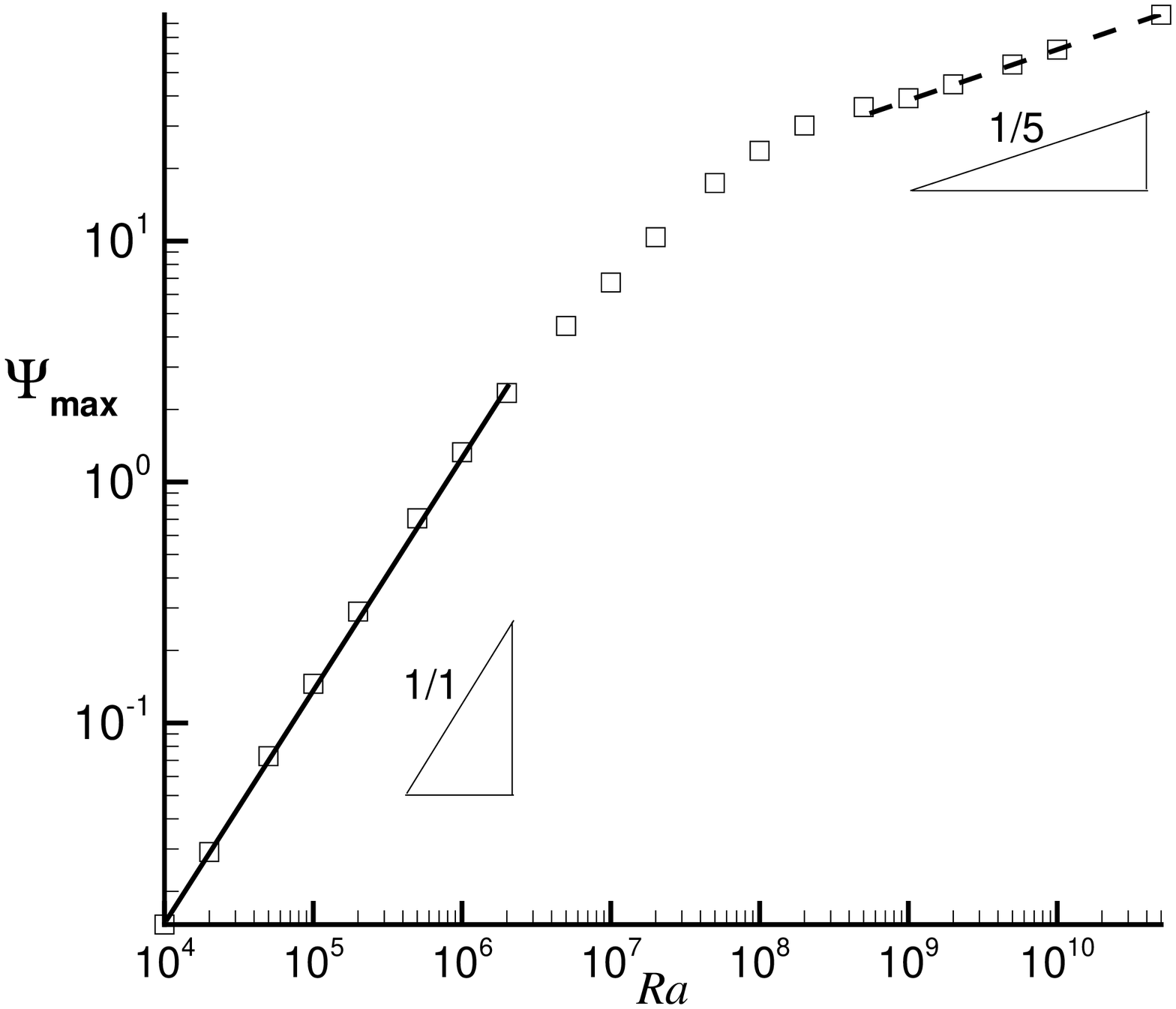}
  \includegraphics[width=8cm]{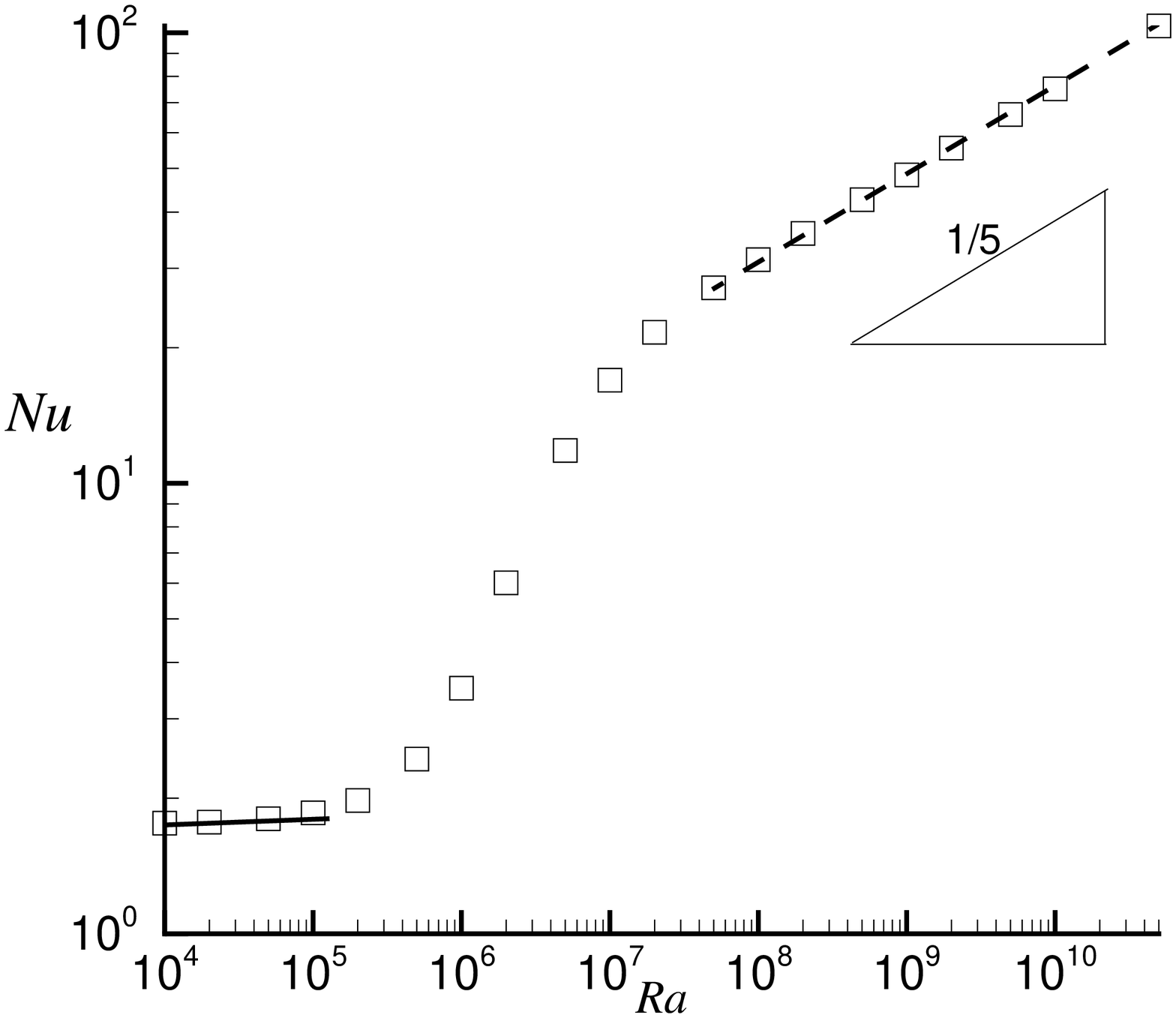}}
  \caption{The flow stream function $\Psi_{max}$ (a) and heat flux (b) vs. Ra. }
\label{Fig:HConv_Pr1H01_Ra_Psimax}
\end{figure}

First, the flow strength vs. Ra is investigated, which is the main
parameter of intensity in the thermohaline circulation. It is from
Fig.\ref{Fig:HConv_Pr1H01_Ra_Psimax}a that the flow strength
$\Psi_{max}$ is linear proper to $Ra$ as $Ra<10^6$, and that
$\Psi_{max}$ is 1/5-power proper to $Ra$ as $Ra>10^9$, between is
the transition regime. The power law of $Ra$ at high Rayleigh
number is,
\begin{equation}
\Psi_{max}=0.62Ra^{1/5}. \label{Eq:HConv_Pr1H01_Psi_Ra}
 \end{equation}
Similar to the numerical simulation by Siggers et al (2004),
$\Psi_{max}\sim Ra^{1/4}$ to $\Psi_{max}\sim Ra^{1/3}$ as
$10^8<Ra<10^9$. But there is an obvious 1/5-power law as Rayleigh
number is higher enough, which implies that this 1/5-power law is
valid only at relatively higher Ra. The power law obtained by
Siggers et al (2004) is only valid for transition regime.

Second, the power law for heat flux is also studied. As in
Fig.\ref{Fig:HConv_Pr1H01_Ra_Psimax}b, the heat flux is seldom
changed as the increase of Ra. In fact, the heat flux is dominated
by heat conduction in this regime \cite{Quon1992,Siggers2004}.
However, this changes to convection dominance as Ra larger than a
critical value $Ra\simeq 10^5$, and $Nu$ increases quickly as
$Ra$. Then, a 1/5-power law emerges as $Ra>10^8$,
\begin{equation}
Nu=0.75Ra^{1/5}. \label{Eq:HConv_Pr1H01_Nu_Ra}
 \end{equation}

Similar to the numerical simulation by Siggers et al (2004), there
is no $1/3$-power law regime in the simulations, though Ra is much
larger here. The simulations supports their first hypothesis that
the theory is overestimation to the heat flux. According to the
analysis by Rossby \cite{Rossby1965}, and Quon \cite{Quon1992},
the thickness of temperature boundary layer is proper to
$1/5$-power of Ra, which implies $Nu\sim Ra^1/5$. Compare the
power laws for $\Psi_{max}$ and $Nu$, there are always three
regimes: linear regime, transition regime and 1/5-power regime.
However, the transition from one regime to an other is not
synchronous for $\Psi_{max}$ and $Nu$, where the transition for
heat flux is leading than for the flow streamfunction.

\subsection{Summary}

In summary, the horizontal convection at high Rayleigh number in a
rectangle cavity with aspect ratio of $1:10$ is numerically
simulated. According to the results within the regime of $10^4
<Ra<10^{11}$, three continues regimes are obtained: linear regime
($10^4<Ra<10^6$), transition regime ($10^6<Ra<10^8$) and 1/5-power
law regime ($10^8<Ra<10^{11}$). For the flow strength, a 1/3-power
law of Ra is fitted when Ra is not high enough ($10^7<Ra<10^8$).
However, a 1/5-power law is obtained as Ra is high enough
($10^8<Ra<10^{11}$). The 1/5-power law confirms Rossby's analysis
and implies that 1/3-power law of Ra for Nusselt number by Siggers
et al. is over estimation.

\section{Onset of Instability}

\subsection{Introduction}
Horizontal convection, in which the water is unevenly heated at
the horizontal surface, was taken as a model of abyssal ocean
circulation. As the abyssal ocean circulation plays an important
role in climate change, the horizontal convection has intensively
been explored in recent years
\cite[]{Paparella2002,Mullarney2004,WangWei2005}. It can be set to
motion by any small temperature gradient, unlike the
Rayleigh-B\'{e}nard convection. But similar to Rayleigh-B\'{e}nard
convection, the horizontal convection may be unsteady at high
Rayleigh numbers $Ra$. There is a critical Rayleigh number $Ra_c$,
and the steady flow is unstable and becomes unsteady when
$Ra>Ra_c$. The unsteady flow in horizontal convection was first
found by numerical simulation \cite[]{Paparella2002}, then was
observed in the experiment at $Ra>10^{12}$ \cite[]{Mullarney2004}.
This unsteady flow is proved to be non-turbulent even as
$Ra\rightarrow \infty$, though the flow field seems to be chaotic
\cite[]{Paparella2002}. The investigation on the unsteady
horizontal convection flow is relatively less, except for
\cite[]{Paparella2002,Mullarney2004,Hughes2007}. However, they
have mainly focused on how the turbulent plume maintains a stable
stratified circulation. Yet how the horizontal convection turned
to be unsteady remains an elusive problem.

To understand this problem, both $Ra_c$ for the onset of unsteady
flow and instability mechanism are of vital. Paparella and Young
\cite{Paparella2002} found $Ra_c\approx 2\times 10^{8}$ at
$\Pran=1$ in their simulations, which is significantly smaller
than others' results. For example, Rossby (1965), Wang and Huang
(2005) found the flow is steady and stable for $Ra<5\times 10^{8}$
in their experiments \cite[]{Rossby1965,WangWei2005}. Yet some
other numerical simulations
\cite[]{Rossby1998,Siggers2004,SunL2006_jhd1} have not found
unsteady flows for $Ra<10^9$. Paparella and Young
\cite{Paparella2002} explained this difference as: (i) lower
aspect ratio ($H/L=1/4$) than the experiments and (ii) middle
plume forcing instead of sidewall plume forcing in the
experiments. Both may lead to destabilization of the flow at lower
Rayleigh numbers. However, their hypotheses have not been
intensely investigated. According to a recent investigation, the
flow is still stable for $Ra<10^{11}$ even at a much lower aspect
ratio ($H/L=1/10$) \cite[]{SunL2007aps}. Thus, it maybe the middle
plume forcing that leads to destabilization at lower Rayleigh
numbers.

Our interest here is to verify their second hypotheses. Is the
flow with middle plume forcing less stable than the sidewall plume
forcing? How the instability occurs? To investigate these
problems, more accurate numerical prediction of $Ra_c$ is need for
both forcing cases, for the spatial resolution of simulation is
very coarse used (e.g. $128\times32$ meshes are used in
\cite{Paparella2002}). Then the flow field under both middle and
sidewall plume forcings are compared, which leads to an
affirmative answer of the above problem.

\begin{figure}
\centerline{
  \includegraphics[width=10cm]{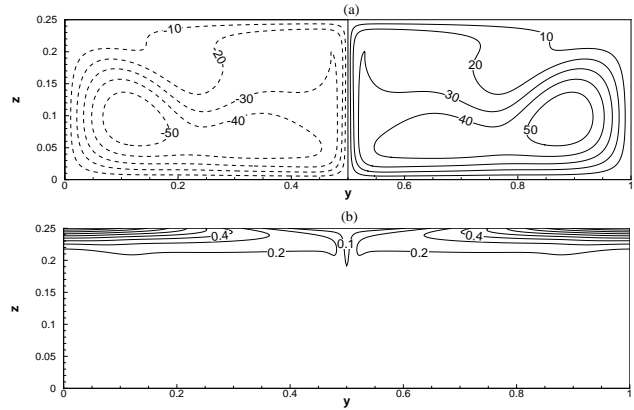}}
  \caption{The flow stream function (a) and temperature field (b) of $Ra=5\times 10^{8}$.
  It is steady and stable and symmetric with middle plume forcing, solid and dashed
  curves for positive and negative values, respectively.}
\label{Fig:HConv_Pr1H025_Ra5E8_PsiDen}
\end{figure}

Similar to the previous investigations, we consider the horizontal
convection flows within the two-dimensional domain, and the
Boussinesq approximation is assumed to be valid for these flows.
As shown in Fig.\ref{Fig:HConv_Pr1H025_Ra5E8_PsiDen}, the
horizontal (y) and vertical (z) regimes are $0\leq y \leq L$ and
$0\leq z\leq H$, respectively. Similar to \cite{Rossby1965}, the
depth $L$ is taken as reference length scale and $A=H/L=1/4$
denotes the aspect ratio. Taking account of nondivergence of
velocity field in Boussinesq approximation, the Lagrangian
streamfunction $\Psi$ and the corresponding vorticity $\omega$ are
introduced. The velocity $\overrightarrow{\mathrm{u}}=(v,w)$,
where horizontal velocity $v=\frac{\parti \Psi}{\parti z}$ and
vertical velocity $w=-\frac{\parti \Psi}{\parti y}$, respectively.
The governing equations in vorticity-streamfunction formulation
are \cite[]{Quon1992,Paparella2002,Siggers2004}:

\begin{subeqnarray} \frac{\partial T}{\partial t} + J(\Psi,T) &=&
(\frac{\parti^2 T }{\parti y^2}+\frac{\parti^2
T }{\parti z^2})\\
\frac{\partial \omega}{\partial t} + J(\Psi,\omega) &=-& \Pran
(\nabla^2 \omega+  \Ra \frac{\parti
T}{\parti y})\\
 \nabla^2  \Psi&=&-\omega
 \label{Eq:thermo_ctl_Horizontal_Onset}
 \end{subeqnarray}
where $J(\Psi,\phi)=\frac{\parti \Psi}{\parti y}\frac{\parti
\phi}{\parti z}-\frac{\parti \phi}{\parti y}\frac{\parti
\Psi}{\parti z}$ denotes the nonlinear advection term. There are
two important dimensionless parameter in
Eq.(\ref{Eq:thermo_ctl_Horizontal_Onset}), i.e. Rayleigh number
$\Ra=\alpha_T \Delta T gL^3/(\kappa \nu)$ and Prandtl number
$\Pran=\nu/\kappa$, where $g$, $\alpha_T$, $\Delta T$, $L$,
$\kappa$ and $\nu$ are gravity acceleration, thermal expansion
coefficient, surface temperature difference, length of horizontal
domain, thermal diffusivity and kinematic viscosity, respectively.
Alternatively, Paparella and Youngs used vertical length $H$ as
length scale, so $Ra=64Ra_H$, where $Ra_H$ is the vertical
Rayleigh number by using vertical length $H$ as unit
\cite[]{Paparella2002}.

More specifically, we consider the horizontal convection in a
rectangle tank at $\Pran=1$. The tank has same velocity boundary
condition as that in \cite{Paparella2002}, i.e. free slip and no
stress at the walls. In addition, two different surface forcings
are used, which are central symmetric. One is middle plume forcing
as $T=[1+\cos(2\pi y)]/2$ \cite{Paparella2002}, the other is
sidewall plume forcing as $T=[1-\cos(2\pi y)]/2$
\cite{Quon1992,SunL2007aps}. Comparing these with one cell forcing
$T=\cos(\pi y/2)$ \cite{Rossby1998}, there are two symmetric cells
in the flow field under such forcings (e.g.
Fig.\ref{Fig:HConv_Pr1H025_Ra5E8_PsiDen} and
Fig.\ref{Fig:HConv_Pr1H025_Ra5E8_PsiDen2}), when the flow is
symmetrically steady and stable. In addition, the middle plume
forcing in the left cell is the same with the sidewall plume
forcing in the right cell (see
Fig.\ref{Fig:HConv_Pr1H025_Ra5E8_PsiDen} behind). Thus in the
steady flows, both forcings will lead to the same flow patterns
except for a position shift, which is proved by the following
investigation.

There are two important quantity describing the circulation, i.e.
the non-dimensional streamfunction maximum and the non-dimensional
heat flux. The non-dimensional streamfunction maximum
$\Psi_{\max}=\Psi^*_{\max}/\nu$, where $\Psi^*_{\max}$ is the
maximum of the dimensional streamfunction.

The above Eq.(\ref{Eq:thermo_ctl_Horizontal_Onset}) is solved with
finite different method in non-uniform grids. Crank-Nicholson
scheme and Arakawa scheme \cite[e.g.][]{Arakawa1966,Olandi2000}
are applied to discretize the linear and nonlinear terms,
respectively. Comparing to the other schemes, Arakawa scheme is
more accuract but more expensive, and it has also been applied to
horizontal convection flows at high Rayleigh number
\cite[]{SunL2006_jhd1,SunL2007aps}. Table
\ref{Table:NatConv_Benchmark} shows the validation of the scheme
with nature convection problem. A fine spatial resolution mesh of
$512\times128$ is used to eliminate numerical instability.

\subsection{Results}

\begin{figure}
\centerline{
  \includegraphics[width=10cm]{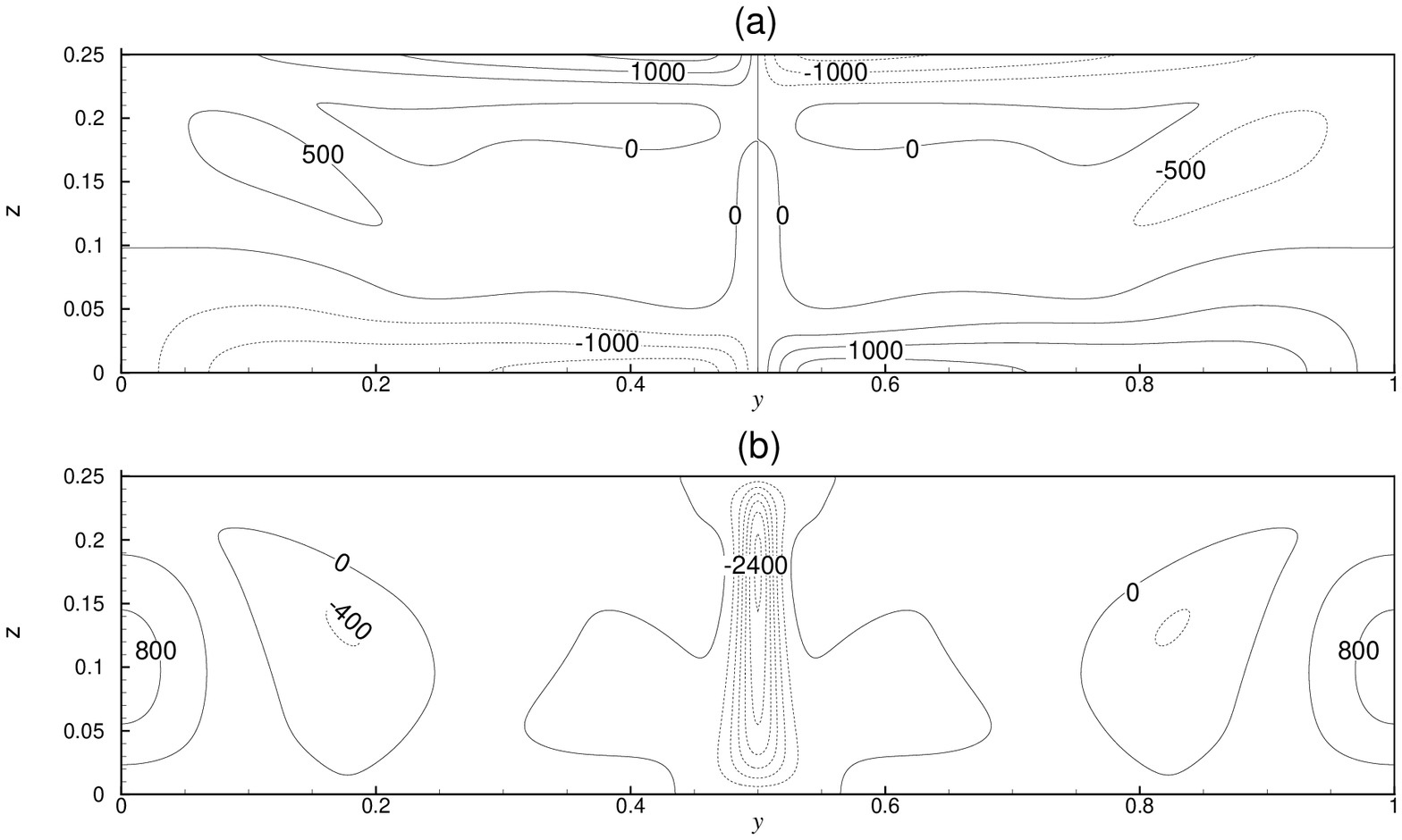}}
  \caption{The horizontal (a) and vertical (b) velocity fields of $Ra=5\times 10^{8}$.
  It is steady and stable and symmetric with middle plume forcing, solid and dashed
  curves for positive and negative values, respectively.}
\label{Fig:HConv_Pr1H025_Ra5E8_VW}
\end{figure}

First, the middle plume forcing is considered, which is steady and
stable for $Ra<5.5\times 10^8$.
Fig.\ref{Fig:HConv_Pr1H025_Ra5E8_PsiDen} shows the flow field (a)
and temperature field (b) of $Ra=5\times 10^{8}$ with
$\Psi_{\max}=59.83$, in which the flow is symmetric, steady and
stable. In this case, the center line symmetrically separates the
flow field into two parts, like a free slip wall. There is a
vigorous downward jet in the center of tank corresponding to the
middle plume forcing (Fig.\ref{Fig:HConv_Pr1H025_Ra5E8_PsiDen}a),
where the vertical velocity field has a minimum of $w=-2513$
(Fig.\ref{Fig:HConv_Pr1H025_Ra5E8_VW}b). The center jet leads to
the clockwise and anticlockwise plume cells in the left and right
part of tank, respectively. In the left circulation cell, the flow
sinks quickly along the center line and upwells clockwise along
the left side wall with relatively slower speed, which can be also
seen from the vertical velocity of the flow
(Fig.\ref{Fig:HConv_Pr1H025_Ra5E8_VW}). Besides, there are two
horizontal jets respectively near top and bottom walls in the left
circulation cell (Fig.\ref{Fig:HConv_Pr1H025_Ra5E8_VW}a). Totally,
there are 2 horizontal jets near wall and a vertical jet at the
center in each cell. Contract to the flow field, the temperature
field is very simple. An obvious boundary layer exists near the
surface in temperature field, which leads to a 1/5-power law of
$Ra$ for heat flux \cite[e.g.][]{Rossby1965,Quon1992,Siggers2004}.
And below the temperature boundary layer, the temperature is
almost homogeneous due to the convection. Thus there is a very
strong stratification near the surface ($\partial T/\partial z\sim
Ra^{1/5}$) but a very weak stratification in other region
($\partial T/\partial z\sim 0$). As the above case is stable, so
the critical Rayleigh number must be larger than $5\times10^{8}$,
which is significantly larger than the value obtained before
\cite{Paparella2002}.

\begin{figure}
\centerline{
  \includegraphics[width=6cm]{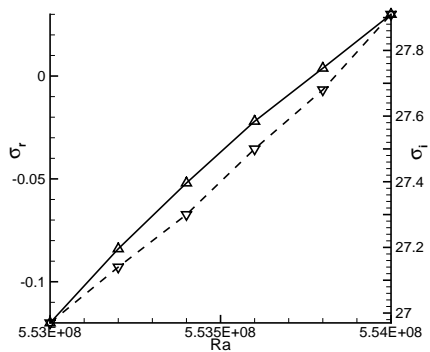}}
  \caption{Growth rate $\sigma_r$ (solid) and $\sigma_i$ (dashed) vs. $Ra$, respectively.
  \label{Fig:HConv_Pr1H025_CrCi_Ra}}
\end{figure}

To find the critical Rayleigh number $Ra_c$, the growth rate of
perturbation $\phi(t)$ is calculated numerically. And $\phi(t)$ is
assumed to satisfy $\phi(t)=e^{\sigma t}\phi(0)$, where
$\sigma=\sigma_r+i\sigma_i$ is the complex growth rate of
disturbance. It is found that the onset of unsteady flow is at
$Ra_c=5.5377\times 10^8$, as shown in
Fig.\ref{Fig:HConv_Pr1H025_CrCi_Ra}. For $Ra=5.53\times 10^8$, the
flow is stable and the growth rate is approximately
$\sigma_r=-0.12$. But the flow is unstable and the growth rate is
approximately $\sigma_r=0.03$ for $Ra=5.54\times 10^8$. Thus the
critical Rayleigh number $Ra_c$ is obtained $5.53\times
10^8<Ra_c<5.54\times 10^8$. The accurate value of
$Ra_c=5.377\times 10^8$ is obtained by interpolating from the
above result. Moreover, the onset of unsteady flow is found to
occur via Hopf bifurcation. As Fig.\ref{Fig:HConv_Pr1H025_CrCi_Ra}
shows, the image part of growth rate is nonzero and the eigenmode
of perturbation is periodic. This Hopf bifurcation of the
horizontal convection has not seen reported yet, and privous
investigations dealt only with chaotic flows.

\begin{figure}
\centerline{
  \includegraphics[width=10cm]{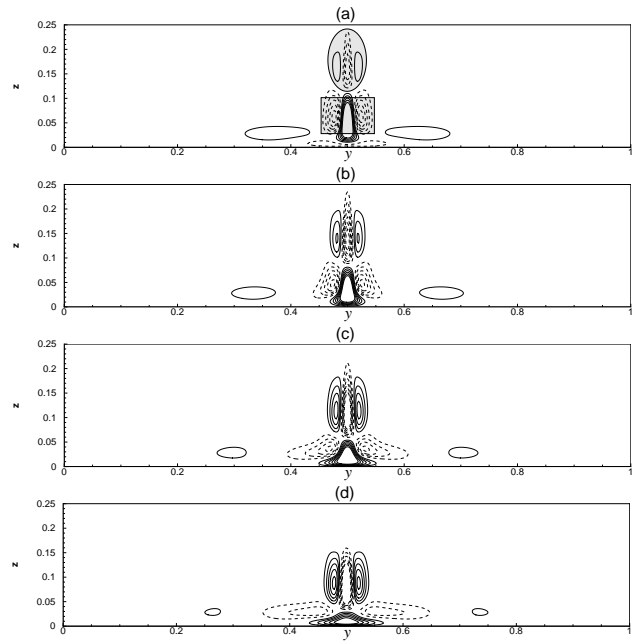}}
  \caption{The perturbational vorticity fields at $t=0$ (a), $t=T/8$ (b),
$t=T/4$ (c) and $t=3T/8$ of $Ra=5.54\times 10^{8}$, solid and
dashed curves for positive and negative values, respectively.}
  \label{Fig:HConv_Pr1H025_Ra540E8_V03}
\end{figure}

\begin{figure}
\centerline{
  \includegraphics[width=10cm]{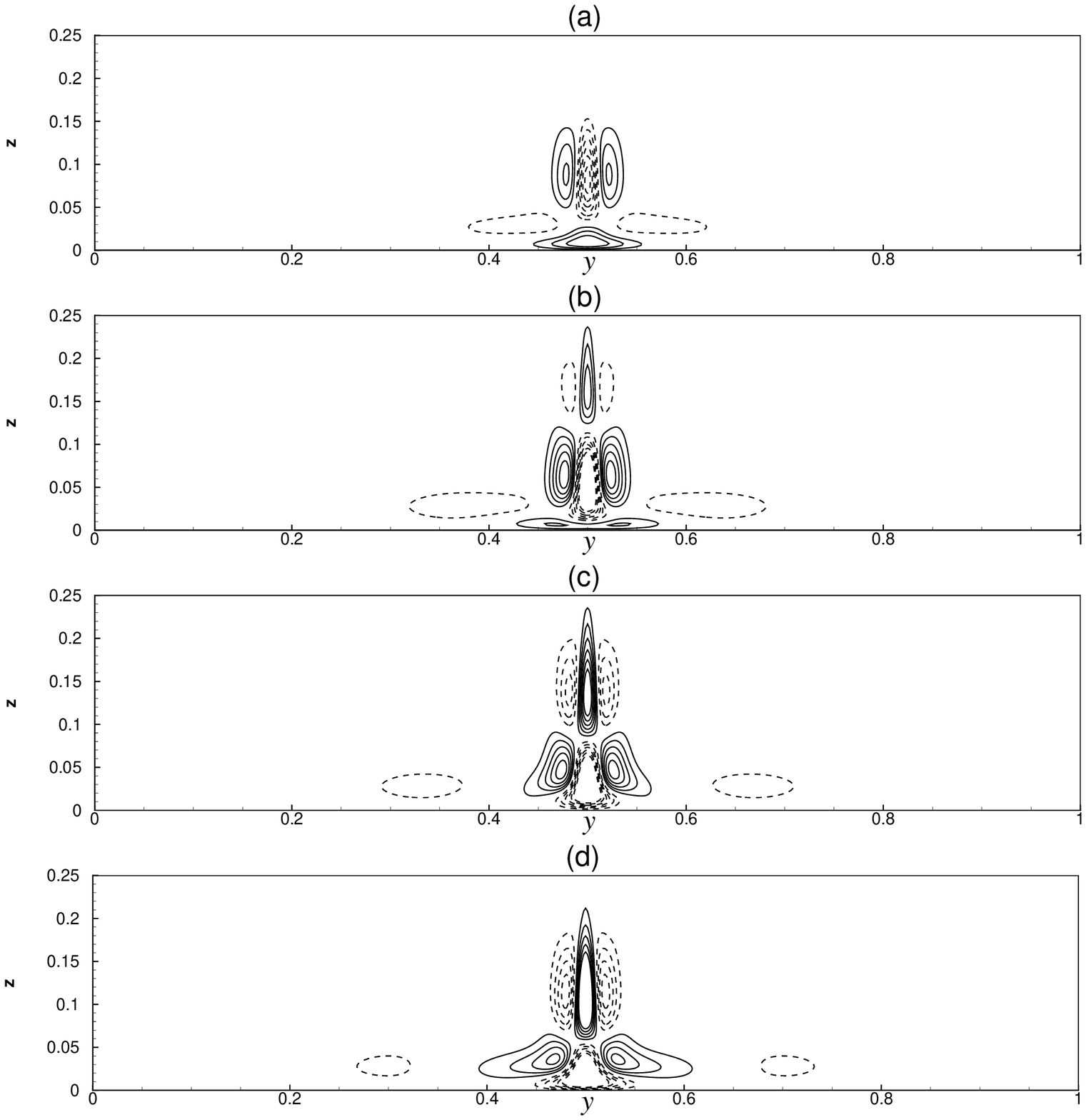}}
  \caption{The perturbational vorticity fields at $t=T/2$ (a), $t=5T/8$ (b),
$t=6T/8$ (c) and $t=7T/8$ of $Ra=5.54\times 10^{8}$, solid and
dashed curves for positive and negative values, respectively.}
  \label{Fig:HConv_Pr1H025_Ra540E8_V47}
\end{figure}

Meanwhile, the evolution of the perturbational vorticity fields
during the fist half period at $t=0$ (a), $t=T/8$ (b), $t=T/4$ (c)
and $t=3T/8$ of $Ra=5.54\times 10^{8}$ are depicted in
Fig.\ref{Fig:HConv_Pr1H025_Ra540E8_V03}, respectively. The
perturbational vorticity fields are symmetric about centerline,
which implies that the horizontal velocity is nonzero at
centerline. It can be seen that the perturbation tripole A (the
shadowed ellipse in Fig.\ref{Fig:HConv_Pr1H025_Ra540E8_V03}a) is
generated from central downward jet, then propagates and amplifies
along the central jet downward to the bottom wall
(Fig.\ref{Fig:HConv_Pr1H025_Ra540E8_V03}b,c,d). When tripole A
approaching to the bottom, it becomes weaker and weaker and breaks
into two parts: the left and the right near the bottom, which can
be seen from the evolution of tripole B (the shadowed rectangle in
Fig.\ref{Fig:HConv_Pr1H025_Ra540E8_V03}a). And the mean flow
advects the broken vortexs horizontally along the bottom wall
(Fig.\ref{Fig:HConv_Pr1H025_Ra540E8_V03}b,c,d). Then in the second
half period, a reverse tripole will generate right the same place
of vortexes A at $t=T/2$, and the same story repeats for it, which
is shown in Fig.\ref{Fig:HConv_Pr1H025_Ra540E8_V47}. In short, the
perturbations generate and amplify in the central vertical jet,
but are propagated and weakened along the horizontal wall.

\begin{figure}
\centerline{
  \includegraphics[width=10cm]{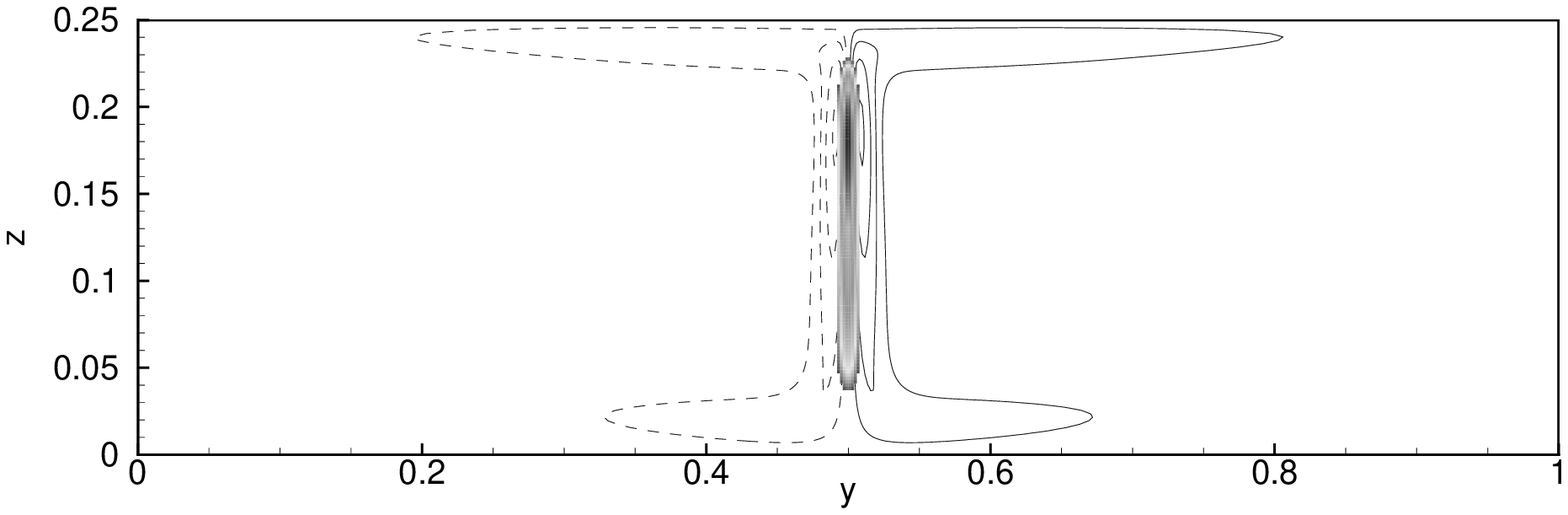}}
  \caption{The vorticity of $Ra=5.54\times 10^{8}$ with vertical velocity $w$ (shadowed  as $w>1800$),
solid and dashed curves for positive and negative values,
respectively.}
  \label{Fig:HConv_Pr1H025_Ra540E8_VorW}
\end{figure}

Further investigation shows that the instability of flow occurs
due to shear. First, as we noted, the instabilities always occur
in the center and propagate along the mean flow. Second, this
trigger place locates in the area where there is a vigorous jet
with strong shear (see e.g. Fig.\ref{Fig:HConv_Pr1H025_Ra5E8_VW}b
and Fig.\ref{Fig:HConv_Pr1H025_Ra540E8_VorW})
 (Fig.\ref{Fig:HConv_Pr1H025_Ra540E8_VorW}).  As the
stratification is very weak here (see e.g.
Fig.\ref{Fig:HConv_Pr1H025_Ra5E8_PsiDen}b), so that the flow in
this region is dominated by momentum dynamics other than thermal
dynamics. All these imply that the onset of instability leading to
unsteady flow is due to shear instability at larger Raleigh
numbers, which is much different from Rayleigh-B\'{e}nard
instability. However, shear is not the only sufficient condition
for instability. For example, the instability near the top surface
is suppressed due to strong stratification
(Fig.\ref{Fig:HConv_Pr1H025_Ra5E8_PsiDen}b), though both the
velocity (Fig.\ref{Fig:HConv_Pr1H025_Ra5E8_VW}a) and the shear
(Fig.\ref{Fig:HConv_Pr1H025_Ra540E8_VorW}) near top surface are
still very large. In words, the onset of instability is due to
velocity shear (shear instability) other than thermally dynamics
(thermal instability).

\begin{figure}
\centerline{
  \includegraphics[width=10cm]{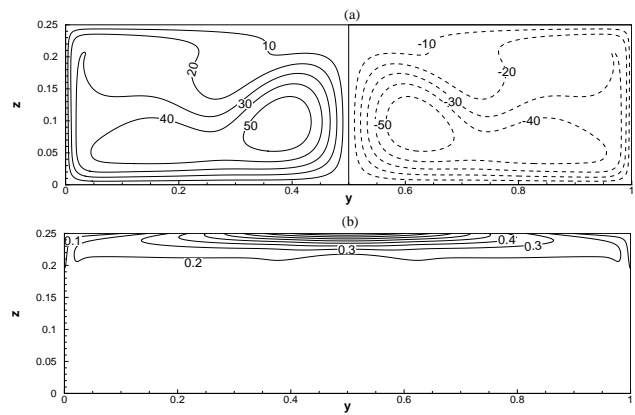}}
  \caption{The flow field (a) and temperature field (b) of $Ra=5\times
  10^{8}$,
  which are steady and stable and symmetric with sidewall plume forcing.}
\label{Fig:HConv_Pr1H025_Ra5E8_PsiDen2}
\end{figure}

Second, the sidewall plume forcing is considered.
Fig.\ref{Fig:HConv_Pr1H025_Ra5E8_PsiDen2} shows the flow field and
temperature field of $Ra=5\times 10^{8}$, in which the flow is
symmetric, steady and stable like that under middle plume forcing.
There are two strong downward jets near the walls corresponding to
the sidewall plume forcing
(Fig.\ref{Fig:HConv_Pr1H025_Ra5E8_PsiDen2}a). As mentioned above,
the sidewall plume forcing will lead to exactly the same flow
pattern as the middle plume forcing does except for a position
shift, which can be seen from
Fig.\ref{Fig:HConv_Pr1H025_Ra5E8_PsiDen} and
Fig.\ref{Fig:HConv_Pr1H025_Ra5E8_PsiDen2}. As the flow is stable,
the center line like a free slip wall symmetrically separates the
two cells. The left cell in
Fig.\ref{Fig:HConv_Pr1H025_Ra5E8_PsiDen} is exactly the same with
the right cell in Fig.\ref{Fig:HConv_Pr1H025_Ra5E8_PsiDen2}.
However, the flow is much more stable with the sidewall plume
forcing. And the critical Rayleigh number is found $Ra_c\approx
1.85\times 10^{10}$ (with $768*192$ meshes) in this case. As noted
above, the flow is much more stable with the sidewall plume
forcing than that with the middle plume forcing, though both
forcings lead to the same flow patterns. This is very interesting,
and can be understood from the mechanism of instability.

It's found that the rigid wall suppresses the perturbation, which
leads a more stable flow with the sidewall plume forcing than that
with the middle plume forcing. As the flow loss stability is due
to strong velocity shear in the center in horizontal convection,
the smaller of the shear the more stable of the flow. In the case
of middle plume forcing, the perturbation with nonzero horizontal
velocity occurs at strongest downward jet. And the perturbed flows
cross the center line and propagate downstream. However, in the
case of sidewall plume forcing, these crossing flows are
suppressed by rigid walls. So that the critical Rayleigh number is
much larger in this case. Paparella and Young (2002) hypothesized
that middle plume forcing may lead to a destabilization of the
flow. Here this hypotheses is proved both physically and
numerically.

\subsection{Conclusion}

In conclusion, the onset of unsteady flow is found to occur via a
Hopf bifurcation in the regime of $Ra>Ra_c=5.5377\times 10^8$ for
the middle plume forcing at $Pr=1$, which is much larger than the
previously obtained value. Besides, the onset of unsteady flow is
due to shear instability of central downward jet. Finally, the
second hypotheses of Paparella and Young (2002) for instability is
numerically approved, i.e. the middle plume forcing can lead to a
destabilization of the flow at relatively lower Rayleigh numbers.

\section*{Acknowledgements}

This work was original from author's dream of understanding the
mechanism of instability in the year 2000, when the author was a
graduated student and learned the course of hydrodynamic stability
by Prof. Yin X-Y at USTC (China). The author thanks Prof. Sun D-J
at USTC (China), Dr. Yue P-T at UBC (Canada) for their help on
preparing the report.

The support of NSFC (No. 40705027 and No. 10602056) and the
National Science Foundation for Post-doctoral Scientists of China
are gratefully acknowledged. The author would like to acknowledge
Professor Wang W at OUC (China) and Professor Huang R. X. at WHOI
(USA) for the useful comments and suggestions on the studies of
horizontal convection during the preparation of the second part of
the report.

\backmatter

\addcontentsline{toc}{chapter}{Reference}

\clearpage
\end{document}